\documentclass[graybox]{svmult}
\usepackage{mathptmx}
\usepackage{amssymb}
\usepackage{helvet}
\usepackage{courier}
\usepackage[pdftex]{graphicx}
\usepackage{makeidx}
\usepackage{multicol}
\usepackage[bottom]{footmisc}
\usepackage{url}
\usepackage{framed}
\usepackage{natbib}
\newcommand{\Ms}{\ensuremath{M_{\odot}}}
\newcommand{\eg}{{\it e.g.}}
\newcommand{\cf}{{\it c.f.~}}
\newcommand{\ie}{{\it i.e.}}
\newcommand{\viz}{{\it viz.}}
\newcommand{\beq}{\begin{equation}}
\newcommand{\eeq}{\end{equation}}
\newcommand{\mtot}{\ensuremath{M_{tot}}}
\newcommand{\mstar}{\ensuremath{M_\ast}}
\newcommand{\mg}{\ensuremath{M_g}}
\newcommand{\vg}{\ensuremath{v_g}}
\newcommand{\tg}{\ensuremath{\tau_g}}
\newcommand{\td}{\ensuremath{\tau_d}}
\newcommand{\tcr}{\ensuremath{\tau_{cr}}}
\newcommand{\mcl}{\ensuremath{M_{cl}}}
\newcommand{\rh}{\ensuremath{r_h}}
\newcommand{\tin}{\ensuremath{t_{in}}}
\newcommand{\tmrg}{\ensuremath{t_{mrg}}}
\newcommand{\tvir}{\ensuremath{\tau_{\rm vir}}}
\newcommand{\tvrx}{\ensuremath{t_{\rm vrx}}}
\newcommand{\sigoned}{\ensuremath{\sigma_{\rm 1d}}}
\newcommand{\mnras}{{\it MNRAS}}
\newcommand{\aap}{{\it A\&A}}
\newcommand{\apj}{{\it ApJ}}
\newcommand{\apjl}{{\it ApJ}}
\newcommand{\apjs}{{\it ApJS}}
\newcommand{\aj}{{\it AJ}}
\newcommand{\araa}{{\it ARA\&A}}
\newcommand{\nat}{{\it Nature}}
\begin{document}

\title*{Formation of Very Young Massive Clusters and implications for
globular clusters}
\author{Sambaran Banerjee \and Pavel Kroupa\\
\vspace{0.8 cm}
To be published in \emph{The Origin of Stellar Clusters}, ed. S. Stahler (Springer)}
\authorrunning{Banerjee \& Kroupa}
\institute{Sambaran Banerjee \at AIfA/HISKP, University of Bonn, Auf dem H\"ugel 71,
D-53121 Bonn, Germany, \email{sambaran@astro.uni-bonn.de}
\and 
Pavel Kroupa \at HISKP, University of Bonn, Auf dem H\"ugel 71,
D-53121 Bonn, Germany, \email{pavel@astro.uni-bonn.de}
}

\maketitle

\abstract{
How Very Young Massive star Clusters (VYMCs; also known as ``starburst'' clusters),
which typically are of $\gtrsim10^4\Ms$ and are a few Myr old, form
out of Giant Molecular Clouds is still largely an open question.
Increasingly detailed observations of young star clusters and star-forming molecular clouds
and computational studies provide clues about their formation
scenarios and the underlying physical processes involved. This chapter is focused
on reviewing the decade-long studies that attempt to computationally reproduce
the well-observed nearby VYMCs, such as the Orion Nebula
Cluster, R136 and NGC 3603 young cluster, thereby shedding light on birth conditions
of massive star clusters, in general. On this regard, focus is given on direct N-body
modeling of real-sized massive star clusters, with
a monolithic structure and undergoing residual gas expulsion,
which have consistently
reproduced the observed characteristics of several VYMCs and
also of young star clusters, in general.
The connection of these relatively simplified model calculations
with the structural richness of dense molecular clouds and the complexity
of hydrodynamic calculations of star cluster formation is presented in
detail. Furthermore, the connections of such VYMCs with
globular clusters, which are nearly as old as our Universe, is discussed.
The chapter is concluded by addressing long-term deeply gas-embedded (at least apparently)
and substructured systems like W3 Main.     
While most of the results are quoted from existing and up-to-date literature,
in an integrated fashion, several new insights and discussions are
provided.  
}

\section{Introduction}\label{intro}

Very Young Massive Clusters (hereafter VYMCs) refer to a sub-category of star clusters
which are $\gtrsim10^4\Ms$ heavy (\ie, massive) and a few Myr old in age,
typically 1-3 Myr (\ie, very young).
\footnote{These objects are also popularly called ``starburst'' clusters. We prefer
to call them Very Young Massive Clusters based on their characteristic properties,
instead of referring to their likely starburst origin. The latter criterion may coincide
with the origins of other types of massive clusters, \eg, globular clusters.
VYMCs constitute the youngest sub-category of Young Massive Clusters (YMCs; \citealt{pz2010}).}
A number of such star clusters are observed in the molecular gas-dominated spiral arms
(\eg, the NGC 3603 Young Cluster) and in the central molecular zone (\eg, the 
Arches and the Quintuplet clusters) of our Galaxy. They are also found in nearby disk
galaxies of the local group (\eg, the R136 cluster of the Large Magellanic Cloud)
and in ``starburst galaxies'' (\eg, in the Antennae Galaxies). An age $\lesssim3$ Myr
would imply that all of the massive stellar members of a VYMC are
in their main sequences (MSs). The key importance of VYMCs is that being newly hatched,
the details of their structure and internal kinematics can constrain the conditions
under which massive star clusters, which are globular clusters
at their infancy \citep{mk2012,kru2014}, form. This allows one to distinguish between the
different scenarios of massive cluster formation \citep{longm2014}.
Note that the above mass and age limits defining VYMCs are meant to be generally true
(see also \citealt{pz2010}) but not
absolutely rigorous, to allow some well studied young systems, like the ONC ($\approx10^3\Ms$),
to be counted in as VYMCs.  

Morphologically, VYMCs are often found as the richest core-halo member cluster of extended
cluster complexes/stellar associations, \eg, the ONC \citep{ab2012}. Young stellar systems are also
found as extended associations of OB stars, \eg, the Cygnus OB2 \citep{kuhn2014,wrt2014}. It
is also common to find them surrounded by HII (ionized hydrogen) gas
(\eg, NGC 3603 and R136; \citealt{pang2013}). As for
the sizes, the half-light radii of VYMCs are $\lesssim1$ pc, \ie, they are
typically a factor of three more compact than Galactic globular clusters.
This is consistent with VYMCs being infant globular clusters
as their subsequent evolution
due to mass segregation, dynamical encounters among stars and stellar binaries and
stellar evolution would expand them. A handful of systems are found near-embedded
in gas and highly compact; RCW 38 is a classic example where the HII-gas
appears to has just begun releasing itself from the cluster (see below), exposing
only the cluster's central part \citep{derose2009}. Interestingly, several VYMCs are
found to contain multiple density centers, \ie, substructures, despite having
an overall spherical core-halo morphology \citep{kuhn2014}.

How (near) spherical parsec-scale VYMCs form out of vast, irregular molecular-hydrogen
clouds is being widely debated for at least the past 10 years. Even without invoking
any specific formation scenario, it can be said that dynamical relaxation
(\ie, statistical energy exchange among stars) must play
a critical role in shaping the spherical core-halo structure of a VYMC. Hence,
any formation channel for VYMCs must allow enough room for dynamical relaxation
of the final (at an age of 1-3 Myr) stellar assembly.
The overall two-body relaxation time, that
determines the secular (near dynamically stable) evolution, is typically
several Gyr for a VYMC, which is much longer than its age. Hence, the 
present-day morphology of a VYMC is primarily dictated by what
is called ``violent relaxation'' \citep{spz}. The latter process refers to
the energy redistribution among stars due to mutual encounters 
and rapid changes of the gravitational potential, leading
to (near) dynamical equilibrium or ``virialization'' of the
system, which happens in the timescale of stellar orbits (or in
the dynamical timescale; \citealt{spz,hh2003}),
\footnote{This timescale is commonly represented by the orbital
``crossing time'' which is the time taken to traverse
the spatial scale of the system (say, its half mass diameter)
by a particle moving radially with the dispersion speed.}
\ie, typically in a fraction of a Myr. The observed \emph{lack}
of an age range among the members of the youngest star clusters
(see, \eg, \citealt{bsv2013,holly2015})
implies that these stars must have formed in a burst
and integrated into a cluster over a short period of time. 

Currently, there exist apparently at
least two distinct scenarios for the formation of VYMCs. The ``monolithic'' or
``episodic'' or ``in-situ'' (top-down) scenario implies
the formation of a compact star cluster in an
essentially single but highly active star-formation episode (a ``starburst'').
The infant cluster of pre-main-sequence (PMS) and
main sequence (MS) stars remains embedded in its parent
molecular gas cloud. The latter eventually gets ionized
by the UV radiation from the massive stars and receives energy from stellar
mass outflows and due to coupling with stellar radiation.
Such energy injection eventually causes the embedding gas
to become gravitationally unbound from the system and to
disperse in a timescale typically comparable
to the dynamical time of the stellar system, \ie, too fast for the stars 
to adjust with the corresponding depletion of the potential well.
This causes the stellar system to expand violently and lose
a fraction of its stars depending on its initial mass and concentration
\citep{lmagd84,adm2000,bokr2003a,bokr2003b,bk2007}. The remaining system
may eventually regain virial or dynamical equilibrium (re-virialization);
hence a given VYMC may or may not be in
equilibrium depending on the time taken to re-virialize and
the the epoch of its observation \citep{sb2013}. Such a monolithic or
top-down cluster formation scenario has successfully explained
the details of well observed VYMCs, \eg, ONC (also the Pleiades;
\citealt{pketl2001}), R136 \citep{sb2013} and the NGC 3603
young cluster \citep{sb2014}.

Alternatively, VYMCs are thought to have formed ``bottom-up'' through 
hierarchical merging of less massive subclusters \citep{longm2014}.
Several of such subclusters fall onto each other
and coalesce to form the final VYMC. The gravitational potential of the background
molecular gas within which these subclusters appear makes the infall faster
(the so-called ``conveyor belt mechanism''; \citealt{longm2014}).
The observational motivation for such a scenario is the apparent
substructures in OB associations and even in VYMCs with
overall core-halo configurations \citep{kuhn2014}.

As of now, star formation has been studied
in hydrodynamic calculations involving
development of seed turbulences, in cubical/spherical gas clouds,
into high-density filaments where star formation
occurs as a result of gravitational collapse and fragmentation \citep{kl1998,bate2004,giri2011}.
In all such smoothed-particle-hydrodynamic (SPH) calculations,
a hydrodynamic ``sink particle''\footnote{A ``sink particle'' is a dense, self-gravitating
region in a fluid field approximated by a point mass for facilitating calculations \citep{kl1998}.
A sink particle can only grow in mass by accreting matter from its surrounding.}
is physically associated with a proto-star.
In these computations, clusters of proto-stars are formed
within high-density filaments and/or filament junctions,
which then fall collectively into the gravitational potential well of the cloud 
to form larger (gas-embedded) clusters (\eg, in \citealt{bate2009} and \citealt{giri2011}).
Different groups have reached the state-of-the-art of such calculations
by including different details of the relevant physical processes but for mass scales
much lighter than VYMCs. Such SPH calculations,
requiring very high particle resolution, is prohibitive for the mass range of VYMCs ($>10^4\Ms$). 

High-resolution (reaching the ``opacity limit'') SPH computations
have so far been done forming stars in spherical gas clouds of up to
$\approx500\Ms$ only \citep{kl1998,bate2004,bate2009,giri2011,giri2012,bate2012}
but without any feedback and hence self-regulation mechanism,
which is critical in determining the star formation efficiency (SFE).
Radiation-magnetohydrodynamic (MHD) calculations
including stellar feedback (radiation and matter outflows)
to the star-forming gas have also been carried out from proto-stellar scales
\citep{mnm2012,bate2013} up to $\approx50\Ms$ gas spheres \citep{PriceBate2010}.
While the latter studies provide insights into the self-regulation mechanisms
in the star formation process and point to an SFE near $30$\%,
the scenario of the ultimate dispersal of the residual gas
still remains superficial. See \citet{krm2014} for an up-to-date review.
Note, however, that
the gas must disperse from the region in the molecular cloud where the
cluster ultimately assembles, to obtain a gas-free young cluster
like what we see today.

Therefore, as it turns out, the majority of the published studies to date
related to the formation and evolution of ``real-sized'' VYMCs 
treat the gas-dispersal phase by including a time-varying external analytical
potential (see Sec.~\ref{gasmodel}) mimicking the residual gas
(\eg, \citealt{adm2000,pketl2001,bk2007,pfkz2013,sb2013,sb2014}).
This captures the essential dynamical effects of the gas dispersal.
The dynamical evolution of the cluster, however, is treated accurately
using direct N-body integration \citep{ars2003}, in most of such works.
This approach has successfully explained several well observed VYMCs,
\eg, the Galactic ONC \citep{pketl2001}
and NGC 3603 young cluster \citep{sb2014} and R136 \citep{sb2013} of the LMC.
Such studies point to a universal SFE of
$\epsilon\approx33$\% and a near-sonic dispersal of the residual HII gas 
(see \citealt{sb2013} and references therein), remarkably reproducing the measured kinematic
and structural properties of these clusters.

On the other hand, the dynamical process of coalescence
of subclusters into more massive clusters has also been studied recently using direct
N-body calculations in both absence (\eg, \citealt{fuji2012}) and presence 
(\eg, \citealt{sm2013}) of
a background gas potential. The role of this process is also investigated in the context
of formation of dwarf galaxies through merger of young massive clusters
\citep{pk1998,fk2005,ams2014}.

The goal of the present chapter is to comprehend the most recent studies
on the formation of VYMCs. Although most of these works treat the star forming
gas indirectly as mentioned above, they incorporate the mass and spatial
scales appropriate for VYMCs and capture the essential physics at the
same time. Extrapolation
of the results from hydrodynamical calculations over a large mass range,
\ie, from the computationally accessible masses (see above) to 
the realistic values, is unreliable since the scalings of all the
relevant physical processes are not known and also they scale differently.
This leaves us with the analytical treatment of the residual gas (Sec.~\ref{gasmodel})
as the only viable option to directly treat VYMC-scale systems, given the
present state of technology. 

In Sec.~\ref{monoform} of this chapter, we discuss the monolithic formation scenario in greater detail. We focus
on those (theoretical) studies that have addressed well observed VYMCs.   
The central young cluster HD 97950 (hereafter HD97950) of the Galactic NGC 3603 star-forming region
is always of particular interest in this regard since, due to its proximity,
it is perhaps the best observed VYMC. Next, in Sec.~\ref{himrg}, we move on to discuss further
on the hierarchical formation of VYMCs. Here, we again focus on HD97950
cluster whose structure is known in detail. This enables us to directly compare
the two formation channels and put constraints on the initial conditions in each case.  
In Sec.~\ref{discuss}, we conclude this chapter by discussing how
VYMCs can be related to embedded clusters.

\begin{framed}
Technology, at present, does not permit self-consistent hydrodynamic calculations
of star cluster formation, with adequate resolution and
including feedback at the same time, for masses relevant for young massive clusters
($\gtrsim10^4\Ms$). 
While such hydrodynamic calculations are doable with gas clouds of much lower masses
(up to $100\Ms$s), a large extrapolation is grossly unreliable since
the physical processes involved are not all well understood (\eg, gravitational
fragmentation, role of magnetic field) and they scale differently.  
An analytic treatment of the gas combined with N-body calculation of the
star cluster is, at present, the only viable way to reach
such mass scales.
\end{framed}

\begin{figure}
\centering
\includegraphics[width=9.5cm]{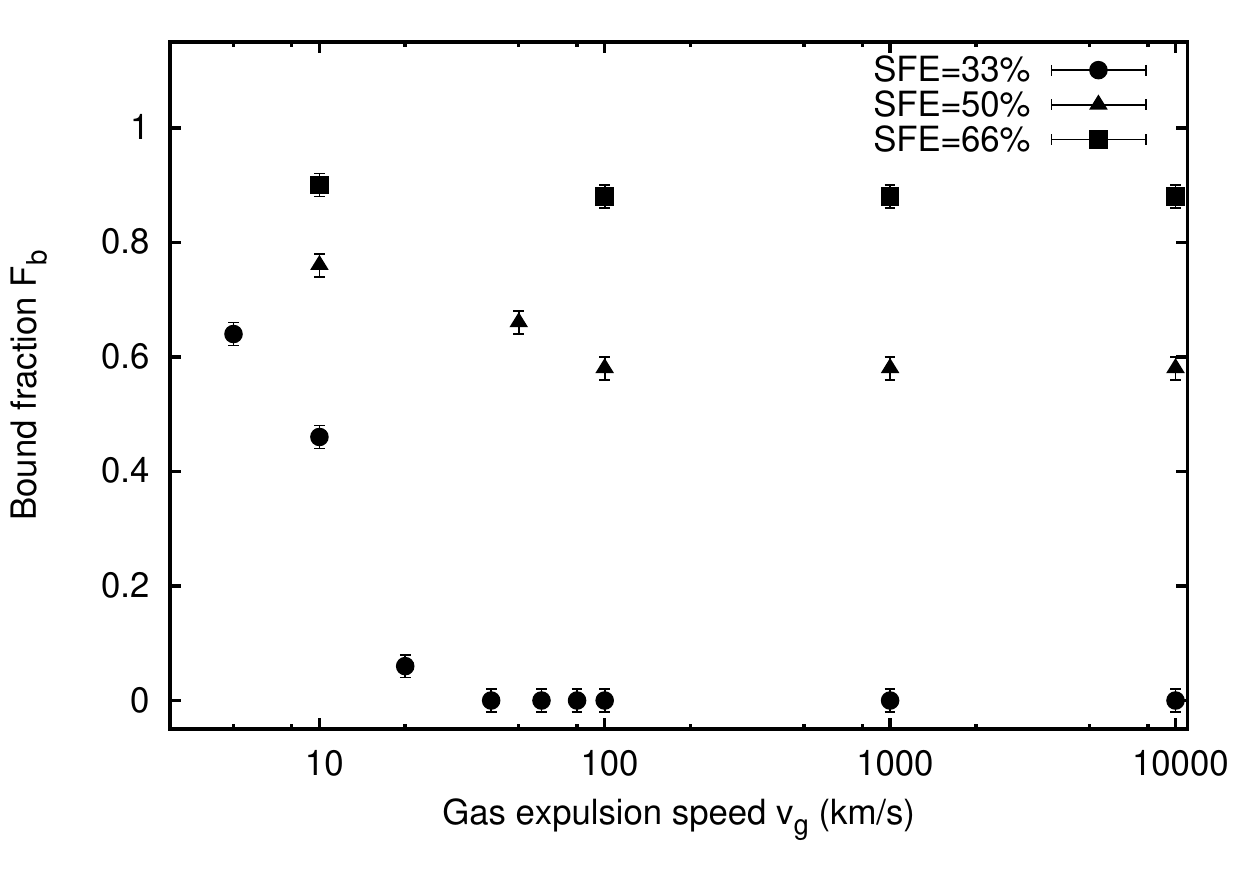}
\includegraphics[width=9.5cm]{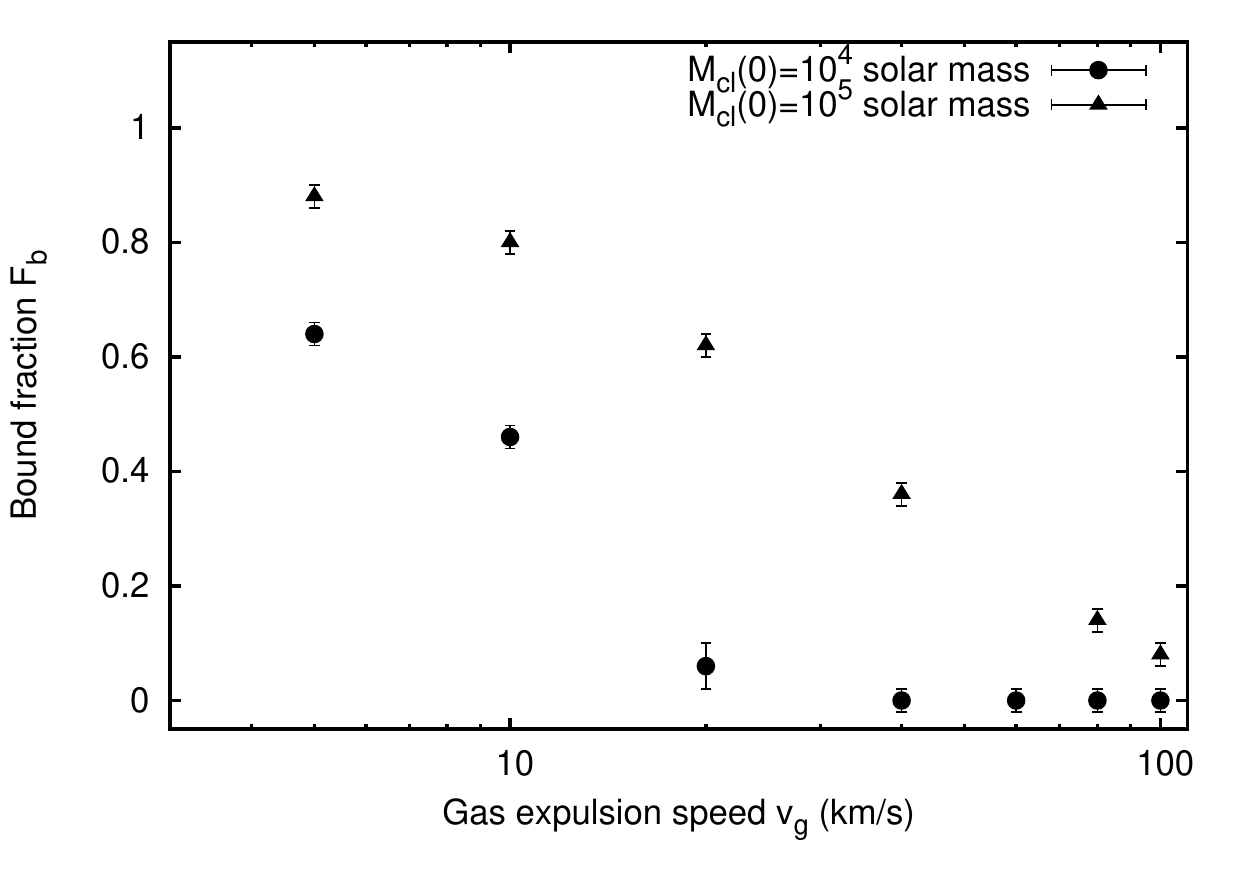}
\caption{Bound fraction, $F_b$, of a star cluster 
as a function of the overall gas removal speed, $v_g$, with varying star formation
efficiency, $\epsilon$ ($\mcl(0)=10^4\Ms$, $\rh(0)=0.3$ pc; top panel), and
initial stellar cluster mass, $\mcl(0)$ ($\epsilon=0.33$, $\rh(0)=0.3$ pc;
bottom panel). All the initial clusters follow a Plummer density distribution.
These results are obtained from {\tt NBODY6} computations. See text for details.
The authors thank Nina Brinkmann of the AIfA, Bonn for providing aid in
preparing this figure.}
\label{fig:bf}
\end{figure}

\begin{figure}
\centering
\includegraphics[width=9.5cm]{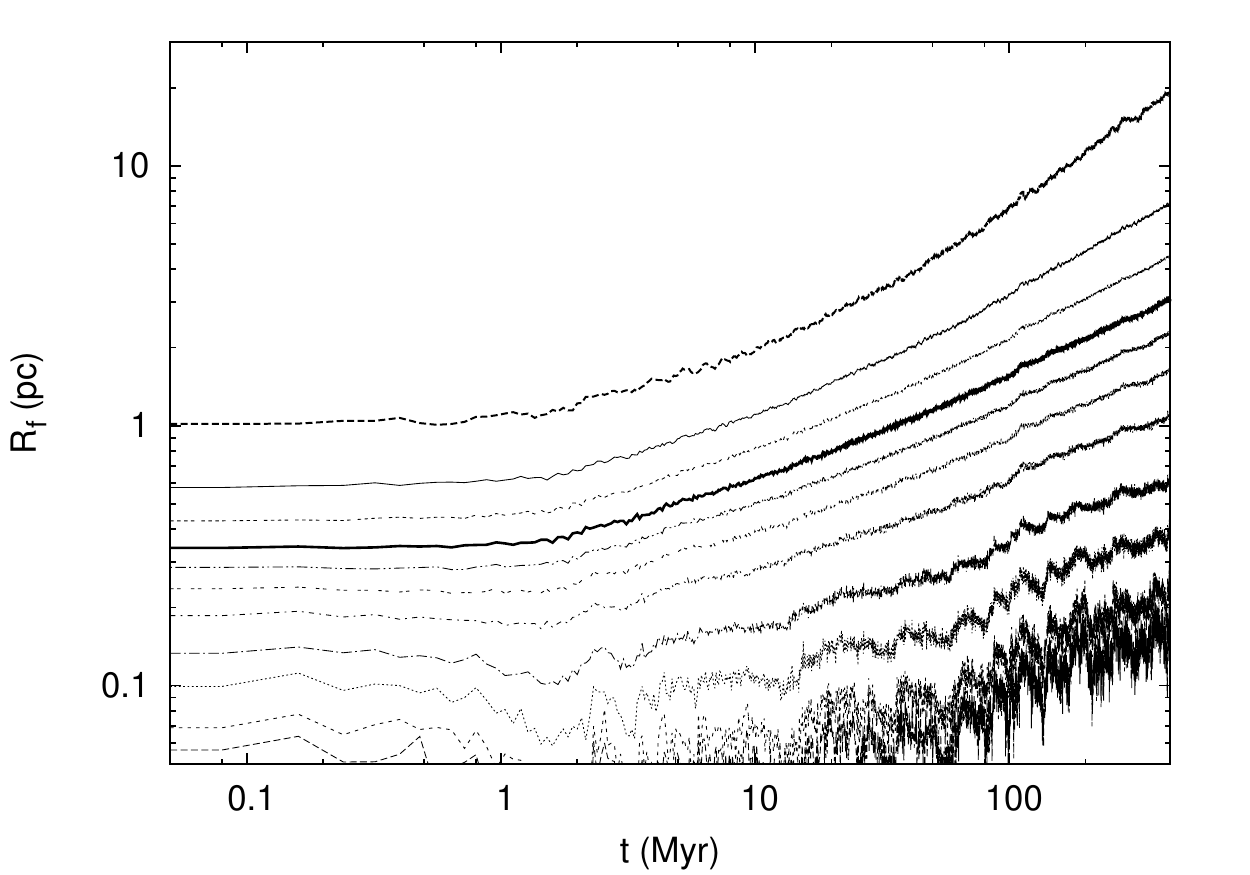}
\includegraphics[width=9.5cm]{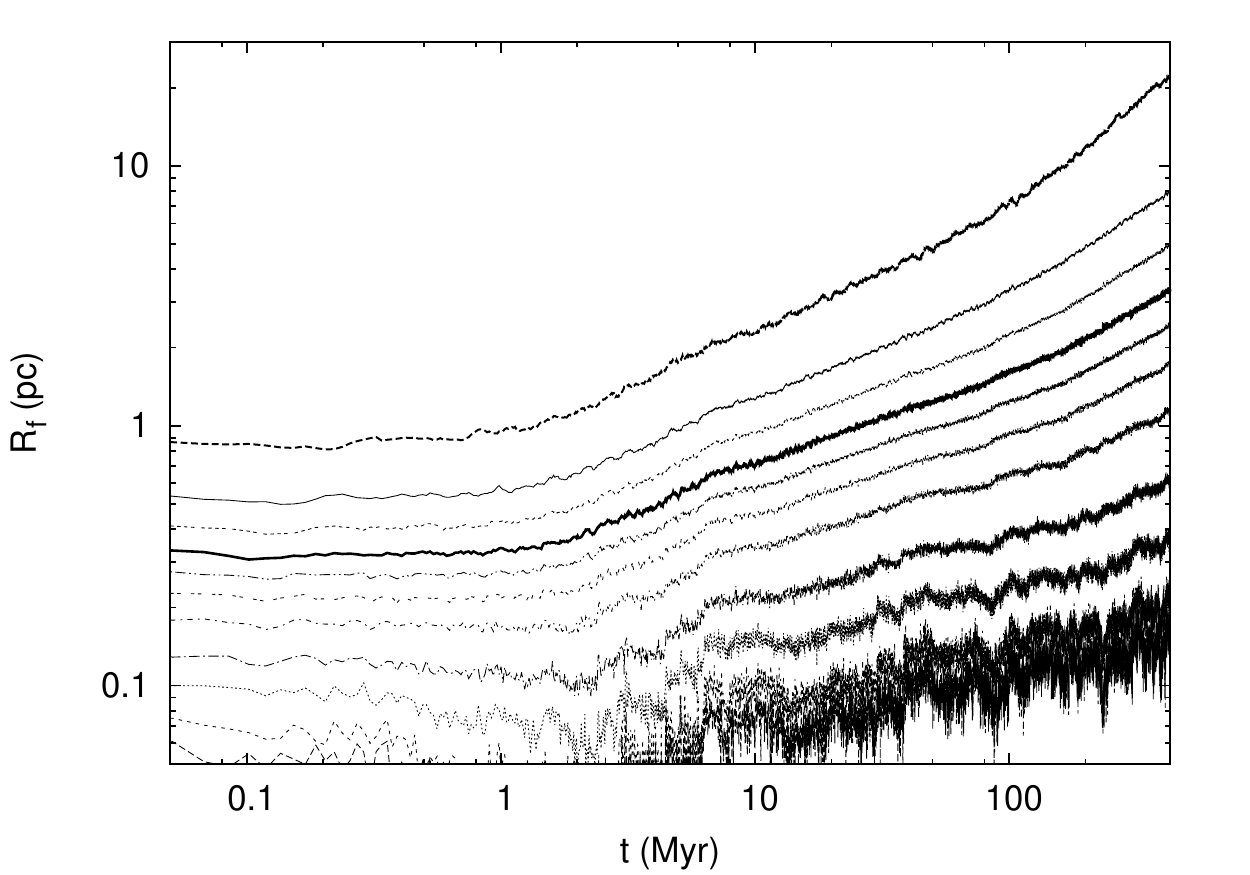}
\caption{Evolution of the overall size of an initial Plummer star cluster, with initial
mass $\mcl(0)\approx1.3\times10^4\Ms$ and half mass radius
$\rh(0)\approx0.3{\rm~pc}$, as shown by the Lagrange radii. 
These clusters are taken to be isolated. The upper
and the lower panels are obtained from {\tt NBODY6} calculations
without and with primordial binaries (\citealt{pk1995b}; see text) respectively. These
computed clusters are \emph{not} subjected to any residual gas expulsion and
their evolution is solely due to dynamical processes (two-body relaxation,
close encounters and ejections) and stellar evolution. These lead to an
overall slow expansion of the cluster due to dynamical heating and
mass loss. In both panels,
the curves are the Lagrange radii $R_f$ of mass fraction $f$ where
$f=$ 0.01, 0.02, 0.05, 0.1, 0.2, 0.3, 0.4, 0.5, 0.625, 0.7 and 0.9
from bottom to top respectively. The thick solid line
is therefore the half mass radius, $\rh(t)$, of the cluster.
The secular evolution hardly expands the initially compact
cluster within 3 Myr age and it takes several 100 Myr to expand
it by several factors (in terms of $\rh$), both in absence or presence
of a realistic population of primordial binaries. 
}
\label{fig:lagrevol}
\end{figure}

\section{Monolithic or episodic formation of Very Young Massive Clusters}\label{monoform}

Before going into any details of modelling, one can obtain preliminary estimates      
that signifies the role of violent relaxation in the formation of VYMCs. For a self-bound
system in dynamical (or virial) equilibrium with total K.E., $T$, and total P.E., $V$, the
virial condition is satisfied \citep{spz}, \ie,
\beq
2T=-V.
\label{eq:vir}
\eeq
For an ``explosive'' gas expulsion (see Sec.~\ref{intro}; \citealt{pk2005}),
\ie, gas removal in a timescale, $\tg$,
smaller or comparable to the crossing time, $\tcr$, of the system (see below), $T$
remains nearly unchanged right after the gas is removed although the P.E.
drops to $V^\prime$. The scale length of the system, $r_h$, usually taken as its
half mass radius, also remains nearly unchanged.

For the overall system to remain bound after the gas expulsion, one must have,
\beq
T+V^\prime\leq0.
\label{eq:inq1}
\eeq
If $M$ and $M^\prime$ are the total systemic masses before and
just after the mass depletion respectively, then $V=-GM^2/\rh$ and $V^\prime=-GM^{\prime^2}/\rh$.
Hence, using Eqn.~\ref{eq:vir}, 
\beq
M^{\prime^2}\geq \frac{M^2}{2},
\label{eq:inq2}
\eeq
or
\beq
M^\prime \gtrsim 0.7M.
\label{eq:inq3}
\eeq
For the present case, $M=\mtot$ is the total mass of the residual gas and the stars in
the cluster, before the gas expulsion, and $M^\prime=\mstar$ is the total stellar mass after
the depletion. Hence,
\beq
\mstar \gtrsim 0.7\mtot.
\label{eq:inqM}
\eeq
In other words, for the cluster to survive the gas expulsion, its SFE should be
$\epsilon=\mstar/\mtot\gtrsim70$\%. This requirement is in contrast with realistic 
values of SFE which is $\epsilon\lesssim30$\% as supported by both observations \citep{lnl2003}
and theoretical studies (\citealt{mnm2012,bate2013}; see below). This alone would
invalidate the monolithic cluster formation scenario since all clusters would
dissolve even for the maximum SFE.

In practice, however, violent relaxation
\footnote{In a self-gravitating system which is in dynamical equilibrium,
the orbital energy exchange among stars by two-body encounters 
occur differentially among similar orbits
(apart from that in occasional close encounters). This ``two-body relaxation'' drives
the overall quasi-static secular evolution of the system
which happens on the timescale of many orbital crossing times.
However, if the system is not in equilibrium, the energy exchange
happens much faster, in crossing times. Such ``violent relaxation''
drives the system (or a fraction of it) towards
dynamical equilibrium (or energy minimum). See, \eg,
\citet{spz} for details.}
among the stars in the expanding cluster generates a ``fallback effect'' which
retains a fraction of gravitationally-bound stars even for
SFE $\lesssim30\%$ (also see \citealt{bokr2002}). The resultant bound fraction, $F_b$,
depends on how efficiently stars in the expanding post-gas-expulsion
cluster exchange energy during this relaxation process.
Hence, $F_b$ is proportional to the central stellar concentration (total stellar mass
vs. size) of the pre-gas-expulsion cluster and to the timescale of the gas
expulsion (time over which the expanding system remains in a dense enough
phase for efficient energy exchange). In other words,
$F_b$ is proportional to the efficiency of violent relaxation which
ultimately scales with the stellar number density. The
latter is governed by the stellar density of the
pre-gas-expulsion cluster. Note that
for a given (fractional) rate of gas removal and a pre-gas-expulsion
stellar density, the resulting efficiency of violent relaxation limits
the expansion of the initial (bound) stellar system and the time in which the bound fraction
returns to equilibrium (or the re-virialization time; see Sec.~\ref{r136}).     

Fig.~\ref{fig:bf} shows the bound fraction as a function
of the initial (pre-gas-expulsion) cluster stellar mass, $\mcl(0)$, and
the effective speed, $\vg$, at which the gas is expelled. $\vg$ directly
translates into the gas expulsion timescale (e-folding time; see below), $\tg$,
since $\tg=\rh(0)/\vg$. The observed trend in Fig.~\ref{fig:bf} is what is expected
from the above discussion.

\begin{figure}
\centering
\includegraphics[width=10.0cm, angle=0]{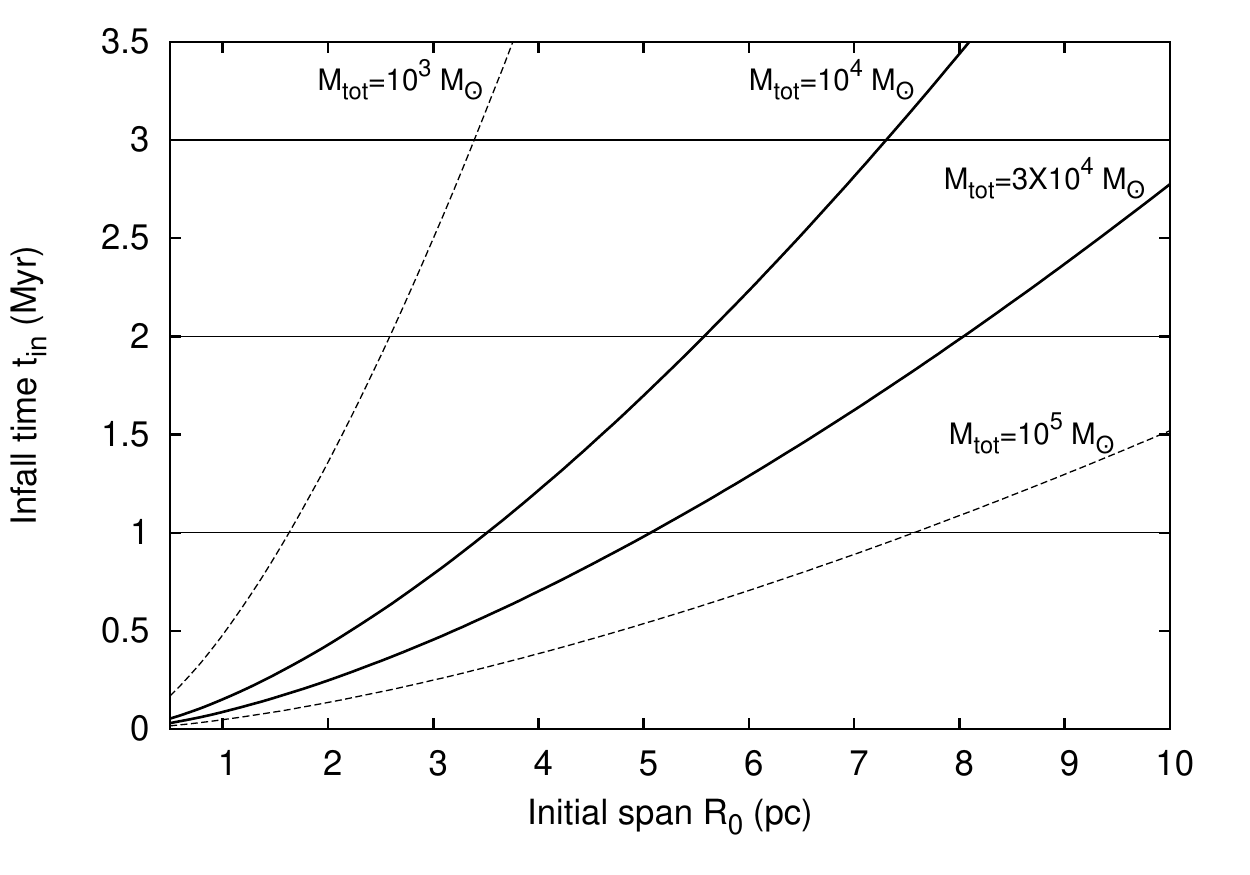}
\caption{The infall time (or the time of first arrival at orbital pericenter)
of the subclusters, $t_{in}$, as a function of the radius, $R_0$,
of the spherical volume over which they are initially distributed. The curves are according
to Eqn.~\ref{eq:tin2} for different systemic mass $M_{tot}$. The
highlighted curves for $M_{tot}=M_\ast=10^4\Ms$ without a residual gas and
$M_{tot}=3M_\ast=3\times10^4\Ms$ correspond to the model calculations
for the NGC 3603 young cluster (see Sec.~\ref{himrg}).
This figure is reproduced from \citet{sb2014b}.}
\label{fig:tin}
\end{figure}

\subsection{Why is an episodic or monolithic mode of cluster formation
necessary?}\label{monoreason}

The conditions in star forming molecular clouds in our Galaxy, which
can be considered representative of star forming environments in gas-rich
galaxies, do not necessarily imply cluster formation in a single go (see Sec.~\ref{intro}).
Such molecular regions typically contain compact,
interconnected filamentary structures as
revealed by detailed observations, \eg, by the \emph{Herschel} space
telescope. These observations \citep{andr2013} reveal clusters of proto-stars forming
within such high-density filaments (or ridges), which
are found to have Plummer-like cross sections with radii of 0.1-0.3 pc, and at their
junctions, implying a highly substructured initial condition
for newborn stellar associations. This is also consistent with
several observed stellar associations that contain individual
stellar groups (\eg, the Taurus-Auriga or T-A association; \citealt{palsth2002}) or
are highly substructured (\eg, Cygnus OB2; \citealt{wrt2014}),
indicating an amorphous and substructured beginning of a star
cluster as is often argued (see, \eg, \citealt{longm2014}).
On the other hand, VYMCs are found with near spherical core-halo
profiles at a few Myr age which does not add up with the above scenario and
calls for a different, episodic regime of cluster formation.

\begin{figure}
\centering
\vspace{1 cm}
\includegraphics[width=13.0cm, angle=0]{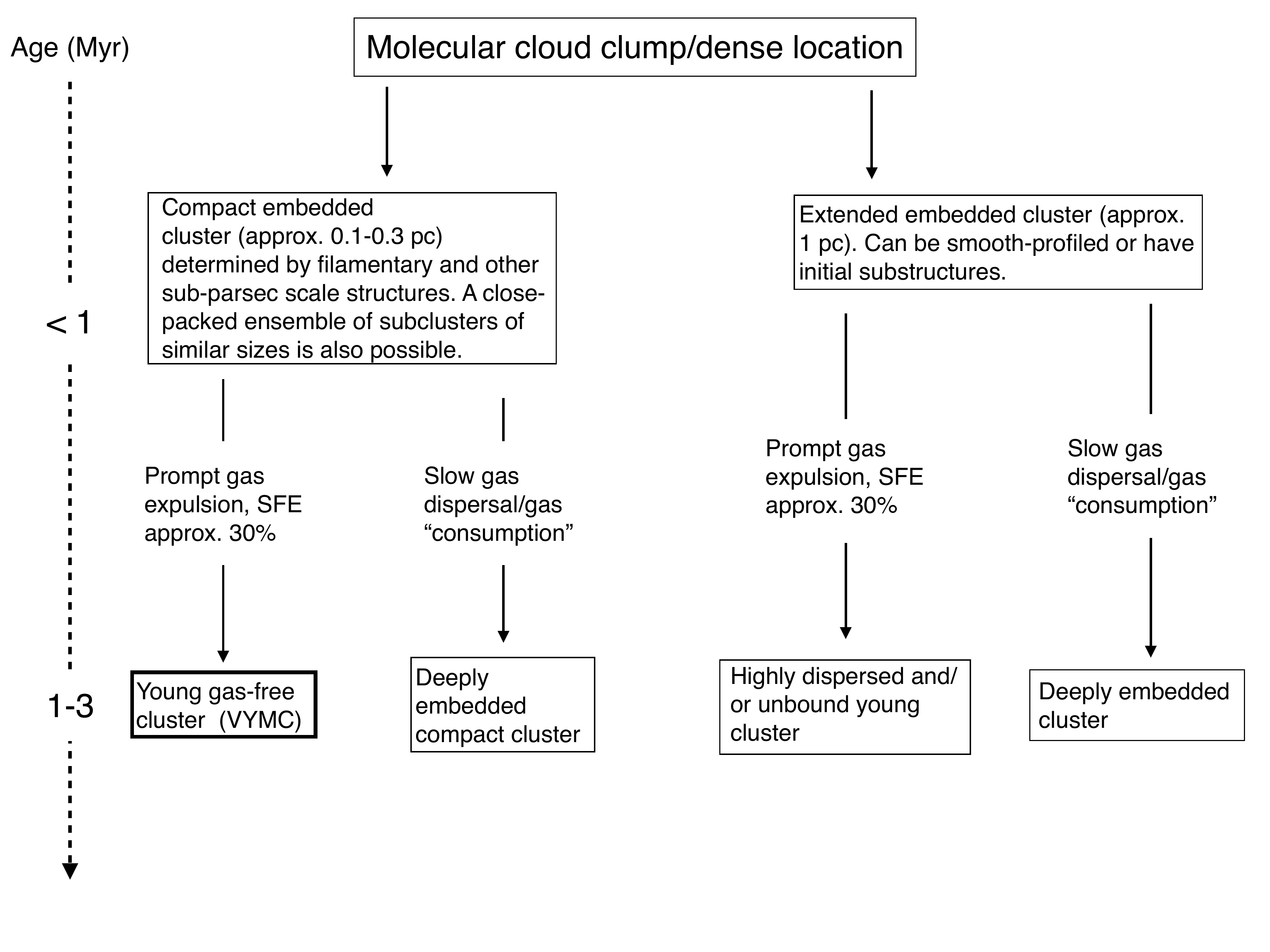}
\caption{Flowchart showing that exposed young massive clusters of 1-3 Myr
age can form through essentially only one channel; a highly compact gas
embedded proto-cluster undergoing a substantial ($\approx70$\% by mass)
and rapid (in a timescale comparable to the crossing time of
the proto-cluster) gas removal.}
\label{fig:VYMCflow}
\end{figure}

Of course, a stellar assembly cannot appear with 100\% SFE since
star formation is quenched by stellar radiative and mechanical feedback.
Hydrodynamic simulations (\eg, \citealt{mnm2012,bate2013}) including
stellar feedback and observations of embedded stellar assemblies in
the solar neighborhood \citep{lnl2003} both indicate
SFE $\lesssim30$\%. Hence a substantial gas dispersal should accompany
such a monolithic cluster formation to expose
the gas-free young cluster. Arguably, several VYMCs, despite
their near spherical monolithic structure contain substructures in the
form of multiple density maxima in their surface stellar density profiles \citep{kuhn2014}.   
Here, such systems will also be called monolithic. 

The present day sizes of gas-free massive, young clusters,
with half mass radii, $\rh$, between 3 - 10 pc
is itself indicative of the importance of gas expulsion in the formation
of such systems \citep{pf2009}. Observations suggest that newborn (\ie, embedded) clusters are
highly compact --- typically with half mass radii $\rh<1$ pc (see below). It is nearly
impossible to expand such compact clusters up to their present day sizes through purely
secular evolution. Fig.~\ref{fig:lagrevol} shows the evolution of the Lagrange radii of a 
cluster with $\mcl(0)\approx1.3\times10^4\Ms$ and $\rh(0)\approx0.3{\rm~pc}$ as obtained
through direct N-body calculation using the state-of-the-art
{\tt NBODY6}
\footnote{Sverre Aarseth's code {\tt NBODY6} 
and its variants \citep{aseth2012} are presently the most advanced and realistic
direct N-body evolution code. The N-body integration engine
computes individual trajectories of all the stars using a fourth-order
Hermite scheme. Close encounters are dealt with two- and multi-body
regularizations. In addition, the code employs the {\tt BSE}
stellar and binary evolution scheme \citep{hur2000} for evolving the individual
stars and mass-transferring binaries. The code also
includes recipes for tidal interactions and stellar collisions.
See \citet{ars2003} for details.}
code. Here, the cluster expands only by a factor of few in $\approx100$ Myr,
due to dynamical interactions and
mass loss due to stellar evolution; it
hardly expands in a few Myr age. Hence, an additional expansion mechanism
is essential to explain the observed cluster sizes. As explained below,
the above initial $\rh(0)\approx0.3{\rm~pc}$ is the most plausible one as supported
by observational and theoretical studies of birth conditions
of star clusters.

The strongest support for an episodic mode of star formation is what can be called
the ``timescale problem'' of cluster assembly. Consider a system of stellar clumps
(or subclusters) of total mass $\mstar$ within a spherical volume of radius $R_0$, which have zero or small
relative speeds (see Sec.~\ref{himrg}).
In other words, they form a ``cold'' system which can fall in and assemble into
a single bound star cluster. An embedding background molecular gas of total mass $\mg$
would accelerate the infall; the so called ``conveyor belt'' mechanism \citep{longm2014}.
The time, $\tin$, for the subclusters
to collide onto each other at the systemic potential minimum is
given by \citep{sb2014b}
\begin{equation}
t_{in}\approx\frac{R_0^{\frac{3}{2}}}{\sqrt{GM_{tot}}}
=0.152\frac{\left(\frac{R_0}{\rm pc}\right)^{\frac{3}{2}}}
           {\left(\frac{M_{tot}}{10^4\Ms}\right)^{\frac{1}{2}}}{\rm~Myr},
\label{eq:tin2}
\end{equation}
where $M_{tot}=M_\ast+M_g$. Fig.~\ref{fig:tin} shows the
dependence of $\tin$ on typical masses and sizes involved in massive stellar associations. 

Note that the above $\tin$ estimates the time taken for the subclusters
to meet each each other \emph{for the first time}, after which they pass through
each other to continue in their orbits. The orbital energy of the subclusters is dissipated
into the orbital energy of the individual stars during their each mutual passage due to
violent relaxation, causing the subclusters' orbits to decay and finally merge into
a single cluster in dynamical equilibrium. Hence, the final merger time, $\tmrg$,
is several $\tin$s as found in N-body calculations (Sec.~\ref{himrg}). Although $\tin$
decreases with increasing background gas mass, $\mg$, this does not necessarily
lead to shorter $\tmrg$ as the subclusters approach faster and hence take
larger number of orbits to dissipate their K.E. As found in N-body
calculations, the background gas actually lengthens $\tmrg$ (see Sec.~\ref{himrg} for the
details) for $R_0\gtrsim2{\rm~pc}$. In other words, the conveyor belt process does
not necessarily accelerate the assembly of the final cluster. Hence, Fig.~\ref{fig:tin}
implies that unless a group of subclusters form too close to each other, \ie, already
within the length scale of a compact star cluster (a few pc), it is practically
impossible to assemble a VYMC by its young age through sequential
mergers of less massive substructures as found in star-forming molecular clouds;
see Sec.~\ref{himrg} for more details.
Therefore, it is far more likely that VYMCs form in cluster- or molecular
clump-scale localized high efficiency star formation episodes, \ie, monolithically.

Interestingly, based on the observed velocity fields of gas clouds
in the neighborhood of several starburst clusters, some authors
\citep{furuk2009,fukui2014,fukui2015}
suggest that these clusters (\eg, Westerlund 2, NGC 3603) form out of
intense starbursts triggered during major cloud-cloud collisions.    
Such conclusions are based on the observed ``broad-bridge''
features \citep{haworth2015} in the velocity-space morphologies of cloud fragments
near these VYMCs. A collision between a pair of massive molecular clouds
lasts for a short time, typically $\approx1$ Myr. Hence, as before,
this points to an episodic formation of these VYMCs
during the cloud-cloud collisions.
After the clouds have crossed each other, the depletion of the background potential
would lead to an expansion of the newly-hatched cluster, as in the
case of internal gas expulsion. 
However, more detailed studies of the internal
velocities of such gas clouds and as well further
theoretical studies of cloud-cloud collisions \citep{duca2011,takahira2014} is necessary
to establish this scenario.   

Finally, one can ask the following question:
\emph{What if a VYMC is simply formed in-situ but with its current
observed size and not being governed by compact molecular filaments
or other compact structures of the molecular clouds, eliminating
the need of a substantial rapid gas dispersal?} In that case, the
cluster must form with a sufficiently high SFE. However, it is unlikely
that SFE can be pushed beyond $\approx30$\% as observational and
theoretical studies suggest (see Sec.~\ref{params}). Hence,
such a scenario is unrealistic. With SFE near 30\%, a typical present-day
sized star cluster would largely become unbound and/or become too
extended, depending on its initial mass. The cluster can, however,
survive if the gas is dispersed slowly; in a timescale longer
than a few crossing times \citep{lmagd84}. The ambient gas can also be
depleted if it is accreted by the (proto-) stars
(``gas consumption''; \citealt{longm2014}), without expanding
or unbinding the cluster. These processes would, however,
take much longer than a few Myr and one would obtain an embedded
cluster instead, like W3 Main. Further discussions follow in Sec.~\ref{discuss}. 
The flowchart in Fig.~\ref{fig:VYMCflow} summarizes the discussions
in this section.

\begin{framed}
The typical present-day density ($10^4-10^5\Ms{\rm~pc}^{-3}$; or size $\approx 1$ pc),
age ($\approx 1-3$ Myr) and (near) spherical core-halo
morphology of gas-free very young massive clusters (VYMCs),
like R136, NGC 3603 and the ONC, dictate an episodic
or monolithic (or near monolithic) formation of such star clusters,
undergoing a violent gas dispersal phase. 
\end{framed}

\subsection{An analytic representation for gas expulsion}\label{gasmodel}

The episodic cluster formation scenario involving gas expulsion,
which is widely used \citep{lmagd84,adm2000,pketl2001,bokr2002,bk2007,sb2013,sb2014,pfkz2013},
has been successful
in explaining the detailed structure and several well observed VYMCs. All these studies 
use a rather straightforward initial condition of a Plummer star cluster of mass, $\mcl(0)$,
\footnote{For monolithic systems we denote here the stellar mass as $\mcl\equiv\mstar$.}
and half mass radius, $\rh(0)$, that is embedded in its spherically symmetric natal gas
of total mass $\mg(0)$.
The latter is assumed to have a constant SFE, $\epsilon$, throughout, \ie, the gas density profile
follows the stellar Plummer density profile. The initial system represents a dense molecular
gas clump with a recent episode of star formation with efficiency $\epsilon$. The initial
mass function (IMF) of the stellar system can be plausibly represented by the canonical
mass function \citep{pk2001,pk2013} although equal mass stars have also been used in the literature for
scalability (\eg, \citealt{bk2007,pfkz2013}).

As discussed above, the gas component is
treated simply as an analytical external (Plummer) potential corresponding to its mass
distribution whereas the stellar system is tracked accurately using direct N-body
calculations, which captures the essential dynamics of the stellar system. The escape of
the gas component is typically modelled as an exponential decay of the gas mass with
e-folding time $\tg$ after a ``delay time'' $\td$, \ie,
\begin{eqnarray}
M_g(t)=& M_g(0) & t \leq \tau_d,\nonumber\\
M_g(t)=& M_g(0)\exp{\left(-\frac{(t-\tau_d)}{\tau_g}\right)} & t > \tau_d.
\label{eq:mdecay}
\end{eqnarray}
$\tg$ is determined by the effective speed, $\vg$, with which the gas
escapes out, \ie, $\tg\approx\rh(0)/\vg$.
The Plummer radius of the gas distribution is kept fixed at $\rh(0)$.
Such an analytic treatment of the gaseous component
is justified by \citet{gb2001} who show that expelling the gas analogously
by detailed SPH calculations (using shock heating) and analytically produce similar effect
on the stellar system.

Note that the corresponding (clump) SFE is $\epsilon=\mcl(0)/(\mcl(0)+\mg(0))$ or
\beq
\mg(0)=\mcl(0)\left(\frac{1}{\epsilon}-1\right)
\label{eq:sfe}
\eeq  

\subsubsection{Parameters for gas expulsion}\label{params}

One key parameter in the above model of the initial phase of cluster formation
is the initial size of the embedded system given by its half mass radius
$\rh(0)$. While in the literature there is no norm in the choice of the
compactness of the gas-filled system, detailed observations of molecular clouds and
embedded proto-stars imply highly compact profiles of proto-stellar clusters.
As seen in the \emph{Herschel} observations, and more recently using the \emph{ALMA},
proto-stellar associations appear in the highly compact filamentary overdensities
and in their junctions \citep{schn2010,schn2012,hill2011,henne2012,2014arXiv1412.1083T} in
giant molecular gas clouds (GMCs). The sections of these filaments are
very compact; typically $\lesssim0.3$ pc and peaked at $\approx0.1$ pc
\citep{andr2011}. The profiles of these filament sections are typically
Plummer-like \citep{mali2012}. This dictates a plausible, idealized initial
embedded cluster to be a highly compact Plummer sphere with
$\rh(0)\lesssim0.3$ pc. Indeed, the embedded associations in the
solar neighborhood \citep{lnl2003,tapia2011,tapia2014} and near embedded young clusters,
\eg, RCW 38 \citep{derose2009,kuhn2014} and the ONC \citep{pketl2001}
are all found to have half mass radii well less than a parsec.

In an independent and semi-analytic study, \citet{mk2012} have investigated
the initial conditions of star clusters that would give rise to
the currently observed binary period distribution in several observed clusters.
Here, the ``inverse dynamical population synthesis'' \citep{pk1995a} is used 
to infer the initial stellar density (and hence the size)
of a given cluster, which would dynamically evolve an
initial universal primordial binary period distribution \citep{pk1995a,pk1995b} to
the present-day distribution.
This study relates the birth mass and the half mass radius of a star cluster as,
\beq
\frac{\rh(0)}{\rm pc}=0.10^{+0.07}_{-0.04}\times\left(\frac{\mcl(0)}{\Ms}\right)^{0.13\pm0.04}.
\label{eq:mrk}
\eeq
This gives comparable initial (embedded) cluster size as above which depends weakly on the
initial mass.    

The observed values of SFE in star forming clouds and embedded associations
ranges widely, from less than a percent \citep{rath2014}
to $\approx30$\% for the embedded stellar associations
in the solar neighborhood \citep{lnl2003}. An appropriate value of SFE
is even more unclear from theoretical studies which depends on a number
of assumptions and inputs that are adopted in the hydrodynamic calculations.  
SPH calculations with spherical (or cubical) gas clouds of
$<100-1000{\rm s~}\Ms$ without any implementation for stellar feedback,
as often done for such masses (\eg, \citealt{kl1998,bate2004,bate2009,giri2011}),
cannot infer any SFE and would eventually let all of the gas be absorbed into the
proto-stars (or sink particles), \ie, give $\approx100$\% SFE as
an artifact.
Self regulation by stellar matter outflow (wind and jet) and radiation \citep{admftz96}
is crucial to arrive at a realistic SFE. Current state-of-the-art SPH studies
(reaching opacity limit and sub-sink-particle resolution)
in this direction incorporate seed magnetic field in the star-forming gas
and diffusive radiation feedback but are limited to individual
proto-stars' scale. Proto-stars or sink particles are found to form with
jet outflows where a self-regulated SFE upto $\approx30$\% is obtained
(\citealt{bate2013}; see also \citealt{mnm2012}). Recently, an independent analytical
study \citep{rb2014} of formation of clump-cores (that would eventually turn into proto-stars)
in gas clumps and of the maximum mass of the cores infers an upper limit
of $\approx30$\% for the clump SFE. 
This is consistent with the hydrodynamic calculations
with self-regulation and observations in the solar neighborhood (see above).
From their SPH calculations of low-resolution but real-sized ($10^5-10^6\Ms$)
turbulent molecular clouds, that include radiative feedback but
no magnetic field, \citet{dale2015}, however, find high SFE
approaching 100\%. This inferred SFE is likely to be an overestimate
due to introduction of too low resolution (where a sink particle represents
a stellar sub-cluster) and partial feedback by excluding magnetic field.   
From this viewpoint, the outcome of the proto-star-scale calculations,
as mentioned above, are much more reliable.
\citet{pfkz2013} also find that $\approx30$\% SFE best describes the
age-mass and age-size correlation in young clusters of $<20$ Myr age. 
It is, therefore, plausible but not entirely obvious to assume that the massive
clusters form with the highest possible SFE, an estimate of which is
$\epsilon\approx30$\%, according to the above mentioned studies. Note that this SFE
refers to the clump (\ie, over the spatial scale of a newborn cluster)
efficiency; the SFE over an entire GMC is only a few percent.

The values of the timescales governing the gas expulsion timescale, \viz,
$\tg$ and $\td$ depend on the complex physics of gas-radiation
interaction. When the gas starts to escape, it should be ionized by
the UV radiation from the massive stars causing efficient coupling of the
stellar radiation with the gas which is one of the primary drivers of the
gas. Hence, one can plausibly use an average gas
velocity of $\vg\approx10$ km s$^{-1}$ which is the sound-speed in
ionized hydrogen (HII) gas. For massive clusters, whose escape
speed (of the stellar system) exceeds the above $\vg$, the coupling of
stellar radiation with the ionized gas
over-pressures the latter and
can even make it radiation pressure dominated (RPD)
for massive enough clusters. During such RPD phase, the gas is driven
at speeds well exceeding its sound-speed \citep{krm2009}. Once the expanding
gas becomes gas pressure dominated (GPD), the outflow continues
with the HII sound speed \citep{hils80}.
Hence, $\tg$ as determined by $\vg\approx10$ km s$^{-1}$,
is an \emph{upper limit};
it can be shorter depending on the duration of the RPD state.
Note that this initial RPD phase is crucial to launch the
gas from massive stellar systems whose escape speed
exceeds the HII sound speed  \citep{krm2009}.

As for the delay-time, a widely used representative
value is $\td\approx0.6$ Myr \citep{pketl2001}.
The correct value of $\td$ is again complicated by radiative gas physics.
Although stellar input to the parent gas has been studied in some
detail for single low mass proto-stars \citep{bate2013},
the phenomena is much less understood over the global scale of
a massive cluster. Nevertheless, an idea
of $\td$ can be obtained from the lifetimes of Ultra Compact HII (UCHII)
regions which can be up to $\approx 10^5$ yr (0.1 Myr; \citealt{chrch2002}).
The highly compact pre-gas-expulsion clusters (see above)
are a factor of $\approx3-4$ larger in size ($\rh(0)$)
than a typical UCHII region ($\approx0.1$ pc).
If one applies a similar Str\"omgren sphere expansion scenario
(see \citealt{chrch2002} and references therein)
to the compact embedded cluster, the estimated delay-time, $\tau_d$, before a
sphere of radius $r_h(0)$
becomes ionized, would also be larger by a similar factor and
close to the above representative value.
Once ionized (\ie, becomes HII from its predominantly neutral molecular or HI state),
the gas couples efficiently with the stellar radiation and launched immediately (see above).
High-velocity jet outflows from proto-stars \citep{pat2005} aid the gas outflow.

For super-massive clusters ($>10^6\Ms$), \ie, for proto-globular clusters,
however, a ``stagnation radius'' can form
within the embedded cluster inside which the
radiation cooling becomes sufficiently efficient to possibly
form second-generation stars \citep{wn2011}. Also, as discussed above,
the gas-outflow can initially be supersonic which generates shock-fronts. Although shocked,
it is unlikely that star formation will occur in such a RPD gas. Later, during the GPD outflow, 
the flow can still be supersonic in the rarer/colder outer parts of the embedded
cluster where the sound speed might be lower than that typical for HII gas.
It is, however, unclear whether the cooling in the
shocked outer regions would be efficient enough to form stars.

Admittedly, the above arguments do not include complications such as unusual morphologies
of UCHIIs and possibly non-spherical ionization front, among others, and only provide basic estimates
of the gas-removal timescales. Observationally,
Galactic $\approx 1$ Myr old gas-free young clusters such as the ONC and the
HD97950 imply that the embedded phase is $\tau_d<1$ Myr for massive clusters.
The above popularly used gas-expulsion model
does capture the essential dynamical response
of the star cluster.

The stellar mass function of the embedded clusters is
typically taken to be canonical \citep{pk2001}. Note that the stellar entities
here are proto-stars which are yet to reach
their hydrogen-burning main sequences. Also, the interplay between gas
accretion and dynamical processes (ejections, mergers)
in the compact embedded cluster continue to shape the global stellar
IMF of the cluster \citep{kl1998}.
This IMF is often observed to be canonical for VYMCs.
This gas accretion and the dynamical processes only influence
the massive tail of the IMF and also sets the maximum stellar mass \citep{wk2004,wp2013},
as indicated in hydrodynamic calculations \citep{kl1998,dib2007,giri2011}.
The overall canonical shape of the IMF as determined by the low mass stars,
which contribute to most of the stellar mass of the system ($>90$\%),
appears primarily due to gravitational fragmentation alone.
There are observational evidences available which suggest that VYMCs posses
a canonical IMF below $1\Ms$ \citep{sh2014}.
This justifies the adoption of the canonical IMF for the
embedded (proto-) star cluster. In more recent studies (\eg,
\citealt{sb2013,sb2014} discussed below) an ``optimal sampling'' of the
canonical IMF is used \citep{pk2013} which automatically terminates
the IMF at the maximum stellar mass \citep{wk2004}.

On a separate note, efficient gas expulsion from star clusters
is indirectly supported by the lack of gas in young and
intermediate-aged clusters, in general. In particular,
a recent survey of the LMC's massive star clusters over wide ranges of mass
($>10^4\Ms$) and age (30-300 Myr) has failed
to identify reserved gas in any of these clusters \citep{bs2014}.
These clusters would have accreted enough surrounding gas
by now for the latter to be detected within them.
This implies that star clusters can, in fact,
disperse their gaseous component efficiently at any age $<300$ Myr
and irrespective of their escape velocities \citep{bs2014}.
However, short ($<1$ Myr) bursts of new star formation episodes
can lead to long-term continued growth of star clusters
as these may accrete gas episodically from the surrounding interstellar
medium \citep{pflm2009}.

\subsection{Matchings with individual very young massive clusters}\label{match}

The best way to validate the monolithic or episodic scenario of VYMC formation is to compare
its computed outcome with the details of well observed VYMCs. There are only a
few VYMCs whose profiles are measured from their centers to their halos
using ground (primarily the \emph{Very Large Telescope} or VLT) and space based 
(the \emph{Hubble Space Telescope} or HST) photometry. To obtain a radial mass density
profile of a dense assembly, proximity is essential as much as low extinction.
This allows reliable estimates of starcounts in well resolved annuli and also
the estimates of the individual stellar masses. Thus (surface) mass density
profiles have been obtained only for nearby and kpc-distance Galactic young
clusters. The stellar velocity dispersion in a young cluster's central region
can also constrain its initial conditions. The (one-dimensional) velocity dispersion
can be obtained from stellar radial velocities from multi-epoch spectroscopy. Proper
motions of the individual stars, as obtained from multi-epoch high-resolution imaging
with sufficient time baseline, can provide the dispersion in the transverse
velocity components. The Galactic NGC 3603 young cluster or HD97950,
being our nearest starburst cluster, is perhaps
the best observed VYMC whose mass density profile (obtained using
the VLT; \citealt{hara2008}) and transverse stellar velocity
dispersions (obtained using HST with a 10 year baseline; \citealt{roch2010,pang2013})
are known out to $\approx3$ pc with reasonable accuracy. Being as young
as $\approx1$ Myr \citep{stol2004} despite being a gas free cluster,
HD97950 acts as a ``smoking gun'' of formation of massive star clusters.   

Table~\ref{tab1} shows model N-body computations in the literature and their corresponding
parameters which have reproduced well observed VYMCs (see Table~\ref{clprop})
beginning from single-cluster initial conditions. The key
results from these works will be discussed below. All these studies utilize certain common
properties and conditions as below:

\begin{itemize}

\item The SFE $\epsilon\approx0.3$ and the gas dispersal is determined by
$\vg\approx10{\rm~km~s}^{-1}$ ($\tg/{\rm~Myr}=(\rh(0)/{\rm~pc})/10$) and $\td\approx0.6$ Myr.

\item The initial stellar and gas distribution follow the same Plummer profile. 

\end{itemize}

\begin{table}[t]
\begin{minipage}{4.8 in}
\renewcommand{\thefootnote}{\alph{footnote}}
\caption{
Initial and gas expulsion parameters, for the computed
models beginning with monolithic initial conditions,
that reproduce well observed very young massive clusters
(see Sec.~\ref{match}). See text for the meanings of the notations.
}
\label{tab1}
\begin{tabular}{lccccccccl}
\hline
Model cluster & $\mcl(0)/\Ms$ & $\mg(0)/\Ms$ & $\rh(0)/{\rm pc}$
 & $\tg/{\rm Myr}$ & $\tcr(0)/{\rm Myr}$ & $\td/{\rm Myr}$
 & $f_{\rm bin}(0)$ & $Z/Z_\odot$ & Reference\\
\hline
ONC-B  & $4.2\times10^3$   & $8.4\times10^3$ & 0.21 &  0.021  & 0.066 & 0.6 & 1.0 & 1.0 &\citet{pketl2001}\\
R136   & $1.0\times10^5$   & $2.0\times10^5$ & 0.45 &  0.045  & 0.021 & 0.6 & 0.0 & 0.5 &\citet{sb2013}\\
HD97950s & $1.0\times10^4$ & $2.0\times10^4$ & 0.25 &  0.025 & 0.029  & 0.6 & 0.0 & 1.0 &\citet{sb2014}\\ 
HD97950b & $1.0\times10^4$ & $2.0\times10^4$ & 0.25 &  0.025 & 0.025  & 0.6 & 1.0 & 1.0 &\citet{sb2014}\\ 
\hline
ONC-A\footnotemark[1]   
         & $3.7\times10^3$ & $7.4\times10^3$ & 0.45 &  0.045 & 0.23   & 0.6 & 0.0 & 1.0 &\citet{pketl2001}\\
NYC\footnotemark[1]      
         & $1.3\times10^4$ & $2.6\times10^4$ & 0.34 &  0.034 & 0.038  & 0.6 & 0.0 & 1.0 &\citet{sb2013}\\ 
\hline     
\end{tabular}

\medskip
The initial gas mass $\mg(0)$ and the gas expulsion timescale
$\tg$ (see Sec.~\ref{gasmodel}) are determined by $\epsilon\approx0.33$ and $\vg\approx10{\rm~km~s}^{-1}$
($\tg=\rh(0)/\vg$) respectively and $\td\approx0.6$ Myr for all these computed models.
The models HD97950s/b refer to the ones with initial single-only stars/primordial binaries
as computed in \citet{sb2014}.

\footnotetext[1]{These computed models are not ``matching'' models but are discussed various in places of this
chapter.}
\end{minipage}
\end{table}

As discussed in Sec.~\ref{params}, these values and conditions
are representatives and idealizations but physically motivated from what we know 
so far from observations
and calculations at smaller scales.
Furthermore, in several of them \citep{pketl2001,sb2014} a primordial binary population is used
according to the ``birth period distribution'' \citep{pk1995b,mk2014}.
While it is more compute intensive, introducing primordial binaries
is more realistic in light of the high multiplicity of PMS stars.
In these cases, a 100\% primordial binary fraction ($f_{\rm bin}(0)=1.0$) is used at $t=0$ \citep{pk1995a}.
The orbital period ($P$) distribution of such binary population
spans over a wide range, between
$1.0 < \log P < 8.43$ where $P$ is in days \citep{pk1995b}.
The binary eccentricities, $e$,
are taken to be thermalized, \ie, distributed as $f(e)\propto e$ \citep{spz}.
With the dynamical evolution (and also due to
the orbital evolution by tidal interaction among PMS stars or the
``eigenevolution''; \citealt{pk1995b}) of the binary population and eventual
disruption of the parent cluster (by the Galactic tidal field),
such a primordial binary population
naturally transforms to the log-normal period distribution observed for low mass 
stellar binaries in the solar neighborhood \citep{pk1995a}. A detailed discussion
of the period distribution of primordial binaries, which is currently a
widely debated topic \citep{pk2013,mk2014,leigh2015}, is beyond the scope of this text.

\begin{table}[t]
\begin{minipage}{4.8 in}
\renewcommand{\thefootnote}{\alph{footnote}}
\caption{
Properties of VYMCs, discussed in this chapter, as inferred from
observations. ``---'' implies that the corresponding value or quantity is ambiguous
or unknown. 
The surface profiles known for these clusters (last column) are either
stellar number or mass density profiles or both (see below). 
}
\label{clprop}
\begin{tabular}{lccccl}
\hline
Cluster Name & Cluster mass ($\mcl/\Ms$) & Age ($t/{\rm Myr}$) & Galactocentric distance ($R_G/{\rm kpc}$)
& Half-light radius ($\rh/{\rm pc}$) & Radial profile\\
\hline  
ONC    &  $\approx 10^3$ & 1-2 & $\approx$ solar & --- & number\\ 
R136   & $\approx 10^5$ & 2-3 & --- & $\lesssim 1$ & ---\\
NGC 3603 & $(1.0-1.6)\times10^4$ & $\approx1$ & $\approx$ solar & $\approx0.75$ & number/mass\\
\hline  
\end{tabular}
\end{minipage}
\end{table}

These studies also incorporate stellar evolution and the associated mass
loss in the N-body calculations using the semi-analytic {\tt BSE} stellar
evolution code \citep{hur2000}. The {\tt BSE}, while available as
standalone, is integrated with the {\tt NBODY6} direct N-body code \citep{ars2003}.
{\tt NBODY6} is currently the most realistic way to associate stellar (and binary)
evolution with dynamics.
In the following, we discuss the key results from the studies mentioned
in Table~\ref{tab1}. For more details the reader is suggested to consult
the respective references.

\subsubsection{The Orion Nebula Cluster: structure and kinematics}\label{onc}

\begin{figure}
\centering
\includegraphics[width=9.5cm, angle=0]{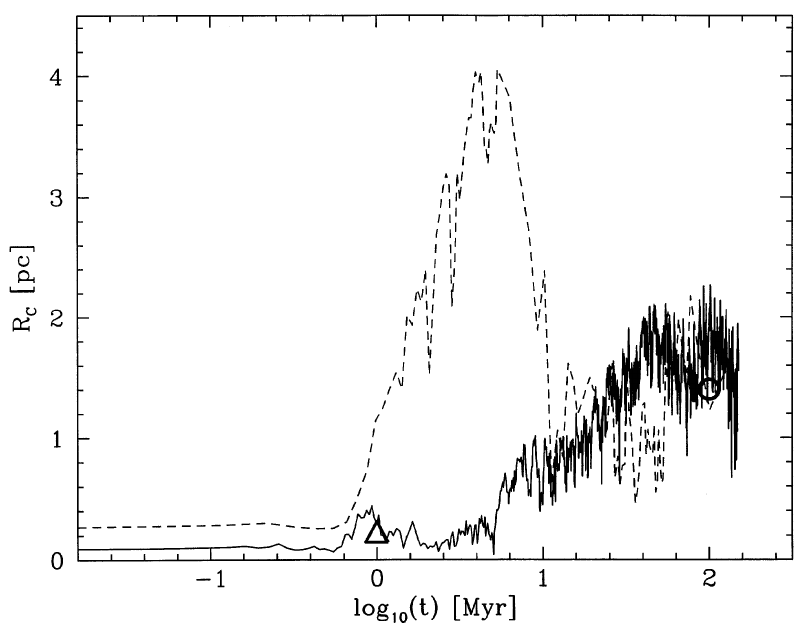}
\caption{The evolution of the core radius, $r_c$, in the ONC A (dashed curve)
and the ONC B (solid curve) model clusters as compued by \citet{pketl2001}. The
open triangle and circle are the observed values of core radius for
ONC \citep{hillen1998} and Pleiades \citep{radmer1998} respectively.
This figure is reproduced from \citet{pketl2001}.}
\label{fig:kah_corerad}
\end{figure}

\begin{figure}
\centering
\includegraphics[width=9.5cm, angle=0]{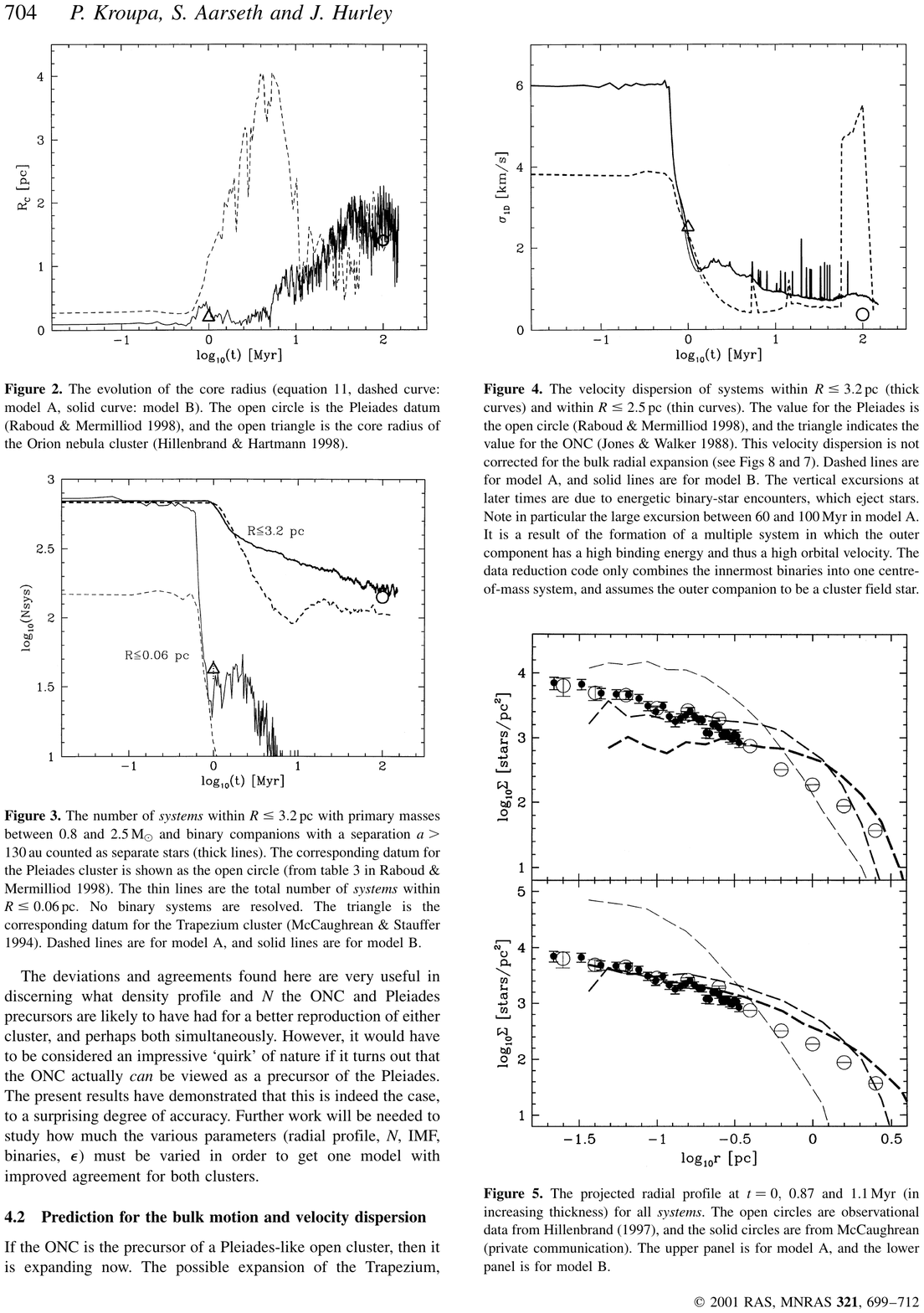}
\caption{
The projected radial stellar number density profile
at t=0, 0.87 and 1.1 Myr (in increasing thickness) for the computed clusters
ONC A (top) and B (bottom). 
The open circles are observed data from \citet{hillen1997}
and the solid circles are from McCaughrean (private communication).
This figure is reproduced from \citet{pketl2001}.
}
\label{fig:kah_starprof}
\end{figure}

\begin{figure}
\centering
\includegraphics[width=9.5cm, angle=0]{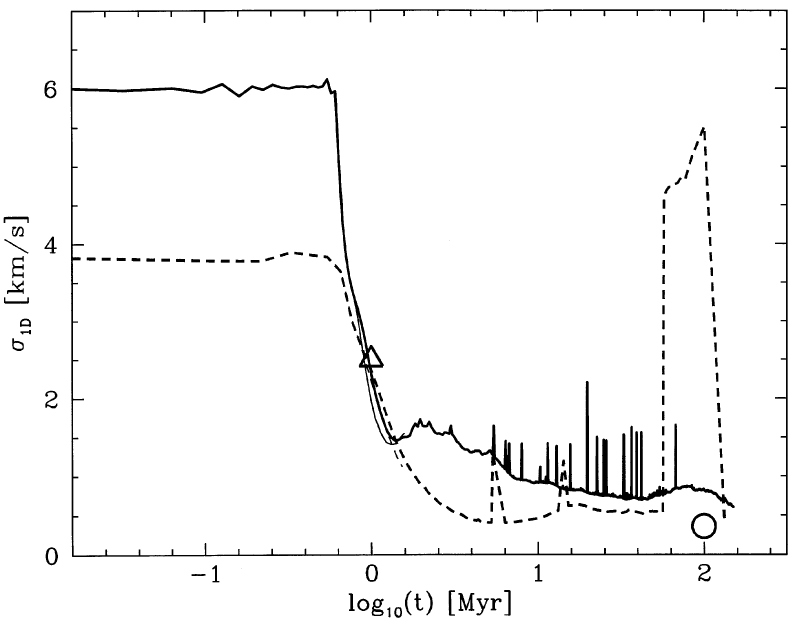}
\caption{The velocity dispersion of systems within $R\leq3.2$ pc (thick curves)
and within $R\leq2.5$ pc (thin curves).
The value for the Pleiades is the open circle \citep{radmer1998},
and the triangle is that for the ONC \citep{jw1988}.
The dashed and the solid lines are for ONC-A and B models
respectively. The vertical excursions at later times are due to energetic
binary-star encounters, which eject stars.
This figure is reproduced from \citet{pketl2001}.
}
\label{fig:kah_v1d}
\end{figure}

\citet{pketl2001} provide a comprehensive study of the Orion's main central cluster
(the ONC). These authors demonstrate that an appropriate monolithic initial state
with the above parameters (Table~\ref{tab1}) well reproduce the key observed
properties of the ONC. An important corollary of this work is that the ONC
would dynamically evolve to a cluster similar to Pleiades in $\approx100$ Myr. 
Hence, a young system like the ONC represents the infant stage of an intermediate age
open cluster like the Pleiades. These
calculations are done using a version of the {\tt NBODY6} code that includes an 
analytic time-varying external
gas potentials (the {\tt GASEX}; \citealt{pketl2001}) as
discussed above. A realistic birth primordial binary population
with 100\% binary fraction is used in these calculations.

Fig.~\ref{fig:kah_corerad} shows that the evolution of the core radius, $r_c$,
for the computed models ONC-A and -B (Table~\ref{tab1}). ONC-B agrees well with
the observed $r_c$ for the ONC at $t\approx1$ Myr age and evolves over
to be agreeable with the observed value for Pleiades at $t\approx100$ Myr.
The core-radius evolution of ONC-A, on the other hand, is far less consistent
with these observed values.

Notably, model B expands to much larger extent at
later times than immediately after its gas expulsion. Such late-time expansion
is common for clusters with a mass spectrum and hard primordial binaries.
This is driven by the mass loss due to the supernovae of the massive stars
that segregate to the cluster's central region by then. The expansion
is further assisted by frequent single star-binary close encounters
in the cluster's central region 
that cause ejections of single and binary stars \citep{sb2012,oh2014}
in super-elastic encounters \citep{heggie1975,hils75}
\footnote{According to ``Heggie-Hills law'', a hard binary (\ie, a binary whose orbital
velocity is higher than the relative velocity of its COM and the intruder)
statistically becomes harder, \ie, gains binding energy, in a gravitational encounter
with a third body. Hence, due to energy conservation, a hard binary-single
encounter would cause gain in the COM energy of the recoiling entities
in the cost of deepening the binary's potential well. This, in turn, results
in an increase of the K.E. in a packed-enough environment, \eg,
the central region of a massive star cluster. Close encounters can
result in the escape of one or both systems if they recoil exceeding
the escape velocity.}.
This boosts the internal K.E. of the cluster's
core due to the associated mass loss and encounter recoils.
The dynamical heating becomes efficient at late times after the bound fraction
of the initial system re-virializes (see beginning of Sec.~\ref{monoform})
and the most massive stars and the binaries segregate towards the cluster's
center, augmenting their density and hence the encounter rates therein. 
The evolutionary course of model A is however different where the
initial expansion due to gas expulsion is more extensive. Being
of lower mass and an initially larger radius,
model A has lower stellar density and hence
less efficient violent relaxation throughout its violent
expansion phase (see beginning of Sec.~\ref{monoform}),
allowing it to expand to larger radii and cause the bound
fraction to take longer to fall back
(\cf dashed line in Fig.~\ref{fig:kah_corerad}).   
The longer duration of expansion and re-collapse covers
a good part of the supernova phase ($t\gtrsim3$ Myr) and makes
binary-single/binary-binary encounters much less frequent. 
Note that for both models, the primordial binaries and the stars
of all masses are distributed initially without any spatial preference,
\ie, without any primordial mass segregation. Also, see
Sec.~\ref{r136}.

Fig.~\ref{fig:kah_starprof} shows the computed evolutions of the stellar
(surface) number-density profiles. Being consistent with the
core-radius evolution, the radial profile for the ONC-B model agrees
reasonably with the observed radial stellar density profile of the ONC,
out to $\approx 3$ pc from the center, at $t\approx1$ Myr (bottom panel).
This is unlike the ONC-A model which is either too dense or too expanded
compared to the observed radial profile (top panel). 

The evolution of computed one-dimensional velocity dispersion, $\sigoned$,
is shown in Fig.~\ref{fig:kah_v1d}. The frequent abrupt jumps in the computed $\sigoned$
at late times is due to binary-single star close encounters which
cause ejections of single and binary systems from the cluster. The recoil
K.E. in the encounter and the mass loss due to the escape heats up the
cluster's core (see above). The long term boost in $\sigoned$ for the ONC-A model
between 60 - 100 Myr is an artifact of the method in which $\sigoned$ is evaluated
in this study which considers only binaries as centers of mass (COMs) but not
higher multiplets. The increased $\sigoned$ here is caused due to the
outer member of a long-lasting triple system that can appear in relatively
low density systems. Given that the excursions in $\sigoned$ is probabilistic
in nature, both A and B models are consistent with the observed values
for ONC and Pleiades (\cf Fig.~\ref{fig:kah_v1d}).

Given the overall consistent agreement with the observed core radius, radial
stellar density profile and one-dimensional velocity dispersion,
ONC-B's initial conditions and parameters (Table~\ref{tab1})
comprise an appropriate initial state of the ONC. In other words,
this model represents a monolithic ``solution'' of the ONC.
It justifies the in-situ formation of this
VYMC from a single, initially bound stellar association
formed in a starburst in a molecular-gas clump, after expelling
the majority of the clump's gas ($\approx70$\% by mass).  
As a corollary, an ONC-like young star cluster would evolve
and expand to a Pleiades-like open cluster as the above
calculations show.

\subsubsection{The Tarantula cluster (R136): central velocity dispersion}\label{r136}

A common criticism put forward against the role of gas expulsion in
the formation of VYMCs is the inferred dynamical equilibrium in several
VYMCs. The central R136 cluster of the Tarantula Nebula of the
LMC is particularly cited in this context \citep{hb2012}. The inferred
total photometric mass of this cluster is $\approx10^5\Ms$ \citep{crow2010} and the
age of the bulk of its stars is $\approx3$ Myr \citep{and2009}. Clearly, 
R136 is quite an outlier by mass in the high side
among the well-studied nearby VYMCs (it is also at least twice
as massive compared to the Arches cluster).
 
\begin{figure}
\centering
\includegraphics[width=9.5cm, angle=0]{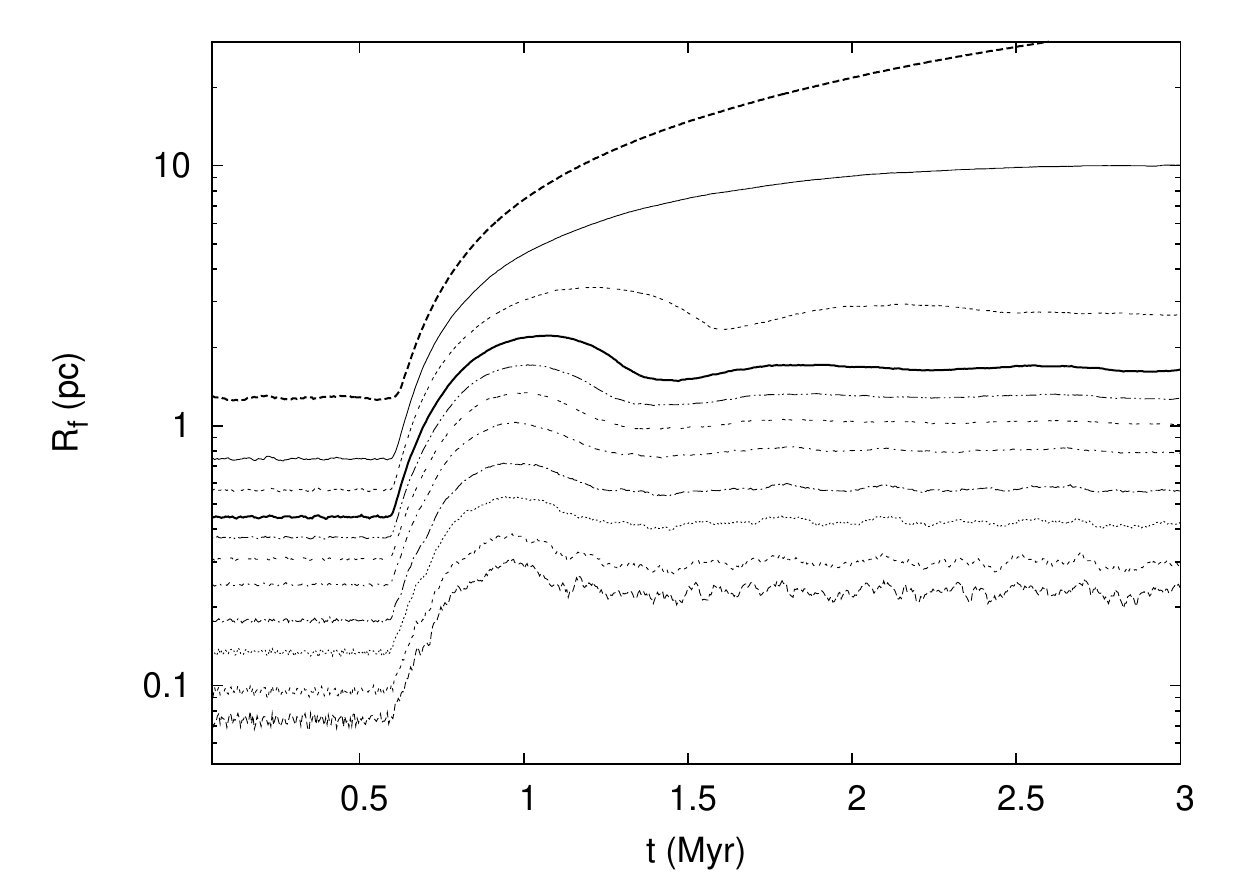}
\includegraphics[width=9.5cm, angle=0]{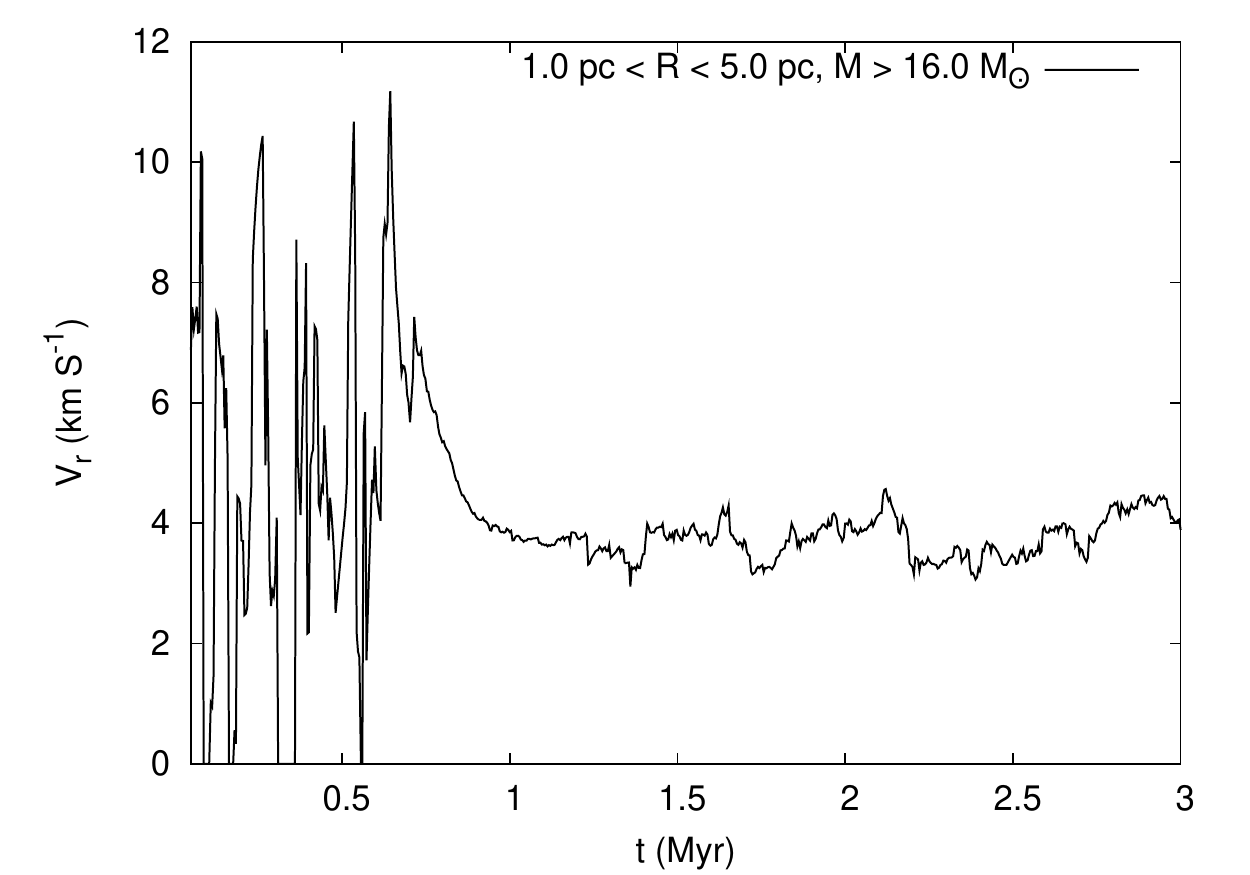}
\caption{
{\bf Top:} The evolution of the Lagrange radii, $R_f$, for stellar cluster mass fractions $f$
for the computed R136 model in \citet{sb2013} (see Table~\ref{tab1}).
The curves, from bottom to top,
correspond to $f=$ 0.01, 0.02, 0.05, 0.1, 0.2, 0.3, 0.4, 0.5, 0.625, 0.7 and 0.9 respectively. The thick solid line
is therefore the half mass radius of the cluster.
{\bf Bottom:} The corresponding evolution of the radial velocity (RV) dispersion,
$V_r$, of the O-stars ($M>16\Ms$), within the projected distances
$1{\rm~pc}< R<5{\rm~pc}$ from the cluster center. 
These panels are reproduced from \citet{sb2013}.
}
\label{fig:sb13_lrad136}
\end{figure}

As a part of the ongoing ``VLT-FLAMES Tarantula Survey'' (VFTS; \citealt{evn2011}), 
\citet{hb2012} measured radial/line-of-sight velocities (RV) of \emph{single} O-stars within
$1{\rm~pc}\lesssim R\lesssim5{\rm~pc}$ projected distance from R136's center.
(Strictly, the distances were
measured from the most massive cluster member star R136a1.
The exact location of the R136's true density center being unknown,
we consider this star be expectedly very close to the cluster's density center.)
They conclude that the RV dispersion, $V_r$, of the \emph{single} O-stars within this
region is $4{\rm ~km~s}^{-1}\lesssim V_r\lesssim5{\rm ~km~s}^{-1}$. Spectroscopy (with FLAMES)
at multiple epochs have been used to eliminate the radial velocities of spectroscopic binaries
\citep{hb2012},
\ie, the above $V_r$ corresponds the COM motion of the stars (and binaries) over the selected
region of the cluster. Given the mass of R136, the
above $V_r$ is consistent with the cluster being in dynamical equilibrium at the
present day.

This is contradictory to the generally accepted notion that young clusters, if
emerged from recent gas dispersal, should presently be expanding. 
This is indeed the case for the ONC \citep{pketl2001}.
\citet{brnd2008} demonstrated that young systems show an overall
increase in size with age. As noted above
(Secs.~\ref{intro} and \ref{monoform}), a recently gas-expelled cluster
may or may not be in dynamical equilibrium at a given age depending on
the effectiveness of violent relaxation in its expanding phase. This,
in turn, depends on its initial density. Hence, if initially massive
and/or compact enough, a VYMC can as well be in dynamical equilibrium
at present even after undergoing a significant amount of gas expulsion. 

Fig.~\ref{fig:sb13_lrad136} (top panel) shows the evolution of the Lagrange radii
for a model of a R136-like massive cluster
($\mcl(0)\approx10^5\Ms$), \viz, model ``R136'' in Table~\ref{tab1}.
The initial half-mass radius, $\rh(0)=0.45$ pc, is chosen according to
Eqn.~\ref{eq:mrk} which is consistent with the size of embedded
clusters and that of the sections of dense filaments in molecular gas
clouds (see Sec.~\ref{params}). These Lagrange radii (escaping and bound stars
are always included) imply that the R136 cluster, under reasonable
conditions (see Sec.~\ref{params}), would re-virialize well within its current
age of 3 Myr. The re-virialization time, in this case, is
$\tvir\approx1$ Myr and the bound fraction after re-virialization
is $F_b\approx0.6$, implying an efficient violent relaxation phase.
This calculation (as well the following one described in this subsection) is done
using the {\tt NBODY6} code (see above). No primordial binaries are used in these
calculations as $\tvir$ and $F_b$ would not get affected by binaries
significantly. Also, primordial binaries are a significant computational
hurdle for direct N-body calculations for such massive clusters
(even for Monte Carlo calculations; \cf \citealt{leigh2015}).

As expected, the corresponding computed
$V_r\approx4.5{\rm~km~s}^{-1}$ between 1 - 3 Myr age, as appropriate
for the remaining bound virialized cluster, is consistent
with the observed value for R136 (see above). This is shown in
Fig.~\ref{fig:sb13_lrad136} (bottom panel).
Note that in Fig.~\ref{fig:sb13_lrad136}, the computed evolution of
$V_r$ corresponds to stars with (zero-age) mass $>16\Ms$ which 
correspond to O-type stars that are used to determine the radial velocity
dispersion in R136 by \citet{hb2012}. Also, $V_r$ is computed
within $1{\rm~pc}\lesssim R\lesssim5{\rm~pc}$ as observed by
the above authors. The initial large fluctuations in $V_r$ 
for $\td<0.6$ Myr are due to the initial mass-segregated condition
used in this calculation \citep{sb2013},
which results in only a few O-stars within $1{\rm~pc}\lesssim R\lesssim5{\rm~pc}$
initially. As shown in \citet{sb2013} by comparing initially segregated
and non-segregated computed models, primordial mass segregation does
not influence the Lagrange radii and the $V_r$ after re-virialization.
In models at these initial densities, the mass segregation time-scale, $\tau_{\rm seg}$,
given by \citep{sb2010}
\beq
\tau_{\rm seg} \approx 15\frac{\langle m \rangle}{m_{\rm massive}}\tau_{\rm rh}(0),
\label{eq:mseg}
\eeq
is very short for the massive end of the stellar IMF. In Eqn.~\ref{eq:mseg},
$\tau_{\rm rh}(0)$ is the initial half-mass relaxation time \citep{spz} of a cluster
with average stellar mass $\langle m \rangle$.

\begin{figure}
\centering
\includegraphics[width=9.5cm, angle=0]{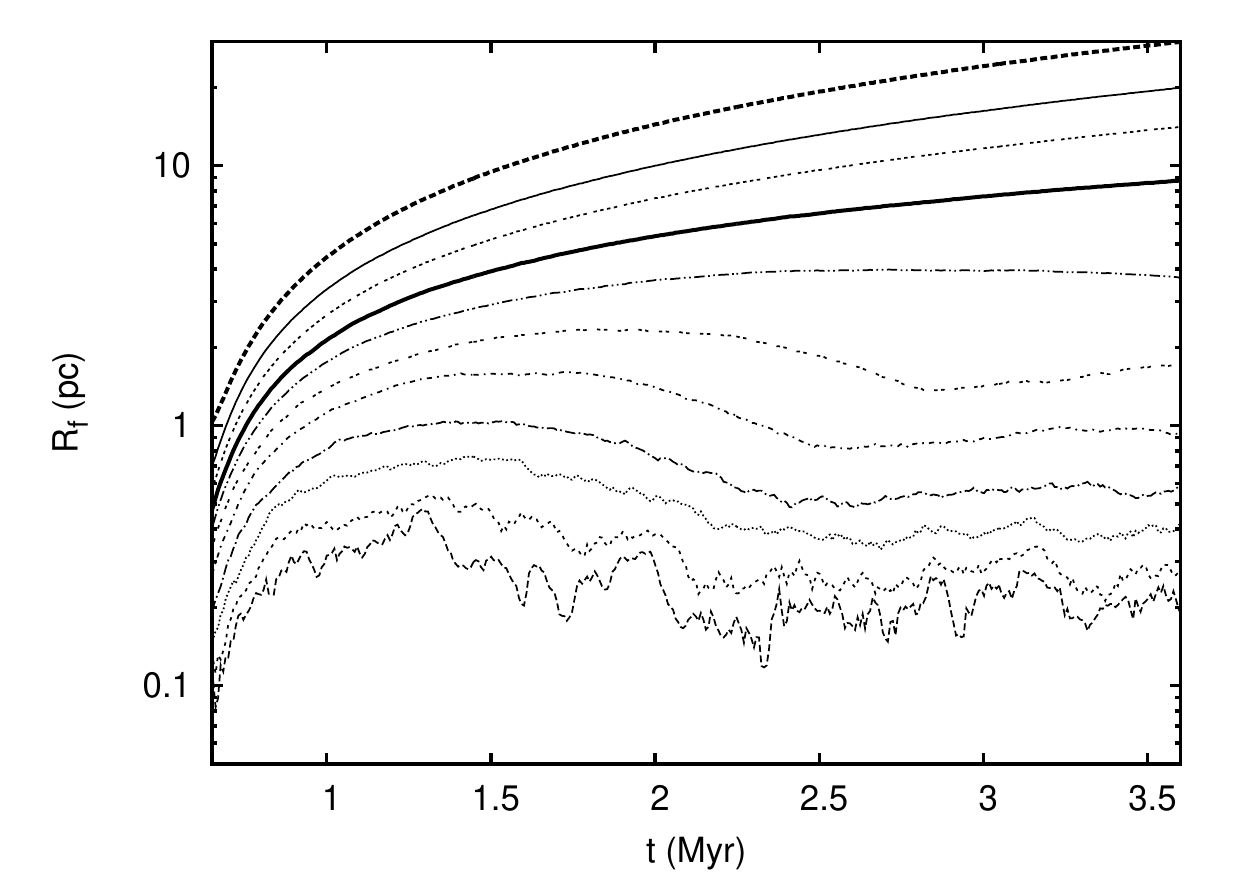}
\caption{
The evolution of the Lagrange radii for the computed NGC3603-like cluster in \citet{sb2013}
(see Table~\ref{tab1}). The legends for the curves are same as in Fig.~\ref{fig:sb13_lrad136}. 
The time axis before the beginning of the gas dispersal (until $\td=0.6$ Myr) is suppressed. 
This panel is reproduced from \citet{sb2013}.
}
\label{fig:sb13_lrad3603}
\end{figure}

Fig.~\ref{fig:sb13_lrad3603} shows the Lagrange radii of a computed model
that is a few factors less massive than the R136 model discussed above.
This model corresponds to the {\tt NBODY6}-computed model ``NYC'' in Table~\ref{tab1}
whose mass is similar to the NGC 3603 Young Cluster (HD97950). Note that
while this model is not a ``matching model'' for NGC 3603 cluster which
are covered in the following subsection (models ``HD97950''s/b of Table~\ref{tab1}),
the initial conditions are similar. Unlike the R136 model, the re-virialization
time is much longer in this case, \viz, $\tvir\approx2$ Myr. Hence at
its present age of $\approx1$ Myr, an NGC3603-like young cluster
would not be in dynamical equilibrium (except in its innermost regions;
see Sec.~\ref{ngc3603}), as one generally expects. As discussed in Sec.~\ref{onc},
a smaller $\mcl(0)$ and hence less initial stellar density causes
the NYC model to take longer to regain dynamical equilibrium with a
reduced bound fraction ($F_b\approx0.3$). 

From the above calculations it can be said that ``an observed dynamical 
equilibrium state of a very young stellar cluster does not 
necessarily dictate that the cluster has not undergone a 
substantial gas-expulsion phase''
\citep{sb2013}. The R136 cluster is very likely a
VYMC that is promptly re-virialized after its gas expulsion
owing to its large mass (hence initial density). Admittedly,
these conclusions depend on the intial and gas expulsion conditions. 
As discussed in \citet{sb2013},
the above conclusions are immune to reasonable variations in the gas expulsion
timescales $\td$ and $\tg$, as long as $\epsilon\approx33$\% which
is reasonable for VYMCs (see Sec.~\ref{params}). 

In passing, it is worthwhile to consider the effect if the SFE possibly varies
radially across the initial cluster \citep{adm2000}. In that case, the central region
of the cluster is likely to have a higher SFE due to higher density there.
The resulting gas removal preferentially from the outer parts of the  
cluster would cause it to expand less than the corresponding 
case of a uniform SFE. This would, in turn, shorten $\tvir$ and
increase $F_b$, \ie, the above conclusions still remain
unchanged.  

\subsubsection{NGC 3603 young cluster: structure and kinematics}\label{ngc3603}

Being our nearest starburst cluster, the central young cluster HD97950
of the Galactic NGC 3603 star-forming region is perhaps the best
observed VYMC. Due to its proximity ($\approx7$ kpc from the Sun) and brightness
(photometric mass $10000\Ms\lesssim\mcl\lesssim20000\Ms$) its
radial stellar mass-density profile for low mass stars \citep{hara2008} (using VLT
observations in NIR) and the number-density
profile for stars up to $100\Ms$ \citep{pang2013} are determined, both
out to 3 pc ($R\approx100''$) from its center. Furthermore,
its central velocity dispersion, within $R\lesssim0.5{\rm~pc}(\approx15'')$,
and the stellar tangential velocities are determined from proper motion
measurements with the HST ($\approx10$ year baseline; \citealt{roch2010,pang2013}).  

\begin{figure}
\centering
\includegraphics[width=9.5cm, angle=0]{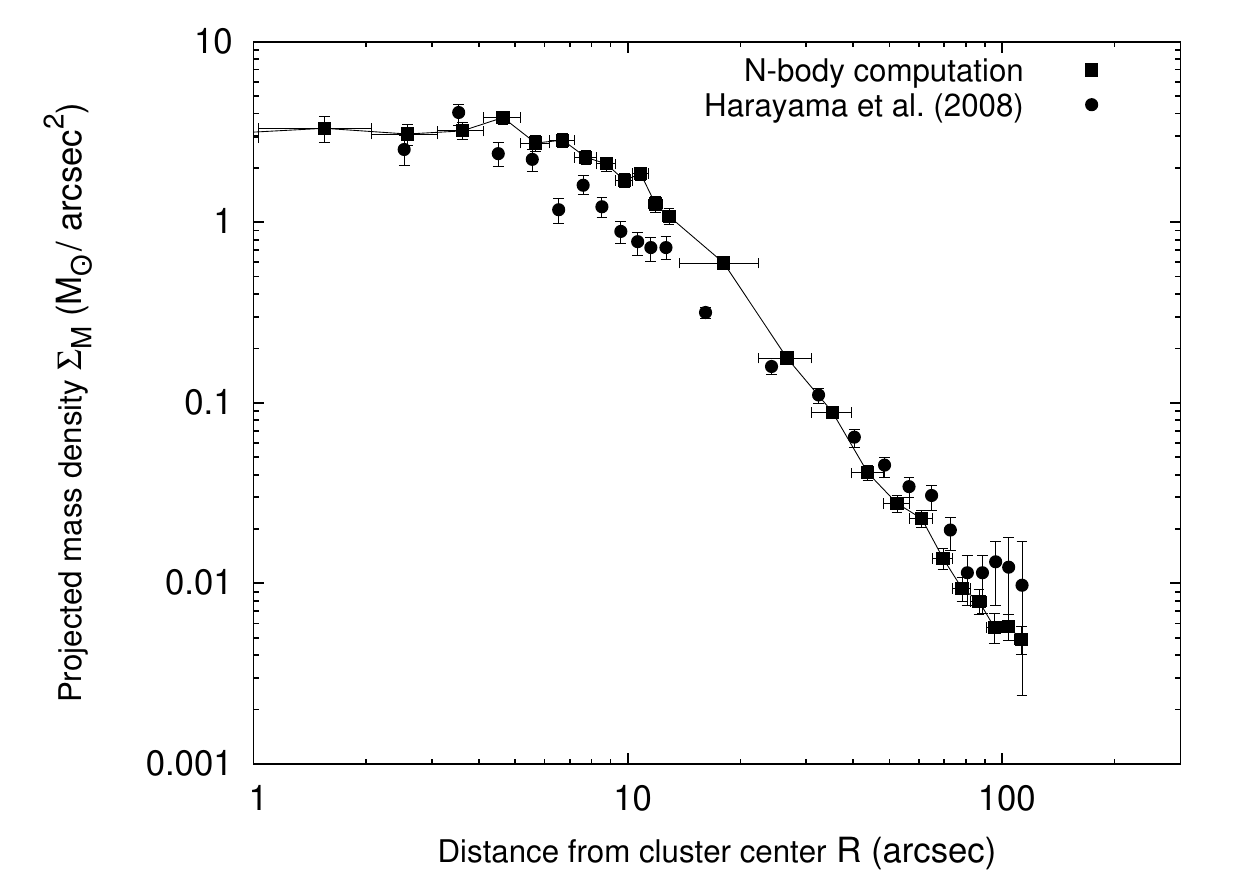}
\caption{
The computed projected mass density profile (filled squares and solid line)
for the stellar mass range $0.5\Ms<m<2.5\Ms$ at $t\approx1.4$ Myr, from
the computed model HD97950b ($\rh(0)\approx0.25$ pc, $\mcl(0)\approx10000\Ms$;
see Table~\ref{tab1}) containing an initial primordial
binary population (see Sec.~\ref{ngc3603}).
This computed profile shows remarkable agreement with the observed profile \citep{hara2008},
for the same stellar mass range, of the central young cluster (HD97950)
of NGC 3603 (filled circles). The angular annuli (the horizontal error bars)
used for computing the projected densities are nearly the same as
those used by \citet{hara2008} to obtain the observed profile.
The vertical error bars are the Poisson errors for the individual annuli.
This panel is reproduced from \citet{sb2014}.
}
\label{fig:sb14_mprof}
\end{figure}

In \citet{sb2014}, a set of initial conditions is presented which remarkably
reproduce the above structural and kinematic data of the HD97950 cluster,
\viz, the models ``HD97950s/b'' of Table~\ref{tab1}. These computed
clusters (using {\tt NBODY6}) have the same initial conditions,
except that HD97950b contains a primordial
binary population. This binary population is taken to be the
birth population \citep{pk1995b}, except that a uniform distribution in $\log_{10} P$
between $0.3 < \log_{10} P < 3.5$ and a mass ratio biased towards unity
(ordered pairing; as introduced by \citealt{oh2012})
is used for stellar masses $m>5\Ms$. This is 
motivated by the observed period distribution of O-star
binaries in nearby O-star rich clusters \citep{sev2011,chini2012}.
It is currently unclear at which stellar mass and how
the orbital period law changes and the above switching of the
$P$-distribution at $m=5\Ms$, therefore, is
somewhat arbitrary. Note that for this $P$-distribution, the
primordial binaries are much tighter and hence energetic in
dynamical encounters for $m>5\Ms$.

\begin{figure}
\centering
\includegraphics[width=9.5cm, angle=0]{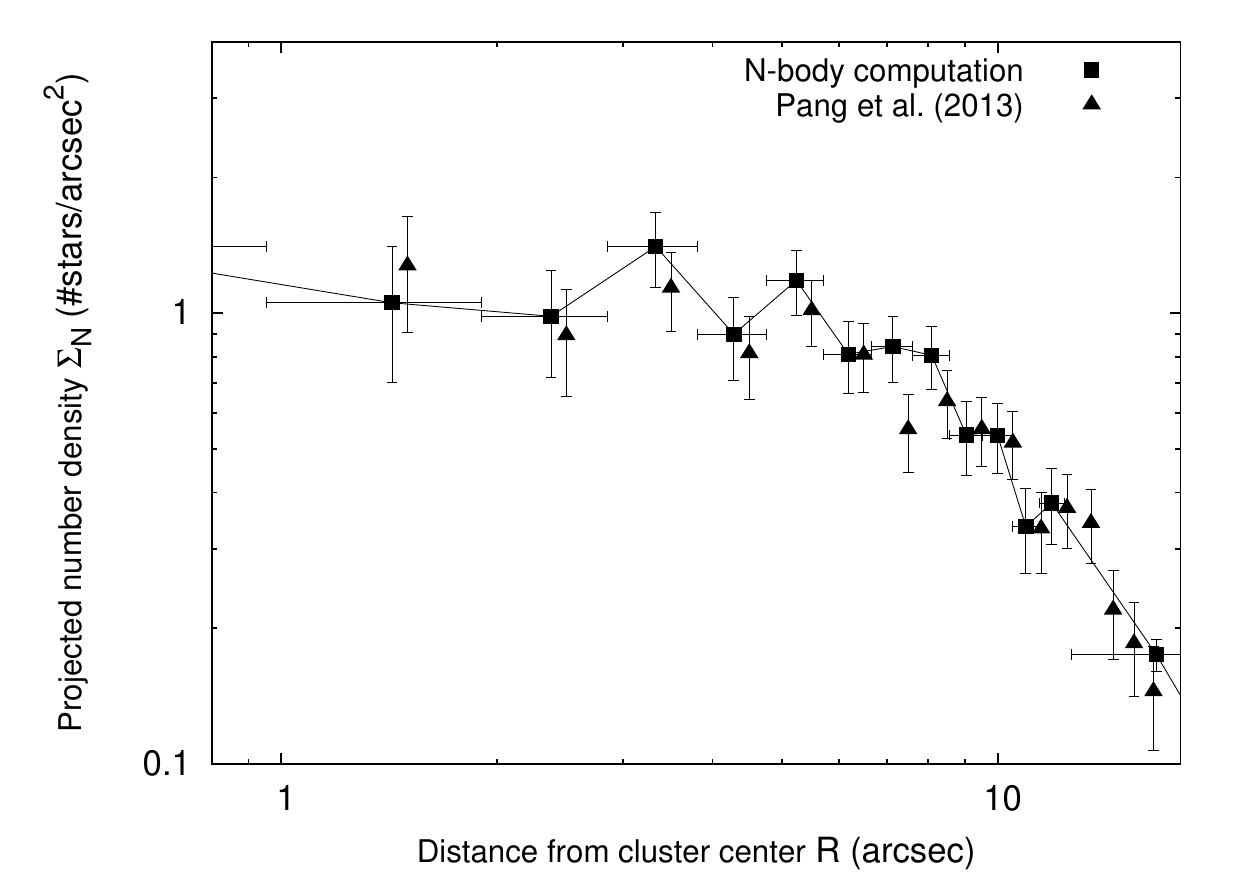}
\caption{
The computed projected stellar number density profile for the calculation 
HD97950b (Table~\ref{tab1}; filled squares and solid line). This
shows a remarkable agreement at $t\approx1.4$ Myr with the same obtained with the stars of
HD97950 from the HST/PC chip (up to the central $15''$; filled triangles;
chip data from \citealt{pang2013}). In constructing the computed profile,
the incompleteness in the detection of the stars is taken into account that depends on the
stellar mass (luminosity) and projected angular annuli on the cluster,
as given in \citet{pang2013}. In this comparison,
similar angular annuli (horizontal error-bars) are used to construct the density profiles.
The vertical error bars are the corresponding Poisson errors.
This panel is reproduced from \citet{sb2014}.
}
\label{fig:sb14_sprof}
\end{figure}

Both the computed models reproduce the HD97950 cluster reasonably but
the one with the above primordial binary distribution (\ie, HD97950b)
does better in terms of matching the central velocity dispersion.
A substantial fraction of tight massive binaries ($\approx50$\%
in this case) augments the central velocity dispersion due to
energetic binary-single interactions (binary heating) and makes
it agree better with the observed value in this case \citep{sb2014}. 
Here, only the HD97950b model is detailed.

Fig.~\ref{fig:sb14_mprof} shows the surface or projected mass
density profile $\Sigma_M$ at $t\approx1.4$ Myr for the HD97950b model
(filled squares joined by solid line).
It matches remarkably with the observed profile in
HD97950 (filled circles; \citealt{hara2008}). 
Note that in this comparison a similar stellar mass range
and annuli as those for the observed profile are used to
construct the density profile from the computed cluster.

Fig.~\ref{fig:sb14_sprof} shows the radial profile of the
incompleteness-limited stellar number density
$\Sigma_N$ from the above computed cluster (filled squares joined by
solid line) at $t\approx1.4$ Myr and that obtained
from the HST \citep{pang2013} (filled triangles) which 
agree remarkably. Here, the computed stellar distribution is
sampled according to the radius and mass-dependent
incompleteness fraction particular to this observation \citep{pang2013}.
This mimics the ``observation'' of the model cluster.
Note that in constructing both of these density profiles
we include only the most massive member (primary) of
a binary which would dominate the detected light from it.

\begin{figure}
\centering
\includegraphics[width=9.5cm, angle=0]{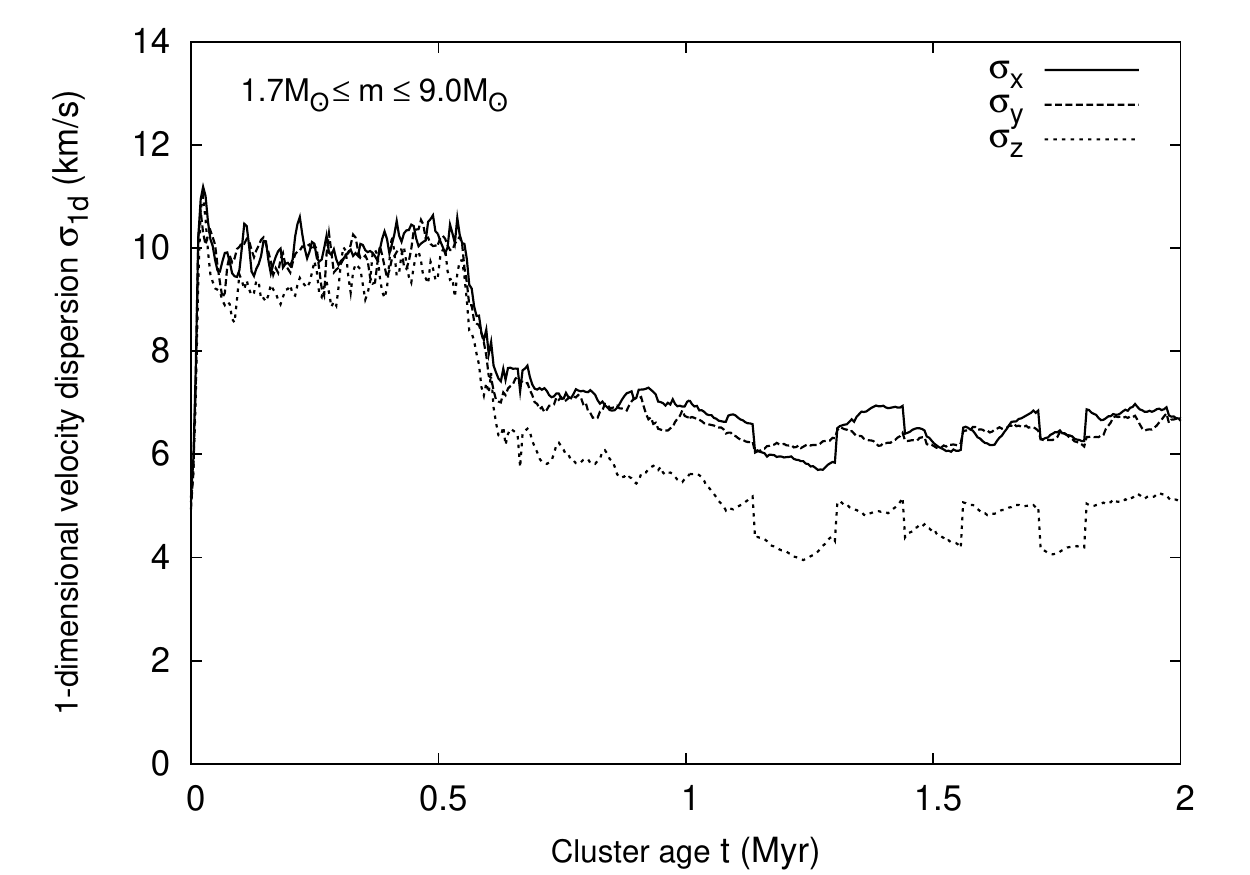}
\includegraphics[width=9.5cm, angle=0]{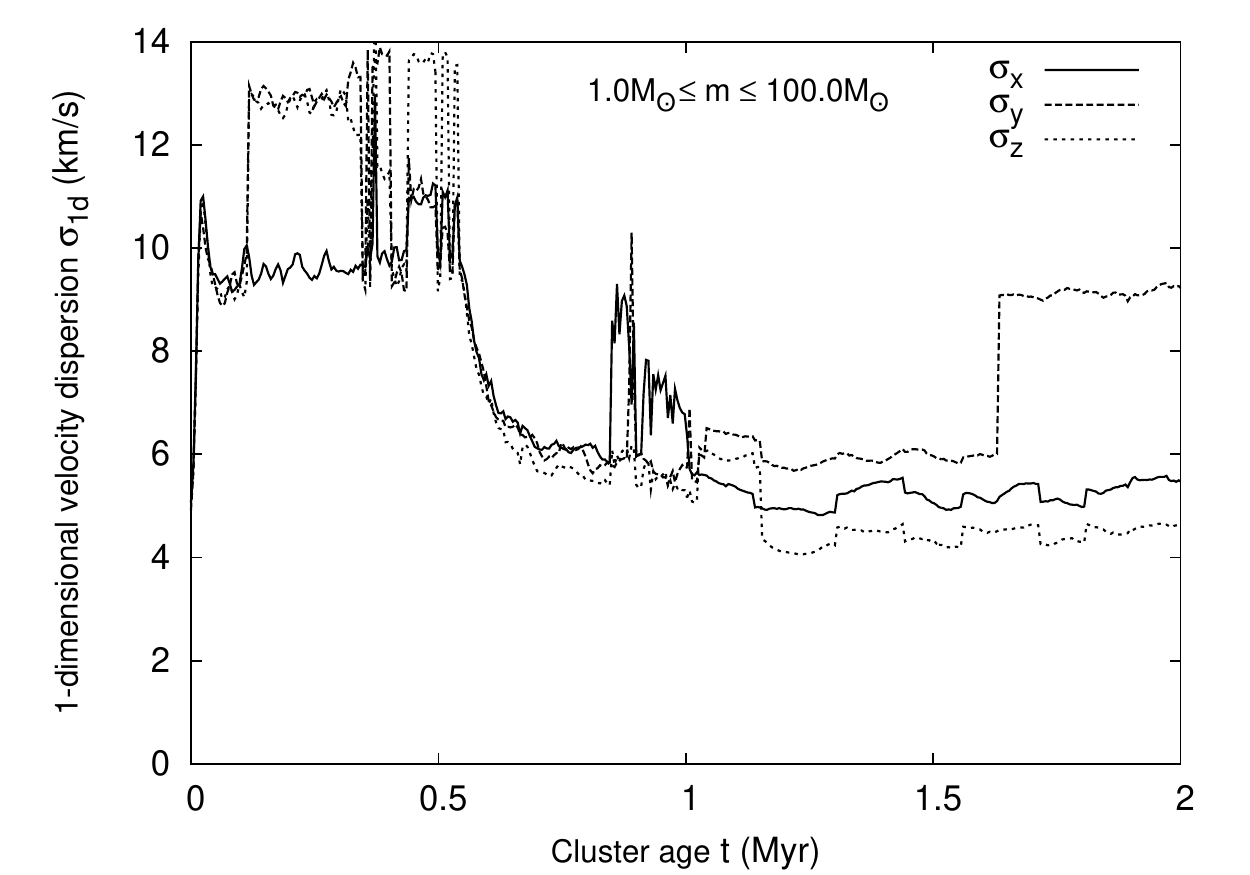}
\caption{
The time evolution of the one-dimensional velocity dispersions, $\sigoned$s,
for the model cluster HD97950b (see Table~\ref{tab1}),
the density profile of which is shown in Fig.~\ref{fig:sb14_mprof}.
They are obtained for $R<0.5$ pc ($\approx15''$) and correspond to
the stellar mass ranges $1.7\Ms<m<9.0\Ms$
(as in \citealt{roch2010}; top panel) and
$1.0\Ms<m<100.0\Ms$ (as in \citealt{pang2013}; bottom panel).
The $\sigoned$s obtained here correspond to the COMs of the
cluster single-stars and binaries. The computed values of
$\sigoned$ (bottom panel) differ in orthogonal directions
as found in observations \citep{pang2013} and span the same range as observed
(4.5-7.0 ${\rm~km~s}^{-1}$) between 1.0 - 1.5 Myr cluster age, implying good agreement.
These panels are reproduced from \citet{sb2014}.
}
\label{fig:sb14_v1d}
\end{figure}

\begin{figure}
\centering
\includegraphics[width=9.5cm, angle=0]{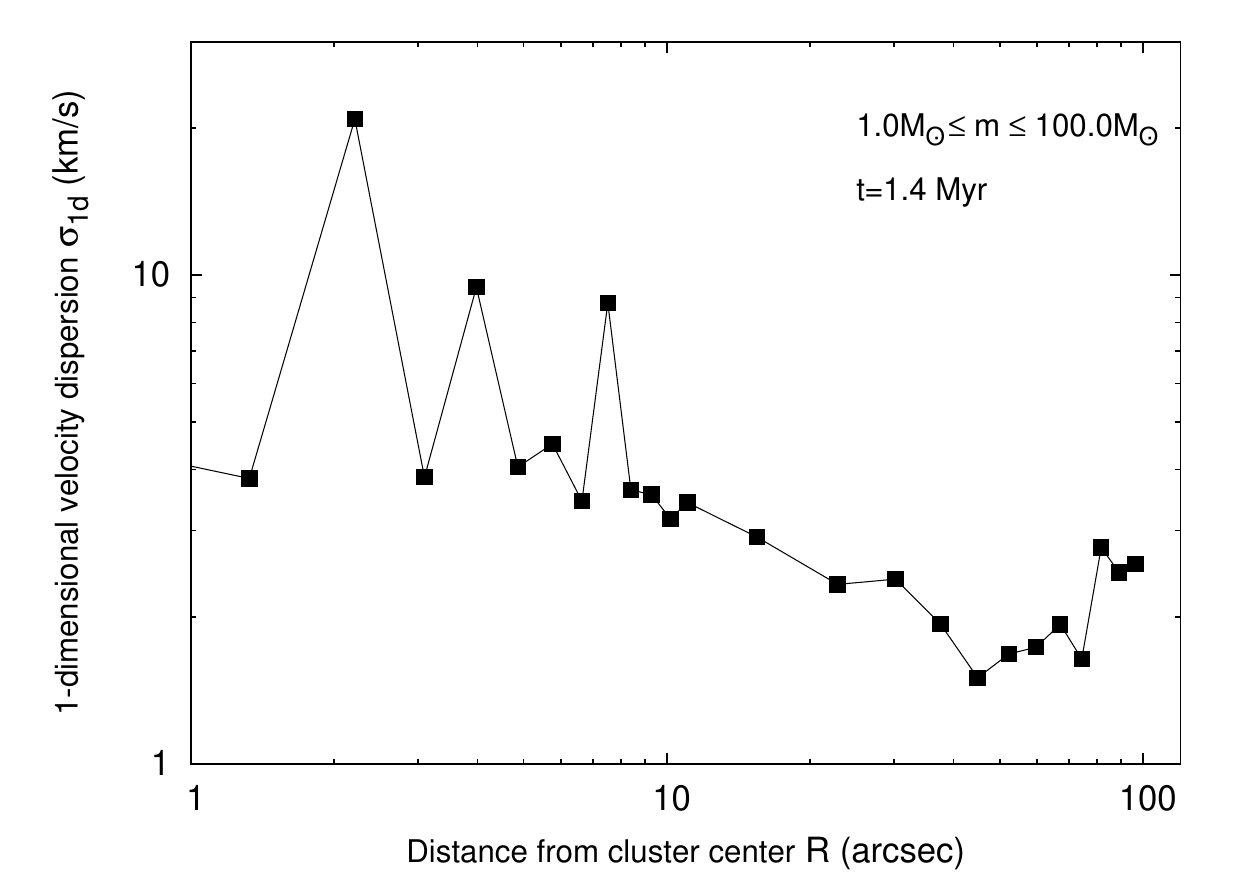}
\caption{
Radial variation of one-dimensional velocity dispersion, $\sigoned$,
for the computed HD97950b model (here in presence of a tidal field)
at $t=1.4$ Myr for stellar mass range $1.0\Ms\leq m\leq100.0\Ms$.
The overall increasing trend of $\sigoned$ with $R$ in the outer regions
($R\gtrsim40''$ in this case) is due to the recent gas expulsion.
The tangential velocities of selected stars in \citet{pang2013}
(for $R\lesssim60''$) do show an increasing trend with $R$.
This panel is reproduced from \citet{sb2014}.
}
\label{fig:sb14_vprof}
\end{figure}

Fig.~\ref{fig:sb14_v1d} (bottom) shows the
evolution of the (one-dimensional) dispersion
of the stellar velocity components, $\sigoned$ $({\rm 1d=x,y,z})$,
for the HD97950b model for $1.0\Ms<m<100.0\Ms$ within $R<0.5$ pc.
The computed $\sigoned$s lie between $4.0<\sigoned<7.0{\rm~km~s}^{-1}$
for $1<t<2$ Myr. The corresponding 
observed one-dimensional velocity dispersions indeed vary
considerably with orthogonal directions \citep{pang2013}
like the computed ones here (see Fig.~\ref{fig:sb14_v1d}; bottom panel)
and their variation well matches the above computed range.
Fig.~\ref{fig:sb14_v1d} (top) shows the $\sigoned$s
corresponding to the stellar mass range $1.7\Ms<m<9.0\Ms$
as in \citet{roch2010}. The corresponding mean $\sigoned$
is consistent with that obtained by \citet{roch2010} from
HST proper motions. The abrupt vertical excursions in
$\sigoned$ in Fig.~\ref{fig:sb14_v1d} (bottom)
are due to energetic two- or few-body encounters which are
most frequent for the most massive stars and binaries
as they centrally segregate the most via two-body relaxation.

Fig.~\ref{fig:sb14_vprof} shows radial profiles of
$\sigoned$, from HD97950b, at $t=1.4$ Myr for $1.0\Ms\leq m\leq100.0\Ms$.
Here, $\sigoned$
tends to increase for $R\gtrsim40''(\approx1.2{\rm~pc})$
which can be attributed to
the recent gas expulsion from the system causing its outer
parts to still expand.
Such a trend, which becomes more pronounced the closer
the epoch of observation is to the gas expulsion, can
be tested by future, more accurate determinations
of stellar proper motions in the outer regions of HD97950,
\eg, by \emph{Gaia}. Notably, the measured tangential
velocities (from HST proper motions) of selected stars in
\citet{pang2013} indeed show an increasing trend with
radial distance in the outer region (measured up to $R\approx60''$)
of HD97950. Note that the inner annuli of the above
computed cluster are already virialized at $t=1.4$ Myr
but the outer region is still
far from re-virialization (see Fig.~1 of \citealt{sb2014}). 
As demonstrated in Sec.~\ref{r136}, the overall re-virialization
time for such a cluster is $\tvir\approx2$ Myr.

Hence, the HD97950 computed cluster well reproduces the structure
and the internal kinetics of the observed HD97950 cluster. In other
words, model HD97950b (also HD97950s to some extent) is a monolithic ``solution'' of
the NGC 3603 cluster; it represents an initial stellar distribution that
would evolve self consistently to make the HD97950 cluster at its
appropriate age.

\begin{framed}
Initial Plummer-profiled and highly compact ($\rh(0)\approx0.2-0.3$ pc)
monolithic embedded clusters, when subjected to residual
gas expulsion, remarkably reproduce the hitherto
known kinematic and structural properties of the well observed young clusters
the ONC, R136 and NGC 3603. The properties of the required gas expulsion are seemingly
universal ($\epsilon\approx0.3$,
$\td\approx0.6$ Myr and $\vg\approx10{\rm~km~s}^{-1}$).
Such computed model clusters are the only ones to date that directly reproduce
these observed clusters.
\end{framed}

\section{Hierarchical formation of young massive clusters: the case of
NGC 3603 young cluster}\label{himrg}

Although an episodic and in-situ formation scenario is well
consistent with the detailed properties of several observed
Galactic or local VYMCs (Sec.~\ref{monoform}), it is still puzzling how
such smooth initial condition can be connected with irregular, substructured
and/or filamentary morphology of GMCs and embedded stellar distributions
(see Sec.~\ref{intro}). Substructures are also found
in several gas-free or near gas-free very young star clusters,
even though they may have an overall core-halo profile \citep{kuhn2014}.  
Computations of gravitational fragmentation in turbulent gas clouds
also point to a highly substructured beginning of a
star cluster (see Sec.~\ref{intro}).

\begin{figure}
\centering
\includegraphics[width=11.0cm, angle=0]{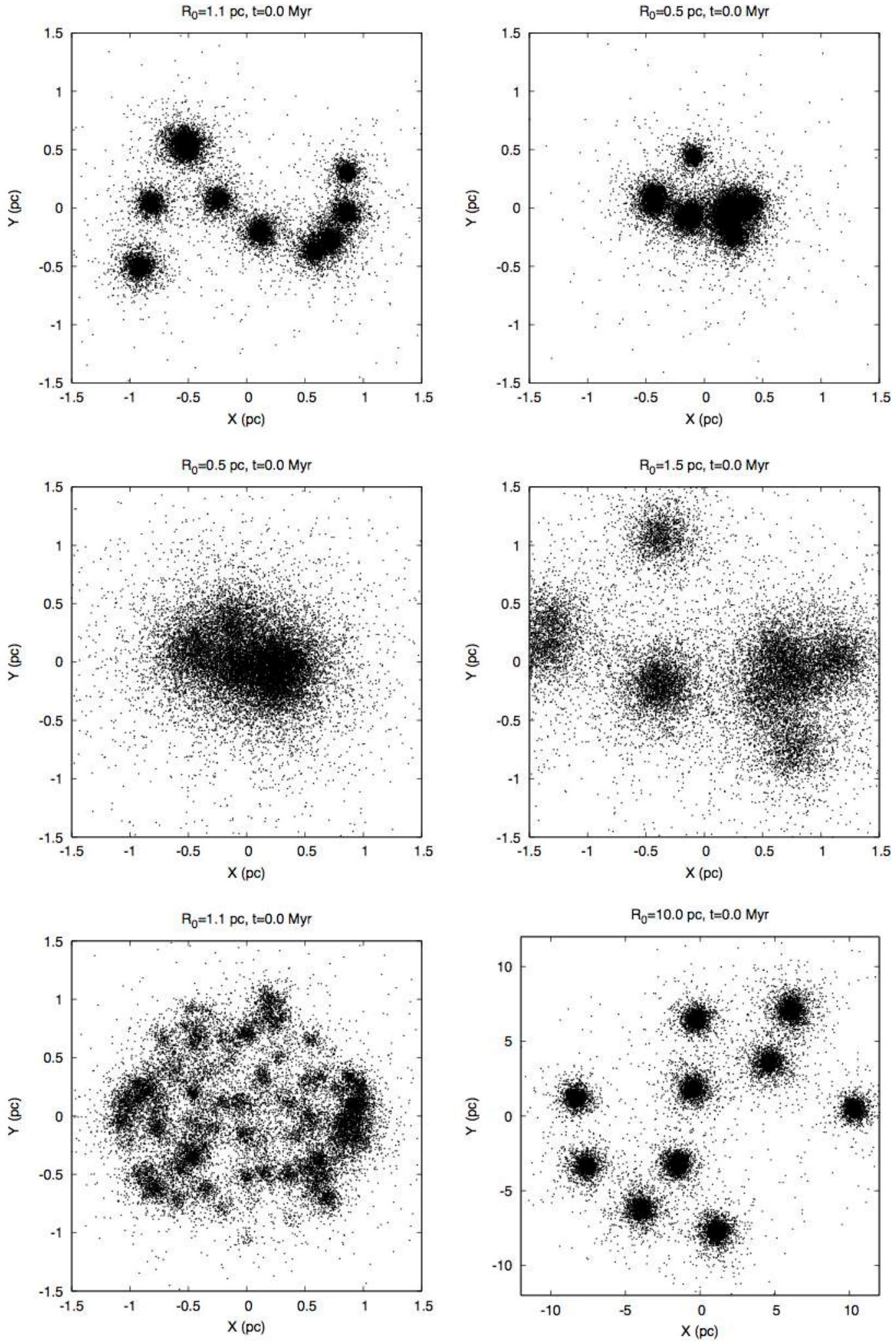}
\caption{
The primary varieties of the \emph{initial} configurations considered in Sec.~\ref{himrg},
shown in projection. Here, the panels are numbered left-to-right, top-to-bottom.
In each case, a set of Plummer spheres (subclusters) are uniformly distributed over a spherical
volume of radius $R_0$, totalling a stellar mass of
$M_\ast\approx10000\Ms$. Panels 1, 2, 3, 4 and 6 are examples of type A or
``blobby'' systems containing 10 subclusters of $m_{cl}(0)\approx10^3\Ms$ each.
With smaller $R_0$, the subclusters overlap more with each other (\cf, panels 1 \& 2
with subcluster half mass radius $r_h(0)\approx0.1$ pc and panels
3 \& 4 with $r_h(0)\approx0.3$ pc). This is also true for increasing $r_h(0)$
(\cf, panels 1 \& 4). Panel 5 is an
example of type B or ``grainy'' initial configuration containing $\approx 150$
subclusters of mass range $10\Ms\lesssim m_{cl}(0) \lesssim 100\Ms$.
While panels 1-5 are examples of ``compact'' configurations,
for which $R_0\leq2.5$ pc, panel 6, with $R_0=10$ pc, represents an ``extended''     
configuration where the subclusters are much more distinct. 
See Sec.~\ref{himrg} for details of the initial setups.
These panels are reproduced from \citet{sb2014b}.}
\label{fig:hi_init}
\end{figure}

\begin{table}
\centering
\begin{minipage}{4.0 in}
\centering
\caption{A basic classification of the different morphologies in the spatial distribution
of stars that can occur in the process of subcluster merging. These morphologies appear
in the models computed here (Sec.~\ref{himrg}).
Note that the distinctions among these morhphologies
are only qualitative and are made for the ease of descriptions.
This table is reproduced from \citet{sb2014b}.}
\label{tab:morphdef}
\begin{tabular}{lc}
\hline
Morphology & Abbreviation \\
\hline 
Substructured                               & SUB  \\
Core + asymmetric and/or substructured halo & CHas \\
Core-halo with near spherical symmetry      & CH   \\
Core + halo containing satellite clusters   & CHsat \\
\hline
\end{tabular}
\end{minipage}
\end{table}

\begin{table}
\centering
\begin{minipage}{4.8 in}
\centering
\renewcommand{\thefootnote}{\alph{footnote}}
\caption{An overview of the evolutionary sequences of the primary systems as computed here
(Sec.~\ref{himrg}). A particular row
indicates how the morphology (see Table~\ref{tab:morphdef})
of the corresponding system evolves with evolutionary time (in Myr as indicated by the numerical
values along the columns 3-5 and 6-8), for systems both without and with a background gas potential (see text).
As expected, the systems, in general, evolve from substructured to a core-halo configuration with a timescale
that increases with increasing initial extent $R_0$. See text for details.
This table is reproduced from \citet{sb2014b}.
}
\label{tab:morphevol}
\begin{tabular}{cclllclll}
\hline
Config. name & Short name
& \multicolumn{3}{c}{Without gas potential} &  &
\multicolumn{3}{c}{With gas potential $(\epsilon\approx0.3)$}\\
\hline
m1000r0.1R1.1N10            
\footnote{m$x$r$y$R$z$N$n$ implies an initial system (at $t=0$)
comprising of N$=n$ Plummer clusters, each of mass m$=m_{cl}(0)=x\Ms$
and half-mass radius r$=r_h(0)=y$ pc, distributed uniformly over a spherical volume of radius R$=R_0=z$ pc.}
&     A-Ia     &        0.2,SUB&0.6,CHas&1.0,CH       &  &     0.2,SUB&0.6,CHas&1.0,CH\\
m1000r0.3R1.1N10            &     A-Ib     &        0.2,SUB&0.6,CHas&1.0,CHas     &  &     0.2,SUB&0.6,CHas&1.0,CHas\\
m1000r0.1R2.5N10            &     A-IIa	   &        0.6,SUB&1.0,SUB&2.0,CH        &  &     0.6,SUB&1.0,CHsat&2.0,CHas\\
m1000r0.3R2.5N10            &     A-IIb    &        0.6,SUB&1.0,SUB&2.0,CHas      &  &     0.6,SUB&1.0,CHsat&2.0,CHas\\
m10-150r0.01-0.1R1.1N150    
\footnote{Further, when a range of values $x1-x2$ is used instead of a single value, it implies that the corresponding
quantity is uniformly distributed over $\left[x1,x2\right]$ at $t=0$.}
&     B-Ic     &        0.2,SUB&0.6,CHas&1.0,CH       &  &     0.2,SUB&0.6,CHas&1.0,CHas\\
m10-150r0.1R1.1N150         &     B-Ia     &        0.2,SUB&0.6,CHas&1.0,CH       &  &     0.2,SUB&0.6,CHas&1.0,CHas\\
m10-150r0.1R2.5N150         &     B-IIa    &        0.6,SUB&1.0,SUB&2.0,CHas      &  &     0.6,SUB&1.0,CHas&2.0,CHas\\
m1000r0.5-1.0R5.0N10        &     A-IIId   &        1.0,SUB&2.0,SUB&3.0,CHas      &  &     1.0,CHsat&2.0,CHsat&3.0,CHsat\\
m1000r0.5-1.0R10.0N10       &     A-IVd    &        1.0,SUB&2.0,SUB&3.0,SUB       &  &     1.0,SUB&2.0,SUB&3.0,CHsat\\
\hline
\end{tabular}
\end{minipage}
\end{table}

One way to ``add up'' these two apparently conflicting pictures of VYMC
formation is indicated by the timescale problem of hierarchical
formation as discussed in Sec.~\ref{monoreason}. 
Essentially, the age, density and velocity dispersion profiles
of observed VYMCs well constrain the admissible initial spatial
scale of any subcluster system from which the VYMC many have formed.
In particular, Fig.~\ref{fig:tin}
implies that substructures can appear and migrate from sufficiently
close separation to possibly form a VYMC within a few Myr. Such
``prompt hierarchical merging'' can connect a monolithic initial condition,
which successfully explains observed VYMCs \citep{pketl2001,sb2013,sb2014}
and general properties
of young clusters \citep{pf2009,pfkz2013}, to the conditions
in dense star-forming molecular regions. The detailed observed
properties of the HD97950 cluster again provides a testbed
for such a scenario as discussed below.  

In \citet{sb2014b}, substructured initial conditions
are generated by distributing
compact Plummer spheres uniformly over a spherical volume of radius $R_0$.
The total stellar mass distributed in this way
is always the lower photometric mass estimate of $M_\ast\approx10^4\Ms$ for HD97950,
as motivated by \citet{sb2014}; see Sec.~\ref{ngc3603}.
This fashion of initial subclustering is an idealization and
extrapolation of what is found in the
largest SPH calculations of cluster
formation to date \citep{bate2009,bate2012,giri2011} (see Sec.~\ref{intro}).
As discussed in Sec.~\ref{params}, the initial half mass radii, $\rh(0)$,
of these Plummer subclusters are taken typically between 0.1-0.3 pc,    
in accordance with the observed widths of these
highly compact molecular-cloud filaments \citep{andr2011,schn2012}. Such
compactness of the subclusters is also
consistent with those observed in stellar complexes, \eg, in
the Taurus-Auriga \citep{palsth2002}. However, in some
calculations, larger $r_h(0)$s are also used (see Table~\ref{tab:morphevol}).

The number of subclusters, $n$, over which the $M_\ast\approx10^4\Ms$ is subdivided 
has to be chosen somewhat arbitrarily.
To keep a broad range of possibilities,
two primary cases of the initial subdivision of
the total stellar mass are considered.
The ``blobby'' (type A) systems comprise 10 subclusters 
of $\mcl(0)\approx10^3\Ms$ each. Panels 1, 2, 4 and 6 of Fig.~\ref{fig:hi_init}
are examples of such initial systems. Note that in
this and all the subsequent figures,
the panels are numbered left-to-right, top-to-bottom, unless stated
otherwise. The ``grainy'' (type B) systems comprise
$\approx 150$ subclusters with mass range $10\Ms\lesssim \mcl(0) \lesssim 100\Ms$
summing up to $M_\ast\approx10^4\Ms$.
The mode of initial subdivision does not influence the key
inferences from these calculations.

The initial spanning radius, $R_0$, is taken over a wide range,
\viz, $0.5{\rm~pc}\lesssim R_0\lesssim 10.0{\rm~pc}$,
to explore the wide range of molecular cloud densities (see below)
and spatial extents as observed in star-forming regions
and stellar complexes. Table~\ref{tab:morphevol}
provide a comprehensive list of the initial conditions
for the computations in \citet{sb2014b}. The detailed nomenclature
of the computed model, in its first column,
is explained in Table~\ref{tab:morphevol} and the
corresponding short names, in the second column, are self-explanatory.

As explained in Sec.~\ref{params}, the proto-stellar mass
function is taken to be canonical but without any upper bound. This
would cause the IMF of the merged cluster, with stellar mass
$\mstar(\equiv\mcl)$, also to be canonical
as often observed in VYMCs.
Note that the gas accretion and the dynamical processes mostly determine
the massive tail of the IMF and also sets the maximum stellar mass,
$m_{\rm max}$, of the \emph{final} cluster,
as seen in hydrodynamic calculations (\eg, \citealt{kl1998,giri2011}).
This gives rise to an $m_{\rm max} - \mcl$ relation that is
consistent with that found from observations \citep{wk2004,wp2013}.
Note that if the $m_{\rm max} - \mcl$ relation applies
to the pre-merger subclusters, then the $m_{\rm max} - \mcl$ relation
for the final cluster will show features arising from the
merging process \citep{wd2010,wp2013}. 

All subclusters are initially at rest w.r.t. the centre
of mass (COM) of the stellar system.
While this condition is again an idealization, 
it is consistent with the results of detailed hydrodynamic
computations in which the system(s) of subclusters formed
is(are) typically sub-virial. Also, for the ease of
computing, primordial binaries are excluded from these calculations.
Test calculations show that primordial binaries
do not influence the subcluster merging process significantly.
The subclusters are generated using the {\tt MCLUSTER} utility
\citep{kup2011} which is integrated in a special program
that generates the overall subcluster system with the
intended parameters.

The dense residual molecular cloud
is represented by a background, external gravitational potential
of a Plummer mass distribution which declines
exponentially as discussed in Sec.~\ref{gasmodel}. In this
way the overall dynamical effect of the molecular cloud is included
(as in the previous studies). In order to compare with
the previous studies \citep{pketl2001,sb2014}, we adopt a local SFE of $\epsilon\approx33$\%
within the span of the subclusters, $R_0$.
Such an SFE is as well consistent with those obtained from self-regulated
hydrodynamic calculations and also with observations of embedded systems
in the solar neighborhood (see Secs.~\ref{intro} \& \ref{params}).  
Hence, the geometric/density centre of the Plummer gas sphere is co-incident with the COM
of the initial stellar system and its half mass radius is equal to $R_0$
which contain $2\mstar$ mass, giving $\epsilon=1/3$ within $R_0$.
For the entire Plummer cloud (of $4\mstar$), $\epsilon=1/5$.
Inserting the adopted value $\mstar=10^4\Ms$ (see above),
one gets $3\times10^4\Ms$ (gas + stars) within $R_0$.
This gives an ONC-like $\rho_g\approx6\times10^3\Ms{\rm~pc}^{-3}$
gas density for $R_0=1.06{\rm~pc}$ and $\approx1/1000$th of this for $R_0=10{\rm~pc}$ which
is appropriate for, \eg, the Taurus-Auriga complex.

\subsection{General evolutionary properties of
subcluster systems}\label{genevol}

\begin{figure}
\centering
\includegraphics[width=11.0cm, angle=0]{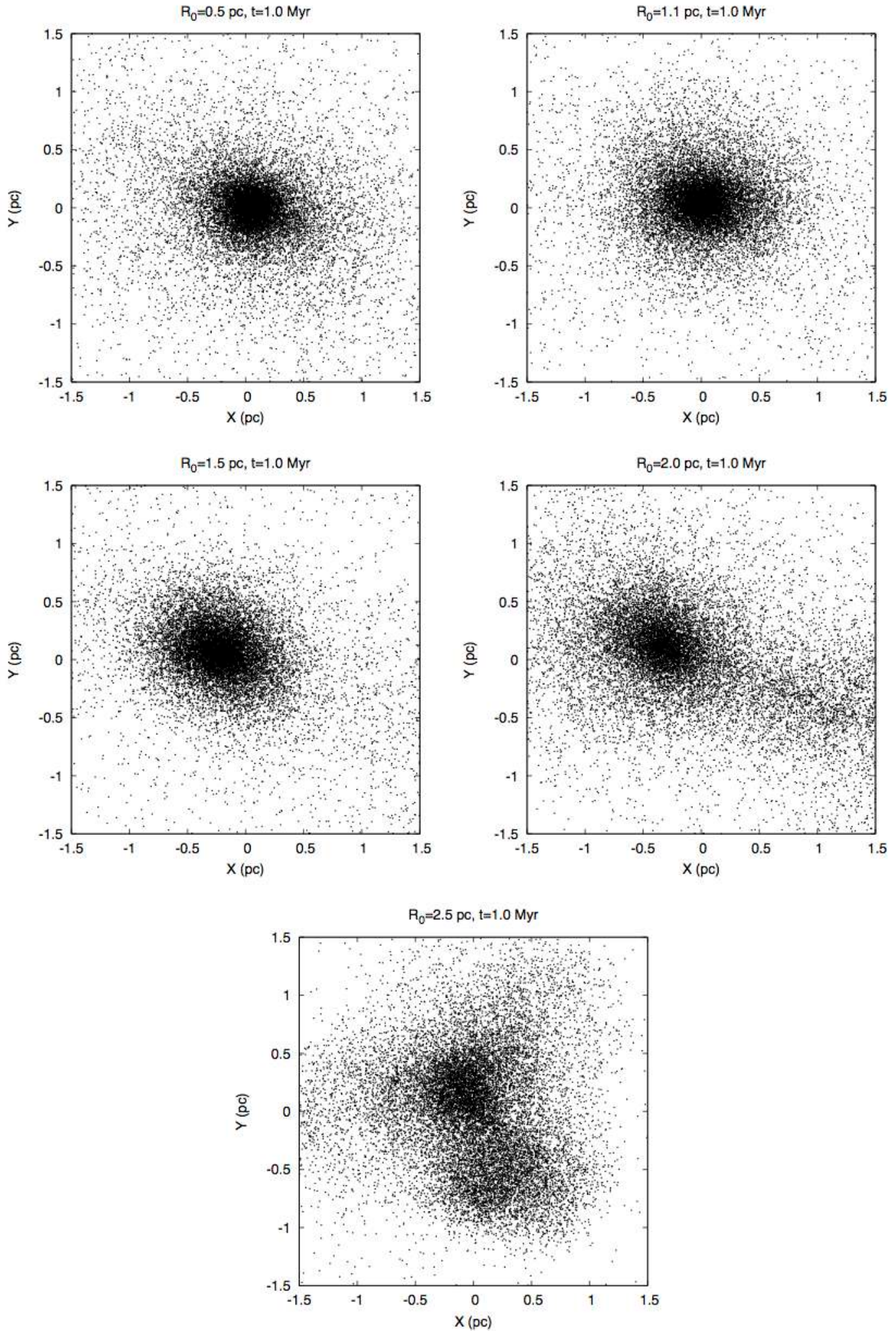}
\caption{
Configurations obtained at $t\approx1$ Myr with increasing initial span $R_0$
(without background gas potential).
With increasing $R_0$, the system's morphology at $t\approx1$ Myr changes from being
near-spherical core-halo (CH; panels 1,2; numbered left-to-right, top-to-bottom),
asymmetric core-halo (CHas; panels 3,4) to substructured
(SUB; panel 5). For $R_0\gtrsim2$ pc (panels 4,5), the stellar system is still well in the process of
merging at $t\approx1$ Myr after the subclusters' first
pericenter crossings (\ie, it is in the violent relaxation phase
$t_{in}<t<t_{in}+t_{vrx}$; see Sec.~\ref{genevol}).
These panels are reproduced from \citet{sb2014b}.
}
\label{fig:morphevol}
\end{figure}

As discussed in Sec.~\ref{monoreason}, the subclusters pass through
each other for the first time at the system's potential minimum
in a time $\tin$ as given by Eqn.~\ref{eq:tin2}. The final merger
of the subclusters, however, is completed after an additional 
violent relaxation time, $\tvrx$, which can be several subcluster
orbital times. During this time, the orbital energy of the subclusters
is dissipated, in multiple mutual passes, in the individual stellar
orbits; this corresponds to the re-virialization process in
initially monolithic systems (see Sec.~\ref{monoform}).
Both $\tin$ and $\tvrx$ increases with the initial span
of the subclusters $R_0$ and hence the time for forming the
final merged, (near) spherical cluster. 

Table~\ref{tab:morphevol} summarizes the evolution (in Myr)
of the subcluster systems  of type A and B (see above)
with increasing $R_0$ and in presence and absence of a gas potential (see above).
Only the primary templates are included here which are computed using {\tt NBODY6}.
For description purposes, the evolving morphology of the stellar system
is divided into four categories as in Table~\ref{tab:morphdef}.
All computed configurations initiate as SUB
and evolve via the intermediate CHas phase to the final CH cluster in dynamical
equilibrium. $R_0\lesssim 1{\rm~pc}$
systems attain a CH structure in $t\lesssim1{\rm~Myr}$ without a gas potential.
On the other hand, initially wider configurations remain SUB
at $t=1{\rm~Myr}$ even with the gas potential and most of them do not attain
the CH phase even in 2 Myr.
In all such calculations, a negligible fraction of stars escape the system
during the infall and the merger process. 
In other words, the total bound stellar mass $M_\ast$
remains nearly unaltered as the system evolves from SUB to CH
configuration.

Fig.~\ref{fig:morphevol} shows the snapshots at $t\approx1{\rm~Myr}$
for a set of A-type configurations (Table~\ref{tab:morphevol}) falling from increasing
$R_0$ (without background gas potentials). With $R_0$,
the morphology at $1{\rm~Myr}$ changes from being CH, CHas to SUB. For $R_0\gtrsim2{\rm~pc}$,
the structure at $1{\rm~Myr}$ substantially deviates from spherical symmetry (and dynamical
equilibrium). 

It is worth noting that due to energy conservation
the size of the final cluster
in dynamical equilibrium can be simply related to
that of the initial subclusters (of equal or similar size and mass) as \citep{sb2014b},
\begin{equation}
\frac{1}{2R_\ast}\approx\frac{1}{2nR_{cl}}+\frac{1}{R_0}.
\label{eq:Rmrg}
\end{equation}
Here, $R_\ast$ is the half mass radius of the final cluster and $R_{cl}$
is that for the initial subclusters. 

\begin{framed}
The morphology of a gravitationally bound stellar population (of a given total mass) at a
given age depends on the initial length scale over which the population
is hatched (\ie, from the mutual separation from which they fall in the resultant
potential well). The dynamical timescale of the stellar population is the
key in determining the morphology and length scale of the stellar distribution at
the epoch of observation. A sufficiently spread-out distribution can remain
highly substructured for 10s of Myr. On the other hand, a spherical massive
star cluster in dynamical equilibrium can form out of a closely distributed (typically $\lesssim2$ pc)
but highly substructured stellar population in $<1$ Myr.
\end{framed}

\subsection{Comparison with NGC 3603 young cluster}\label{nycomp}

\begin{figure}
\centering
\includegraphics[width=5.7cm, angle=0]{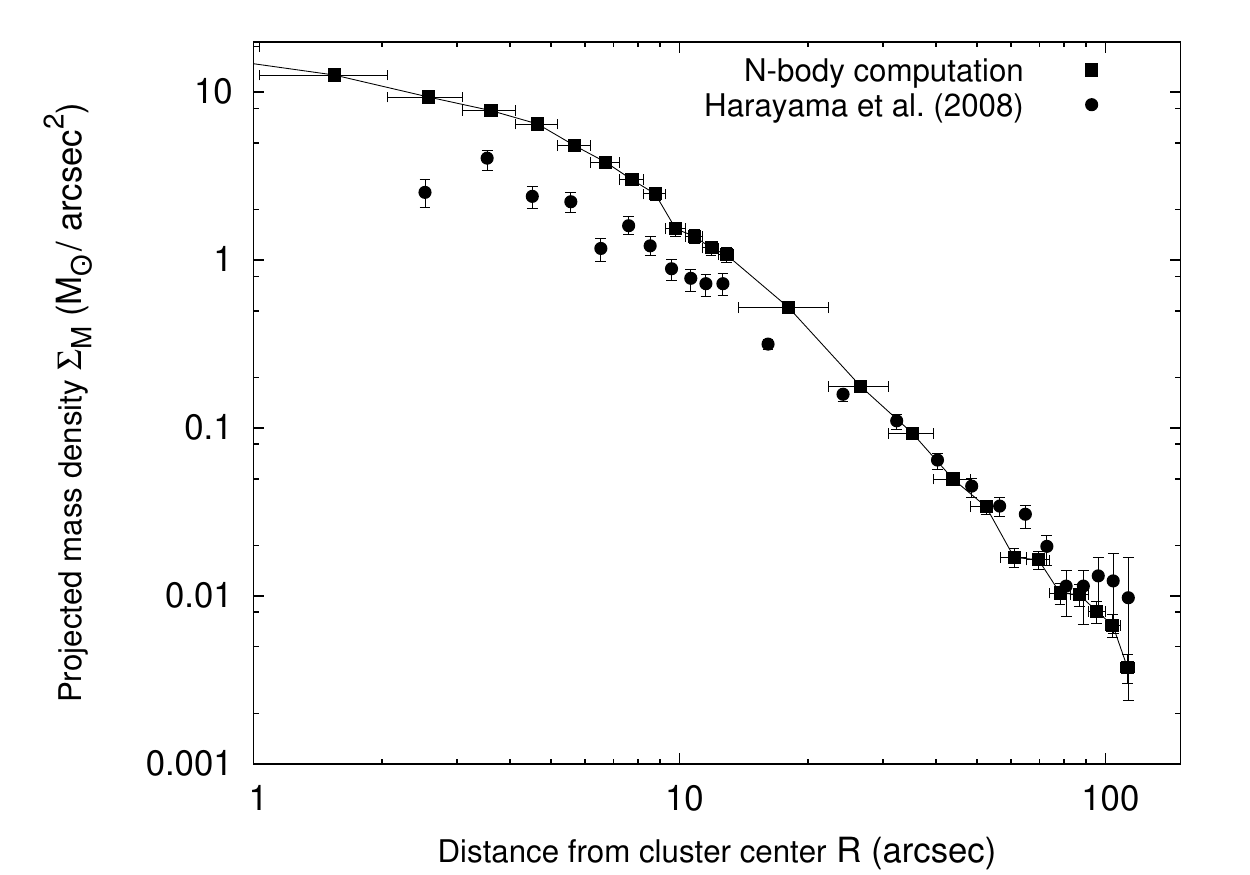}
\includegraphics[width=5.7cm, angle=0]{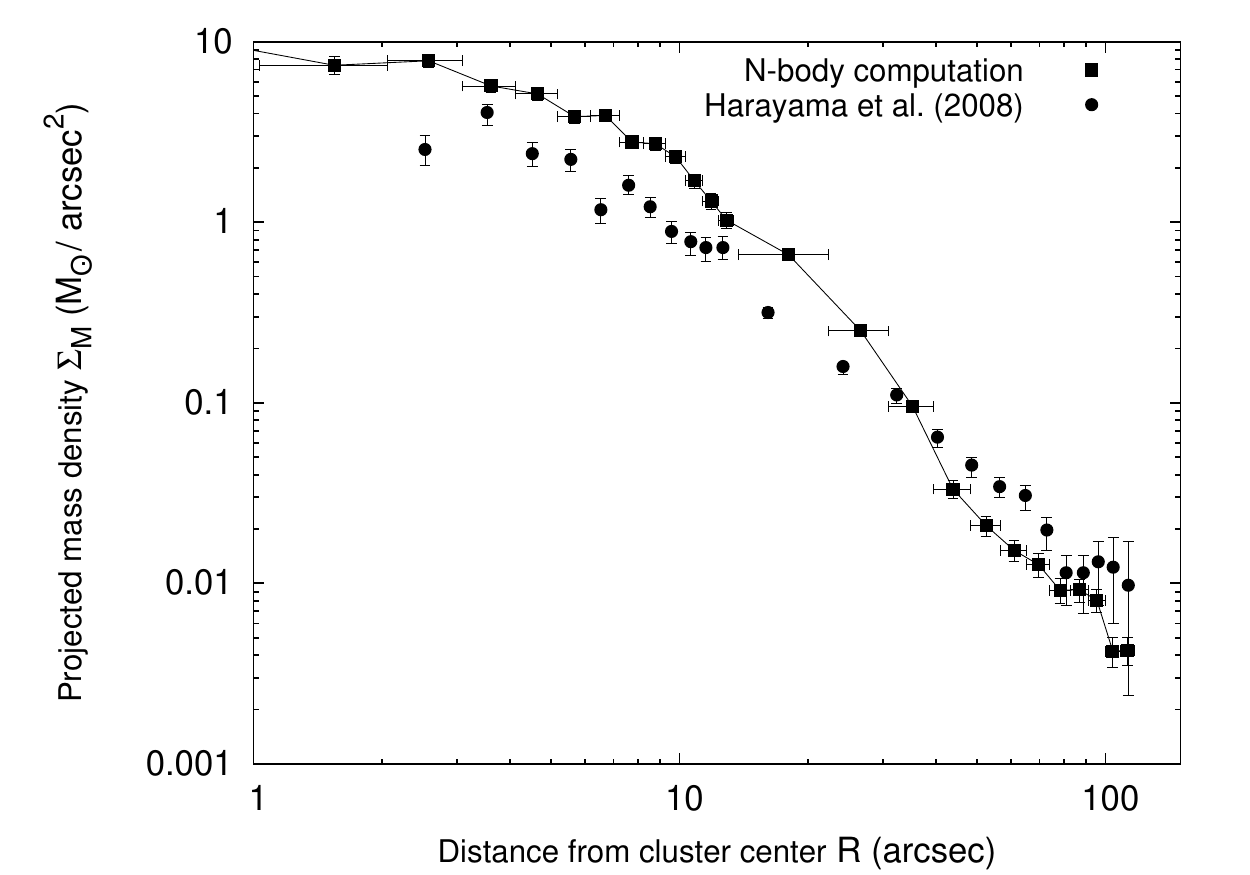}
\includegraphics[width=5.7cm, angle=0]{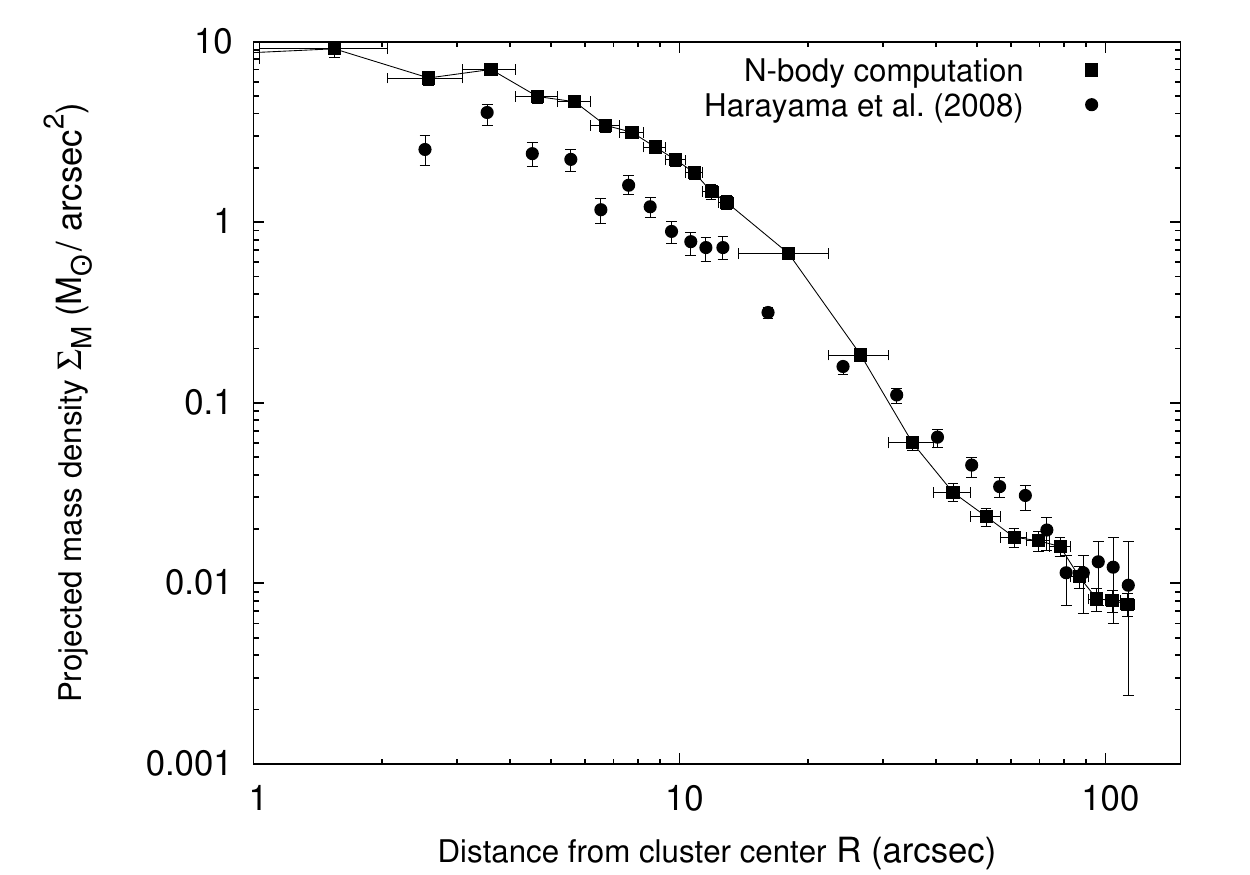}
\includegraphics[width=5.7cm, angle=0]{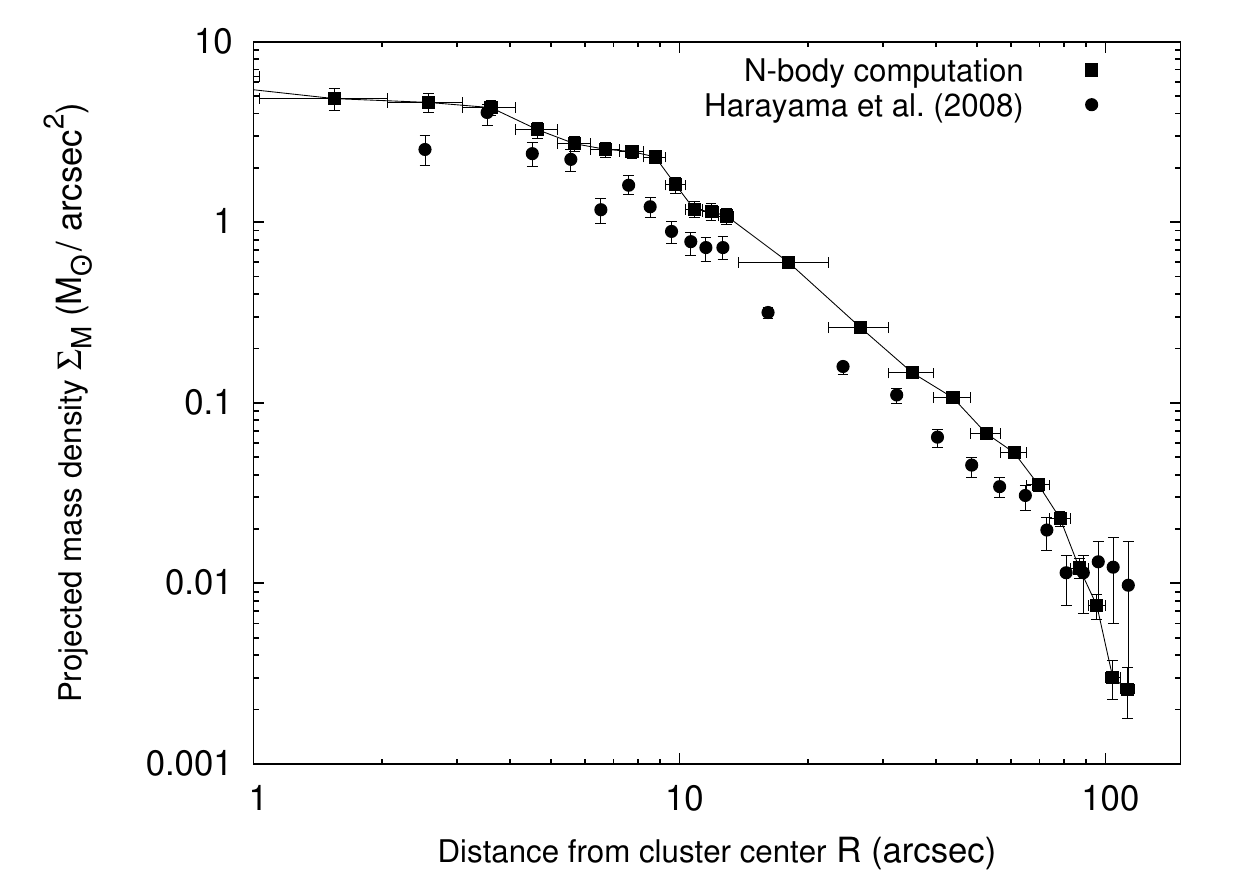}
\caption{
Examples of surface mass-density profiles at $t\approx1$ Myr for those
computed configurations (filled squares connected with solid line)
which evolve to form a star cluster with near-spherical core-halo structure
(the CH-type morphology; see Table~\ref{tab:morphdef}) within $t<1$ Myr,
in absence of a background gas potential.
These computed profiles are significantly more
compact and centrally overdense
than that observed in HD97950 (\citealt{hara2008}; filled circles).
For HD97950 $1''\approx0.03$ pc.
This panels are reproduced from \citet{sb2014b}.
}
\label{fig:dryprofs_1myr}
\end{figure}

\begin{figure}
\centering
\includegraphics[width=5.7cm, angle=0]{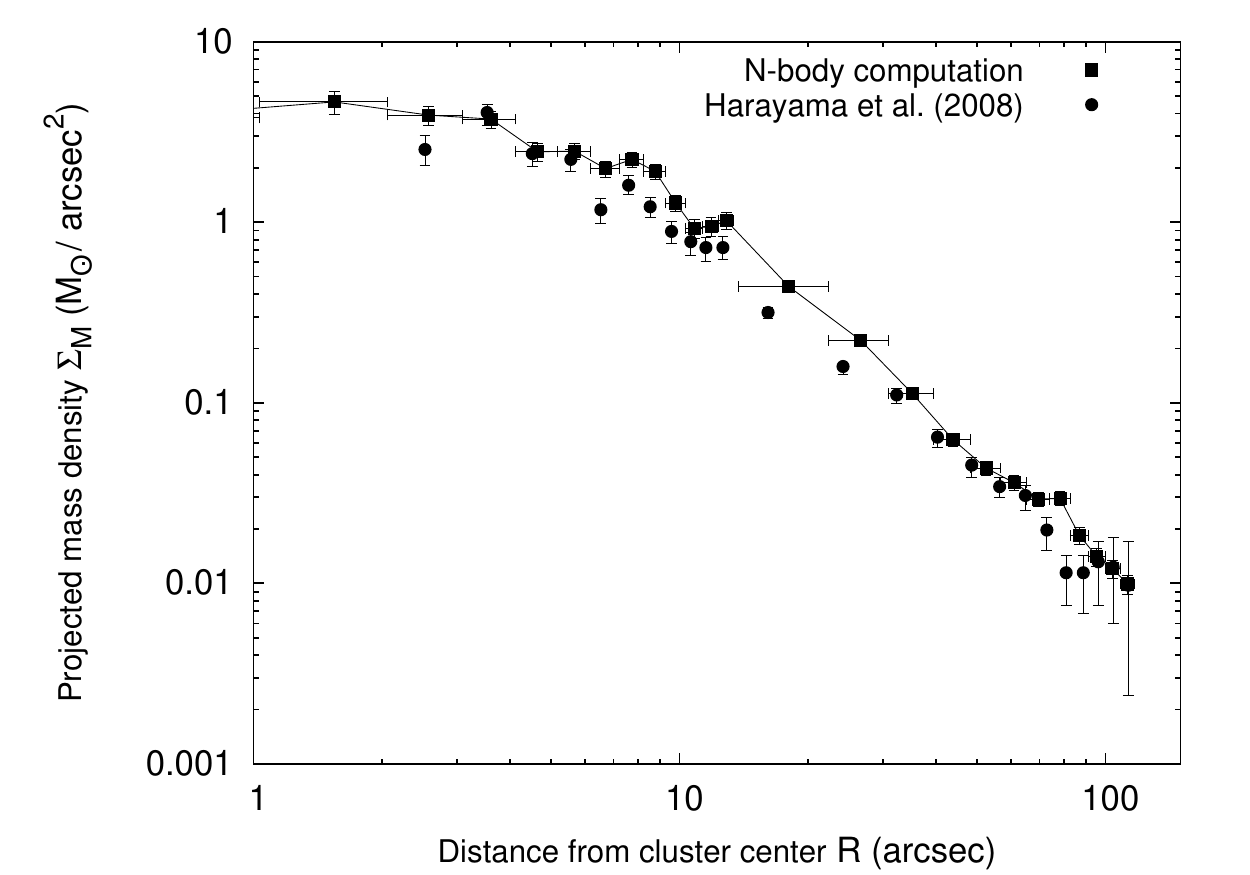}
\includegraphics[width=5.7cm, angle=0]{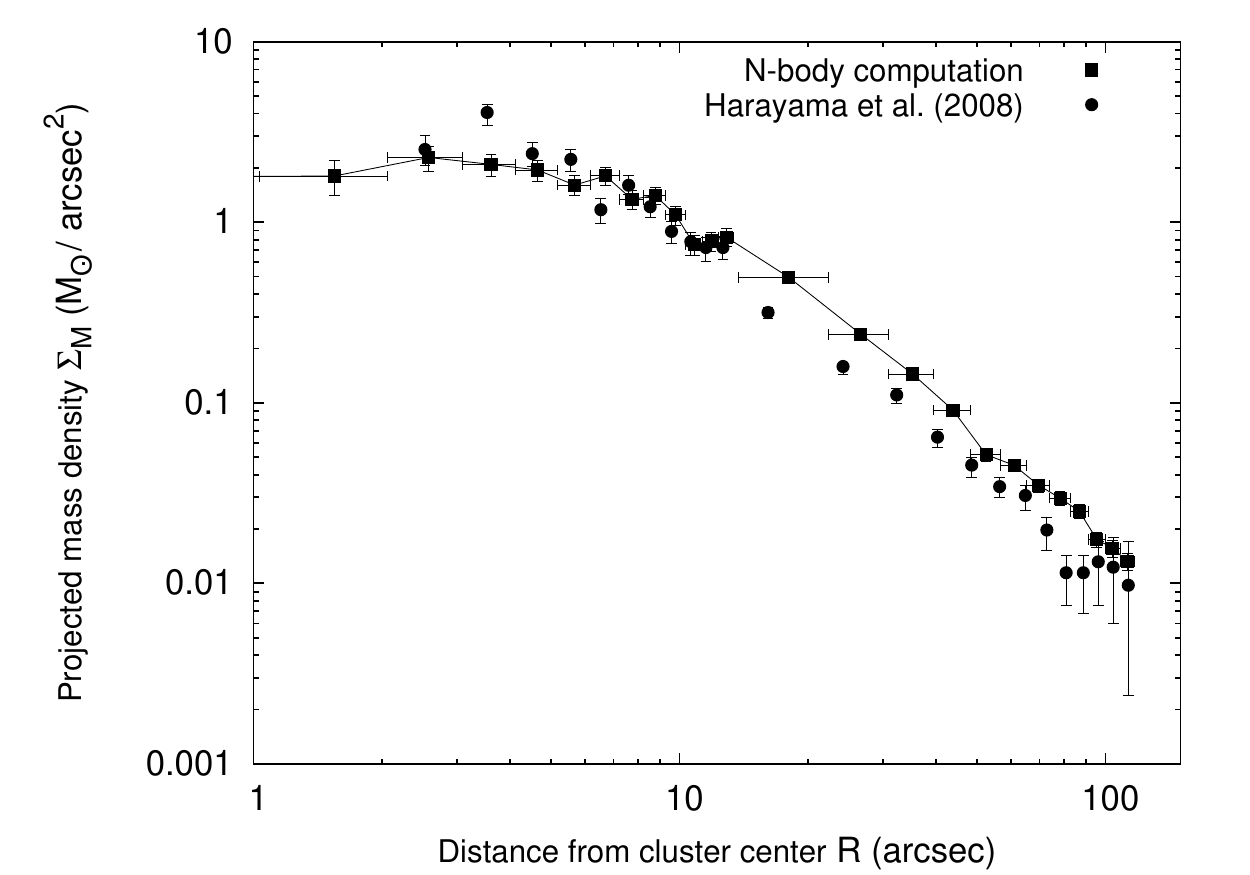}
\includegraphics[width=5.7cm, angle=0]{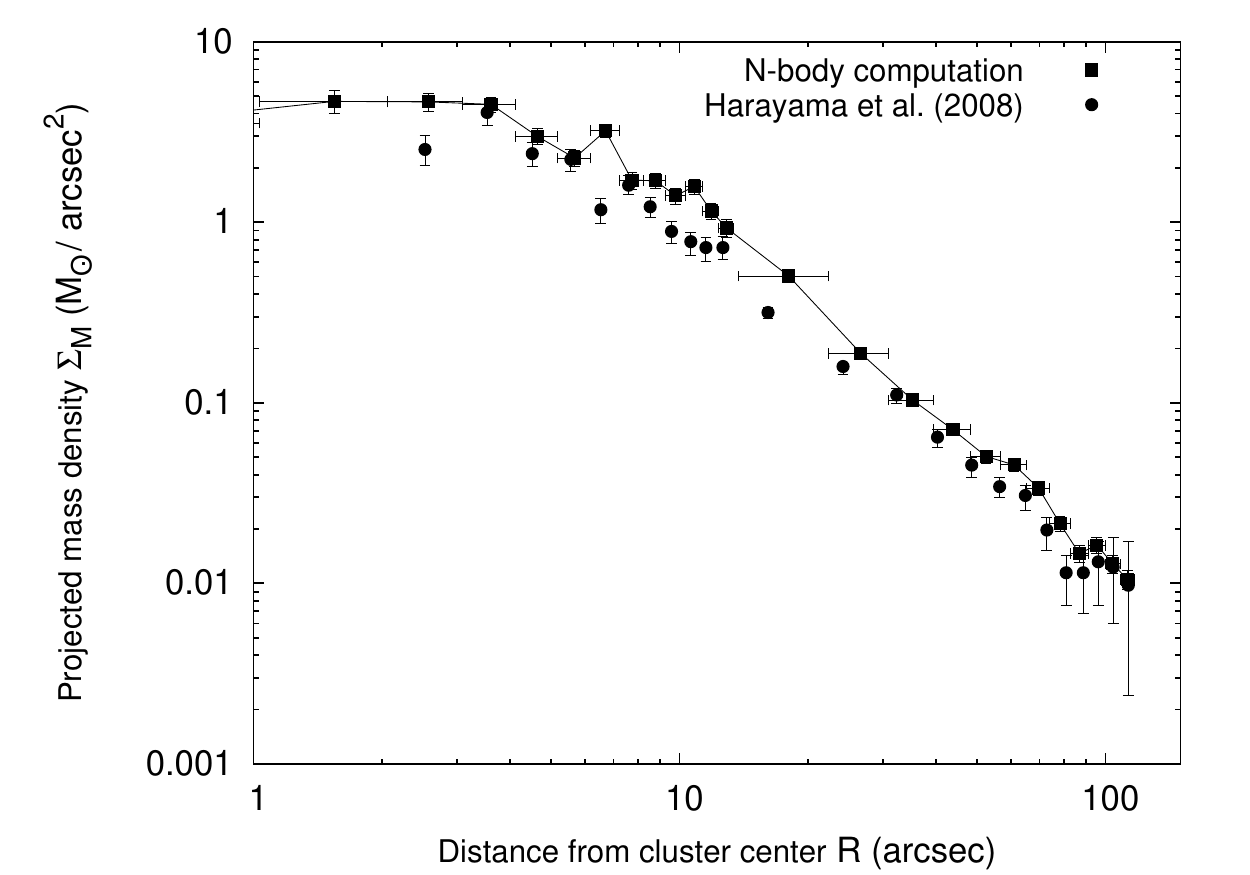}
\includegraphics[width=5.7cm, angle=0]{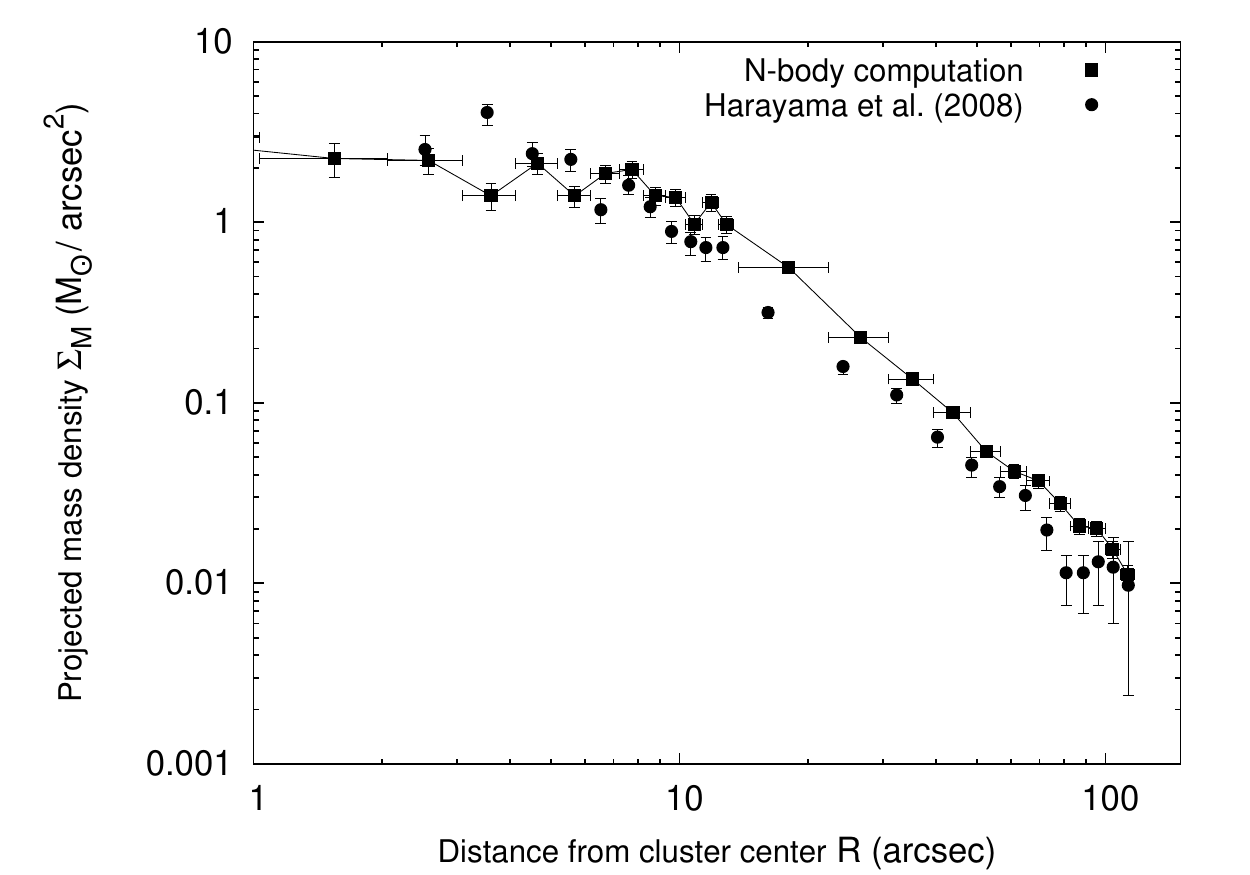}
\caption{
Examples of surface mass-density profiles at $t\approx1$ Myr for computed
post-gas-expulsion configurations.
Here, the systems evolve in a background residual gas potential (see Sec.~\ref{himrg})
to form a star cluster with near-spherical core-halo structure
(the CH-type morphology; see Table~\ref{tab:morphdef})
within $t<1$ Myr followed by residual gas dispersal at
$\td\approx0.6$ Myr. The legends are the same
as in Fig.~\ref{fig:dryprofs_1myr}. 
All these computed profiles agree well with the observed one for HD97950.
These panels are reproduced from \citet{sb2014b}. 
}
\label{fig:GPprofs_1myr}
\end{figure}

To assemble a HD97950-like star cluster by hierarchical merging of subclusters, the necessary
but not sufficient condition is to arrive at a CH configuration in $t\lesssim1{\rm~Myr}$.
The above calculations imply that
to have a CH morphology at 1 Myr, one should have $R_0\lesssim2{\rm~pc}$ 
with or without a gas potential (\cf Fig.~\ref{fig:morphevol}).
As discussed in Sec.~\ref{monoreason}, the gas potential would actually delay the approach to a
CH configuration. How does this final cluster compare with the observed HD97950 cluster?

All the configurations, \emph{without} the gas potential,
that become CH in $t\lesssim1{\rm~Myr}$ are found to form clusters that are
much more dense and compact compared to the observed HD97950 profile \citep{hara2008}    
at $t\approx1$ Myr. This is found to be true irrespective of the
mode of subdivision of the initial stellar mass $\mstar$ (\cf Table~\ref{tab:morphevol}).
The latter fact can be expected from Eqn.~\ref{eq:Rmrg}.
This is demonstrated in Fig.~\ref{fig:dryprofs_1myr}. Note that
all these calculations are for $M_\ast\approx10000\Ms$
which is the lower mass limit of HD97950. For larger
$\mstar$, the assembled system will be even more overdense    
since the length scale of the final merged cluster, 
$R_\ast$, is nearly independent of the total stellar mass $M_\ast$ (\cf Eqn.~\ref{eq:Rmrg}).
From test calculations, it is also found that the ``heating effect''
of primordial binaries and mass loss due to stellar winds 
do \emph{not} expand and dilute the merged cluster's center
sufficiently.  

One way to dramatically expand a star cluster, however,
is to subject it to a substantial gas expulsion
on a timescale of the order of its dynamical time, as discussed 
in the above sections. Fig.~\ref{fig:GPprofs_1myr} shows the computed
stellar mass-density profiles at $t\approx1$ Myr for
similar calculations but including gas expulsion with parameters as
discussed in Sec.~\ref{params}. As in Sec.~\ref{ngc3603}, they agree reasonably
with the observed profile of HD97950 \citep{hara2008}, particularly in the inner regions.
Note that in Fig.~\ref{fig:GPprofs_1myr}, the ``natural'' matchings
with the observed profile are obtained
\emph{by simply overlaying it with the computed profiles at 1 Myr
without any scaling or fitting}, as in \citet{sb2014}. The King-fit
parameters to the observed and the computed profiles are also found to
agree fairly.  

We can now summarize the key inferences in the
study discussed in this section, in logical sequence, as follows:

\begin{itemize}

\item A system of subclusters of total stellar mass $M_\ast\approx10^4\Ms$
assemble into a (near) spherical core-halo
star cluster by the age of HD97950 (\ie, in $t<1$ Myr) provided these
subclusters are largely
born over a region of scale length more compact than $R_0\lesssim2$ pc.
This can happen, \eg, in an intense starburst
event at a dense ``spot'' in a molecular cloud.

\item The initial
sizes of the subclusters are constrained by the
compact sections of molecular gas filaments or filament
junctions which, in turn, determines the compactness of the final assembled cluster.
Therefore, the mass density
over the central region (within a virial radius) of the merged cluster
is determined by the total stellar mass that is involved in its assembly.

\item A ``dry'' merger of subclusters, \ie, infall in absence of any residual molecular gas
(all gas consumed into stars)
always leads to a star cluster that is centrally overdense w.r.t. HD97950,
even for the observed lower mass
limit $M_\ast\approx10^4\Ms$. This holds irrespective of the initial mode of subclustering.

\item A substantial residual gas expulsion ($\approx70$\%) after the formation of the merged system
expands the latter to obtain a cluster profile
that is consistent with the observed HD97950. With the lower stellar mass limit $M_\ast\approx10^4\Ms$
and an SFE of $\epsilon\approx30$\%, the observed surface mass density profile of HD97950
can be fairly and optimally reproduced.
 
\end{itemize}

Notably, recent multi-wavelength observations of the Pismis 24 cluster of NGC 6357
\citep{massi2014} indicate that this cluster (age 1-3 Myr)
contains distinct substructures
which must have formed out of dense gas clumps packed within $\approx1$ pc radius.
A similarly close-packed stellar substructures are found in the W3 complex \citep{rz2015}. 
\citet{jah2014} also find that the stellar distribution in
young clusters (1-3 Myr) tend to smoothen out with age and local stellar density.
This indicates an appearance of these systems as closely packed stellar
overdensities which disappear on a dynamical timescale as seen above.

In the context of any scenario that involves infall and merger of subclusters
in their parent gas cloud, it is important to keep in mind that the \emph{effective} SFE at the
location of the merger (deepest part of the potential well, see above) can be
much higher than 30\%, depending on the stellar concentration in
the newly merged cluster. This is true for the calculations presented in this
section also. Note that this enhanced SFE is essentially a ``population effect''
and does not reflect that with which the star formation has actually taken place
(\ie, SFE within the gas clump(s)).
The latter would be much smaller; $\lesssim30$\% (see Sec.~\ref{params}).      

\begin{framed}
NGC 3603 young cluster (HD97950) has formed essentially monolithically
followed by a substantial and violent gas dispersal.
The initial monolithic stellar distribution has either formed in situ or has been assembled
``promptly'' (in $\lesssim1$ Myr) from closely packed (within $\lesssim2$ pc)
less massive stellar clusters (subclusters).
Both scenarios are consistent with the formation of HD97950's entire stellar population in a single starburst
of very short ($\lesssim10^5$ Myr) duration.
\end{framed}

\section{Globular clusters and the stellar IMF}\label{gcimf}

The Milky Way (hereafter MW) contains approximately 160 globular clusters
(hereafter GCs; see \citealt{1996AJ....112.1487H} for a compilation)
most of which are nearly as old as the Universe. A major hot topic in astrophysics
is to understand their birth and initial conditions. Here a discussion is provided which is
consistent with and which is also based
on the information gleaned from VYMCs as discussed in the previous sections.

Globular clusters appear to form a separate population of star clusters from those discussed above.
However, \citet{2002IAUS..207..421L} has shown that star cluster formation extends from low stellar
masses, $\mcl\lesssim10^3\Ms$, to the most massive very young clusters observed in the nearby
universe in interacting galaxies ($\mcl\gtrsim10^5\Ms$) without a detectable change in their
distribution with luminosity. It appears that star cluster formation is a continuous process
in cluster mass, and that this mass distribution extends to high, GC-type masses in galaxies with high
star formation rates (SFRs; \citealt{2004MNRAS.350.1503W,2013ApJ...775L..38R}).
However, three mass ranges are evident within each of which generically different physical processes
play a role when clusters form \citep{2002MNRAS.336.1188K}: low mass clusters ($<$ few $10^2\Ms$ in stars)
do not have O-stars \citep{wp2013} such that they are more likely to loose their residual gas
adiabatically. Intermediate-mass clusters (few $10^2\Ms$ to $\approx10^5\Ms$) contain one to many
O-stars \citep{wp2013} which photoionize the residual gas and expel it probably explosively with 
a disruptive effect on the stellar component, while very massive clusters ($M>10^5\Ms$) have,
with increasing mass, an increasingly deep potential such that even photo-ionized plasma may
not be able to leave the cluster within many initial crossing times \citep{2008MNRAS.384.1231B}.
These three regimes lead to different reactions of the clusters to the blow-out of the residual
gas, such that an initially power-law mass function of embedded clusters evolves to a form with
a turnover or flattening near $10^5\Ms$. Globular cluster-mass young clusters survive with a larger
fraction of their stars and do not dissolve within a Hubble time
\citep{1997ApJ...480..235E,2002MNRAS.336.1188K,2008MNRAS.384.1231B}.

Such GC-precursors would have formed with very high densities, as the study by
\citep{2010MNRAS.406.2000M} suggests. This work is based on inferring the initial
conditions of GCs subject to the constraint that their IMF for stars
$\lesssim 1\Ms$ is canonical. Since GCs with a low concentration are
observed to have a deficit of low-mass stars \citep{demarch2007}, and because neither a theoretical
explanation through star formation is evident for this observation nor can it be
explained with secular cluster evolution (fig.4 in \citealt{2013MNRAS.436.3399L}),
it is possible
to infer that GCs may have formed mass segregated and increasingly compact with decreasing
metallicity and that their initial radii are broadly consistent with Eqn.~\ref{eq:mrk}.

\begin{figure}
\centering
\includegraphics[width=9.5cm, angle=0]{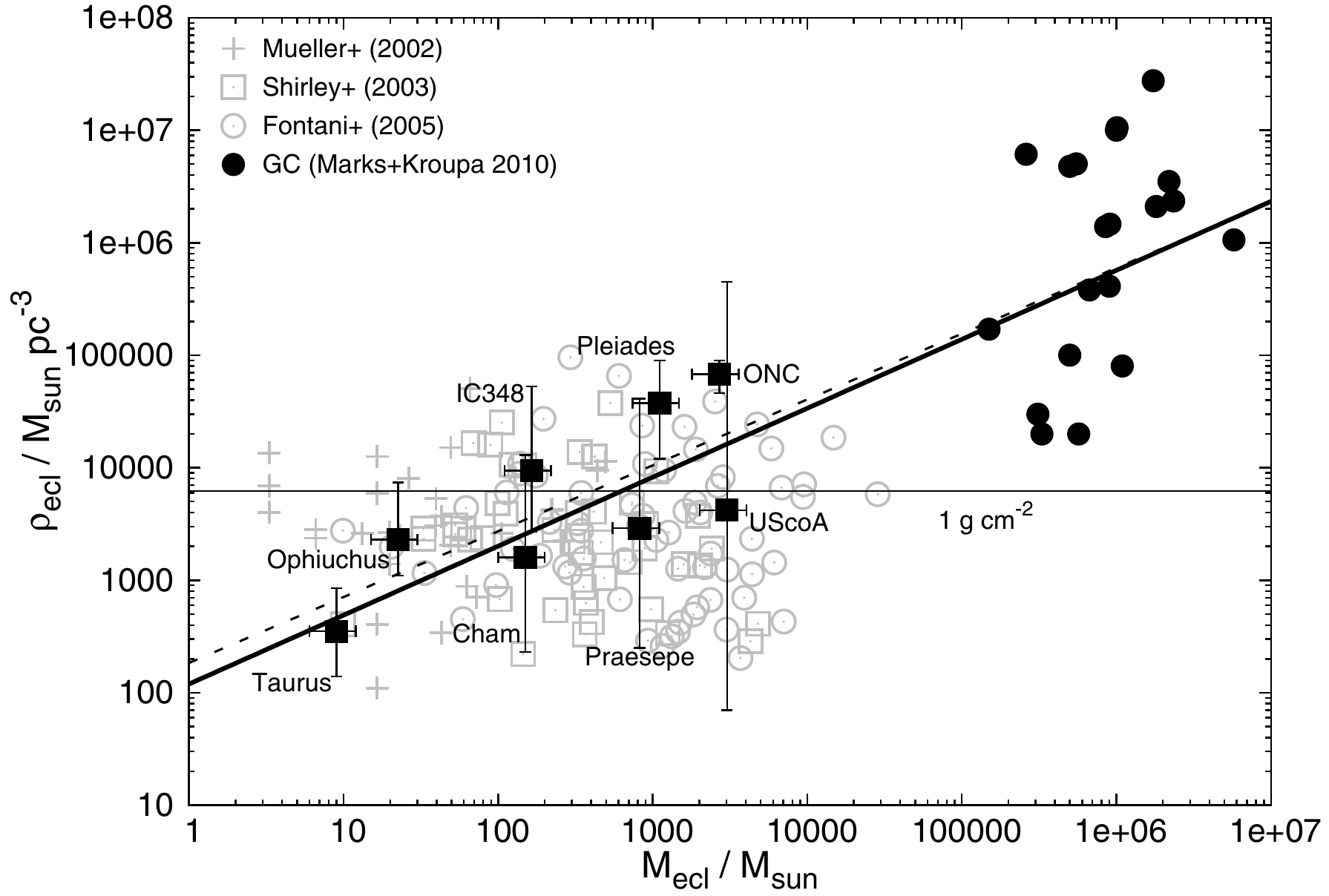}
\caption{
Constraints on the initial volume-densities within the half-mass radius
for the seven clusters (filled squares) versus the initial stellar mass as
obtained by \citet{mk2012}. The indicated errors in mass correspond to the
observationally inferred present-day mass on the left end of a bar and
two times the present-day mass on its right end, to be understood as an
estimator of the possible range of initial mass. The filled circles are Galactic GCs.
Underlaid as grey symbols are data of molecular cloud clumps
as collated by \citet{park2011}.
The clump masses are multiplied with a SFE of 1/3 to be comparable
with the stellar masses and densities inferred in \citet{mk2012}.
The thin solid line is the threshold for massive star formation \citep{krmac2008}.
The thick solid line is a least-squares fit to both the young cluster and GC values,
implying that there exists a mass-radius relation for star cluster-forming cloud clumps.
The dashed line shows the same when the GCs are excluded from the fit.
This figure is reproduced from \citet{mk2012}.
}
\label{fig:initdens}
\end{figure}

Consistency is also found independently by the first-ever N-body
computation of two low-concentration GCs by
\citet{2011MNRAS.411.1989Z,2014MNRAS.440.3172Z}.
This work confirms that such GCs must have formed mass segregated and must have lost
a substantial part of their low-mass stellar population during emergence from their
embedded state. This leads to an understanding of the star-formation events
within $<1$ Gyr, as the first proto-Galactic gas cloud collapsed to form the
MW population~II halo and its associated present-day GCs at a time when the
MW did not exist as the Galaxy \citep{2010MNRAS.406.2000M}.

But the results also imply that in order for the young GCs to emerge with a ``damaged''
(low-mass depleted) stellar mass function and low concentration,
the IMF may have been increasingly
top heavy with increasing density and decreasing metallicity \citep{2012MNRAS.422.2246M}.
The remarkable result here is that these dependencies are well-consistent with the
entirely independent results on the dependence of the IMF with density from the dynamical
mass-to-light ratios and, independently, from the X-ray-source  population of ultracompact
dwarf galaxies (UCDs) by \citet{2009MNRAS.394.1529D,2010MNRAS.403.1054D,2012ApJ...747...72D}.
According to these results,
the stellar IMF is canonical for SFR densities ${\rm SFRD}<0.1\Ms/({\rm pc}^3\,{\rm yr})$
and becomes increasingly top-heavy for stellar mass $\gtrsim1\Ms$ with increasing
density and decreasing metallicity (see eqn. 4.65 in \citealt{pk2013}).
These constraints on the
variation of the stellar IMF with the physical conditions of star-forming cloud cores on scales
of a parsec and time-scales of a Myr yield dependencies of M/L
ratios of stellar populations which
appear to be consistent with the observed values for
the GCs of the Andromeda \citep{2011AJ....142....8S},
as well as the high rate of type~II supernovae in ULIRGs \citep{2012ApJ...747...72D},
as well as with the observed top-heavy galaxy-wide IMFs in star-forming galaxies within
the IGIMF theory \citep{2011MNRAS.415.1647G,2013MNRAS.436.3309W}.
\footnote{From purely dynamical considerations, \citet{sbpk2012} also infer
that VYMCs, in particular R136, should have born with the massive end
of their IMF top-heavy. Continued dynamical interactions, resulting in mergers
among the most massive stellar members in a massive cluster like R136 ($\approx10^5\Ms$),
can make the stellar mass function top-heavy even at the present day \citep{sb2012b}.}
\citet{leigh2015} has demonstrated that the initial population
of binary stars in very young GCs must have been very similar to the presently occurring
binary population in the Milky Way, and the currently available result that the stellar IMF
below a few $\Ms$ was also largely canonical \citep{2012MNRAS.422.2246M}, is consistent with
this result and thus with the notion of an inherently largely universal process and
outcome of star formation.

The generic formation of extremely massive, GC-like very young clusters is thus being
understood increasingly better, but the detailed physical processes within these extremely
dense (see Fig.~\ref{fig:initdens})
star-burst clusters remain a subject of intense study with major unsolved problems.
In particular, the processes acting during the first few Myr in the
highly dense plasma-star mixture, which may be pressurized from the ambient interstellar
medium (ISM) in strongly star-bursting pre-galactic environments or in interacting galaxies,
remains largely not understood (\eg, \citealt{2013A&A...552A.121K}).
The fact that increasingly massive
GCs show evidence for increasingly complex metal abundance anti-correlations and spreads are
rather certainly due to these extreme physical conditions
(\citealt{2009MNRAS.396.1075G,2014AA...569L...6C};
also see \citealt{2014ApJ...789...88J} and references therein).
But even the Hubble-time long dynamical evolution of GCs with a significant initial binary population
may lead to hitherto not appreciated effects on the present-day chemical
properties of GC stars \citep{2014ApJ...789...88J}, while even intermediate-mass clusters and
in particular very massive clusters may re-accrete ISM into their
potential wells and this may occur repeatedly depending on the orbits of the clusters
and of the ISM \citep{pflm2009}.

\section{Concluding remarks: embedded vs. exposed young clusters}\label{discuss}

How VYMCs form is still an open question and even may not have a unique answer.
As seen above, there are two primary directions of research that seek answer(s)
to this question. The most physically detailed hydrodynamic calculations so far
(including feedback; see Sec.~\ref{intro}), that led to star formation, currently range
from a single proto-star to about $50\Ms$ gas spheres, \ie, less massive than
the heaviest stellar member found in VYMCs (typically $>100\Ms$). Scaling the
inferences all the way to the VYMC mass $(>10^4\Ms)$ and size $(\approx {\rm~pc})$ is grossly
unreliable since the physical behaviour of the feedback processes
(radiation, magnetic field) are scarcely understood.

On the other hand, dynamical modelling of monolithic stellar systems,
using direct N-body calculations can
reach up to several $10^5\Ms$, with the current technology. As seen in the
above sections, the main drawback in such studies is the simplified treatment
of the residual gas (see Sec.~\ref{gasmodel}). It is impressive, though,
that such an approach remarkably reproduces the detailed observed properties
of several young star clusters (see Sec.~\ref{match}).
This is also true if VYMCs (and young clusters in general)
can be considered as formed via
``prompt merging'' of smaller subclusters in a background gas potential
(see Sec.~\ref{himrg}). In other words, a 1-3 Myr old VYMC  
can as well form from a highly subclustered stellar
distribution, far from having spherical symmetry, provided
the dynamical time (infall + violent relaxation;
see Sec.~\ref{monoreason}) for the initial stellar system is
sufficiently small. Note that the studies mentioned in
Secs.~\ref{match} and \ref{himrg} are currently the only theoretical
studies in the massive cluster scale that provide direct and detailed agreements
with observed clusters. 

These results imply that
once ``blasted off'', the details of the gas hydrodynamics do not critically influence
the dynamical evolution of the stellar system as long as cluster ages
of the order of Myr are concerned. The key point is whether a rapid gas dispersal,
with time-scale of the order of stellar crossing times (\cf Table~\ref{tab1}),
should always occur in reality. The issue arises since compact
embedded systems of several Myr age appear to be found throughout our Galaxy.
An interesting example, in this respect, is the embedded W3 Main (hereafter W3M) cluster
\citep{bik2014,fglsn2008}.
Located $\approx2$ kpc from the Sun, this is the most embedded region of the
W3/W4/W5 star forming complex in the outer Galaxy, containing hyper-compact
to extended HII regions.
From the age estimates of the most massive (O-)star (IRS2), it
can be inferred that W3M is forming stars since 2-3 Myr
at least \citep{bik2014}. It is argued that triggering as well as
dynamical effects might keep the star formation
ongoing in this system \citep{fglsn2008}. In particular, there are
observational evidences that W3M is triggered externally
by an expanding bubble from an OB association \citep{oey2005}
and also internally by swept away material from OB stars \citep{wang2012}.
There are several similar-aged embedded
clusters (\eg, W33 complex, \citealt{messi2015}),
but W3M is among the most well studied ones. 

The existence of such systems appear to imply that not all forming stellar
assemblies expel their residual gas promptly in an ``explosive''
\citep{pk2005} manner. For
those cases this does happen, a VYMC like NGC 3603 is born, otherwise
one is left with a ``VYMC-aged'' embedded system like W3M.
For the latter case, the SFE
would keep increasing with time as an increasing amount of gas is converted into and/or accreted
onto proto-stars. It is currently unclear which physical processes (or absence of them)
would delay the prompt gas expulsion. It could be the inherent properties of the natal gas
(\eg, its chemical abundances) and as well the surroundings that determines
the fate of the star-forming clump.
The presence or absence of ionizing O-stars
could have been a plausible determining factor at a first glance,
but several OB stars are found in W3M and other embedded clusters.
The key question is as follows:
\emph{Despite the presence of ionizing OB stars in either case, why are some
young star clusters (nearly) gas-free at an age of a few Myr and why are
some others apparently deeply embedded in molecular gas at similar ages?}

For an explosive gas expulsion, the (compact) HII regions should rapidly
engulf the densest (most populous) parts of the molecular clouds
within their lifetimes; typically $<1$ Myr (see Sec.~\ref{params};
\citealt{sb2013}). 
Hence, the embedded proto-stellar population (or a
significant part of it) should form in one episode (\ie, with a small age spread)
and within a compact enough region of the molecular cloud,
typically $<1$ pc (see Sec.~\ref{params}; \citealt{sb2013}). 
The propagation of the (overlapped) HII region can be additionally powered by
simultaneous presence of OB stars over a small region, further
supporting compact and episodic star formation for VYMCs. An interesting
example in this regard is the RCW 38 cluster \citep{derose2009,kuhn2014}, which is as compact
as $\approx0.1$ pc (Eric Feigelson: private communication) and where
only the central region is gas free, indicating that the residual gas might have just
begun being expelled from the cluster. Indeed,
as seen in Sec.~\ref{match}, N-body computations that reproduce present-day
VYMCs begin from such initial conditions. The stellar distribution in W3M,
on the other hand, is more extended, $>1$ pc, and contains
several compact HII zones \citep{bik2014}. This may also
indicate that this region has formed a number of compact embedded clusters
containing O/B-stars (see, \eg, \citealt{testi1999}).
In future, more detailed observations will help to
illuminate this possibility. Improved age estimates of the stellar
population in W3M will be particularly important on this regard.

In other words, the fate of
a newly hatched cluster might be governed by its scale length at birth.
If the embedded stellar distribution is sufficiently compact (\eg, if formed
at a molecular gas filament junction; see Sec.~\ref{params}), the natal
gas can be blown apart when the UCHII region(s) engulfs (and hence ionizes)
the scale length \citep{chrch2002,krm2009,sb2014}.
On the other hand, if the initial proto-stars are more widely
distributed the gas may not be efficiently expelled and one is left with
an embedded cluster. Of course, scale length may not be the only factor
that determines if the gas expulsion ``fails''. The gas expulsion can
fail if the propagation of the HII region from around OB stars is stalled
for any reason so that the embedding gas is ionized at most locally.
Of course, if the newborn assembly is not massive enough so that no
OB stars are formed, the gas would not be expelled even if the assembly
is highly compact. This could be the case with several compact, low-mass
embedded clusters (see, \eg, \citealt{tapia2011}). Note that
a bimodal regime of cluster formation has been proposed earlier
by \citet{bokr2002} but in a somewhat different context.

If the fate of a newborn stellar assembly happens to depend on its length scale,
a possibility as discussed above, the wider embedded systems can be expected to
contain substructures. This is because, as seen in Sec.~\ref{himrg},
the lifetime until which the substructures are erased to form a stellar
cluster in dynamical equilibrium depends on the initial spatial span of
the stellar distribution. This is consistent with the fact that substructures
are found in many embedded clusters. If the embedded phase continues
for a sufficiently long time the stellar system can virialize while
remaining deeply embedded, as found recently \citep{foster2014}. On the other hand,
for a sufficiently extended system, the substructure can remain even
if the gas is largely dispersed, \eg, as seen in Cygnus OB2 \citep{wrt2014}. 

Of course, it is important to re-consider the observationally
inferred ages of the stellar members of the
W3 Main and other embedded clusters \emph{vis-\'a-vis} their distances, since
age plays a crucial role in inferences such as above. This is
true for gas-free VYMCs as well. For embedded clusters, the age
estimates can be improved through even deeper and highly resolved
observations, say, using \emph{ALMA}. Such observations of
dense molecular filaments and their junctions (or other sub-parsec
scale dense structures) would reveal embedded clusters in them (or
lack of them). For VYMCs, better parallax measurements with, \eg,
\emph{Gaia} would provide independent distance measurements that
would improve their age estimates. It is as well important to better
establish the depth of embedding of W3 Main and other embedded systems,
in particular, whether the stellar system is actually embedded
in gas or if it is a gas-free system inside or aside a molecular cloud, like, \eg,
the ONC and NGC 3603. Increased detection of HII regions and as well mapping
the gas velocities over and surrounding embedded clusters would
help to resolve this. For lightly-embedded and exposed star clusters
as well as for embedded clusters,
ongoing surveys such as the \emph{MYStIX}
(Massive Young Star-Forming Complex Study in Infrared and X-ray; \citealt{fglsn2013})
would improve their age estimates, where X-ray observations
provide additional constraints on the individual stellar ages \citep{gtm2014}. 
More detailed observations of both embedded
and exposed clusters is the key to answer the fundamental
question raised above, without which no concrete
conclusions regarding the differences between embedded and
exposed young clusters can be drawn. 

\begin{framed}
The relation between exposed and embedded young clusters is currently unclear;
the length scale over which the stellar population is hatched might play a role
in determining whether the natal
gas is expelled early (in $<1$ Myr) or the stellar population remains
embedded (slow gas dispersal and/or ``gas consumption''). 
The detailed physical processes responsible for failing gas blow out
in presence of the ionizing OB stars remain unclear. More detailed
observations of embedded clusters like W3 Main is necessary to resolve
this.
\end{framed}

In summary, star formation is an intrinsically spatially and temporally correlated
process which is evident to the astronomer as embedded clusters.
The star formation process can be viewed as leading to a continuous distribution of
embedded cluster masses where the most massive clusters, that can form, are constrained
by the galaxy-wide SFR which, in turn, is controlled by the depth of the potential of
the galaxy and thus the pressure in the turbulent interstellar medium
\citep{1997ApJ...480..235E,2001ApJ...556..837K,2011MNRAS.410.2339B}.
VYMCs are a particularly shining part of this range of events, and GCs are the 
evolutionary fate of the most
extreme VYMCs. The physics of the formation and of the emergence of VYMCs from their
natal clouds carries through to the formation of GCs, but the extreme densities
involved for them pose new and largely not-at-present-understood challenges.
It is therefore, broadly speaking, not surprising that GCs show complex properties
which are not evident in VYMCs. Finally, it is thanks to the continued decade-long
algorithmic and mathematical progress achieved by Sverre Aarseth and Seppo Mikkola
that the astrophysical community has now access to realistic
N-body codes which allow us to address the issues discussed here. With the continuously
improving and universally adaptable software platform like the ``AMUSE'' \citep{pz2008},
such N-body calculations can be made even more realistic (\eg, introduce full
hydrodynamical treatment) in foreseeable future.

\end{document}